\documentclass{eas}
\usepackage{graphicx}
\usepackage[authoryear]{natbib} 
\usepackage{times}
\usepackage{xspace}
\usepackage{epsfig}
\usepackage{natbib}
\usepackage{rotating}
\usepackage{dcolumn}
\usepackage{txfonts}
\usepackage{cite} 
\usepackage{graphicx}
\usepackage{lscape}
\usepackage{txfonts}
\begin{document}

%%-----------------------------
%%      the top matter
%%-----------------------------
\title{Clusters: age scales for stellar physics} 
\author{David Barrado}\address{Depto. Astrof\'{\i}sica, Centro de Astrobiolog\'{\i}a (INTA-CSIC),  ESAC campus,
  Camino Bajo del Castillo s/n,
  E-28692 Villanueva de la Ca\~nada, Spain }
%\author{...}\address{...}
%\author{...}\address{...}
%
%
\begin{abstract}
Ages are key to truly understand  a large plethora of astrophysical phenomena.
On the other hand, stellar clusters are open windows to understand stellar evolution, specifically, the change with time and mass of different stellar properties. As such, they are our laboratories where different theories can be tested, but without accurate ages, our 
knowledge would impaired. We revisit here a large number of age-dating techniques and discuss their advantages and draw-backs. In addition, a step-by step process is suggested in order to built a coherent age scale ladder, minimizing the error budget and the sources of uncertainty.
\end{abstract}
\maketitle
%%-----------------------------
%%      your text
%%-----------------------------

%%%%%%%%%%%%%%%%%%%%%%%%%%%%%%%%%%%%%%%%%%%%%%%%%%%%%%%%%%%%%%
%%%%%%%%%%%%%%%%%%%%%%%%%%%%%%%%%%%%%%%%%%%%%%%%%%%%%%%%%%%%%%
%%%%%%%%%%%%%%%%%%%%%%%%%%%%%%%%%%%%%%%%%%%%%%%%%%%%%%%%%%%%%%
%%%%%%%%%%%%%%%%%%%%%%%%%%%%%%%%%%%%%%%%%%%%%%%%%%%%%%%%%%%%%%
%%%%%%%%%%%%%%%%%%%%%%%%%%%%%%%%%%%%%%%%%%%%%%%%%%%%%%%%%%%%%%
\section{Introduction}\label{sect:intro}

This lecture was imparted during several days at the ``Ecole Evry Schatzman'', which with the title ``Stellar Clusters: benchmarks of stellar physics and galactic evolution'', was held on  Banyuls sur Mer, France, on  4-9 October 2015.
The rationale of the schools declared: 
``{\it Clusters are also cornerstones for understanding stellar physics and for constraining increasingly sophisticated models of stellar structure and evolution. In the modeling, it is becoming possible and necessary  to account for various processes related to interactions between neighbor stars and with their environment (disc, planets, interstellar medium). These interactions, which strongly depend on the cluster density, affect the early evolution of stars by modifying their mass during accretion phases, their multiplicity rate, their disc, and, in fine, planetary formation and the initial mass function.}'' 
Underlying these properties were a great unknown: stellar ages, which obviously are essential to understand the evolution of any stellar property.

The age scale (or better, age scales, in plural) is very important for different reasons and we will try to revisit these  here. But to illustrate how relevant it is, we can look back to the history of science, in particular to the XIX century. It was assumed that the known, conventional energy sources could not account for the solar output. The Kelvin-Helmholtz mechanism, based on gravitational contraction, proposed by William Thomson and Herman von Helmholtz, could only account by  about 22 Myr. However,  geology imposed some limits since it was already known that Earth was much older, and the energy  emission rate of the Sun was therefore not sustainable. Therefore, some new mechanism laid behind.  What physicists did not know was that  complete new branch of physics was at hand: relativity and quantum mechanics. Eventually the nuclear reactions were identified as the culprit (\citealt{Bethe1939-EnergyStars}). This is an important lesson: pieces which do not match  in this enormous puzzle (our understanding of natural  phenomena) do provide, eventually, very interesting hints for new avenues and discoveries. Age and related issues belong to this category.

In the recent years the age problem has summarized  quite comprehensively. To name a few,
 \citet{Mamajek2008-Ages} --the result of a splinter session held during the 14th Cambridge Workshop on Cool Stars, Stellar Systems, and the Sun;
\citet{Soderblom2010_AgeReview} --a general review, but mostly pre-main sequence and late-type dwarfs; 
\citet{Soderblom2014-Ages} --centered on stars younger than 100 Myr; and
\citet{Jeffries2014-Ages} --focusing on low-mass stars-- 
have discussed different techniques from some how diverse approaches,
 in some cases focusing on specific perspectives. 
 On the other hand, it has been shown that some  age scales are not converging in some key cases, such as
the cosmological age and the value derived for the older globular clusters  (\citealt{DAntona1997-Age-GlobularCluster}).
Our main goal here is not only to provide an update, but to emphasize the limits and the caveats related to different methods,
 and to include some thought from a historical perspective.

%%%%%%%%%%%%%%%%%%%%%%%%%%%%%%%%%%%%%%%   FIGURE 
\begin{figure}
\center
\includegraphics[width=1.0\textwidth,scale=1.0]{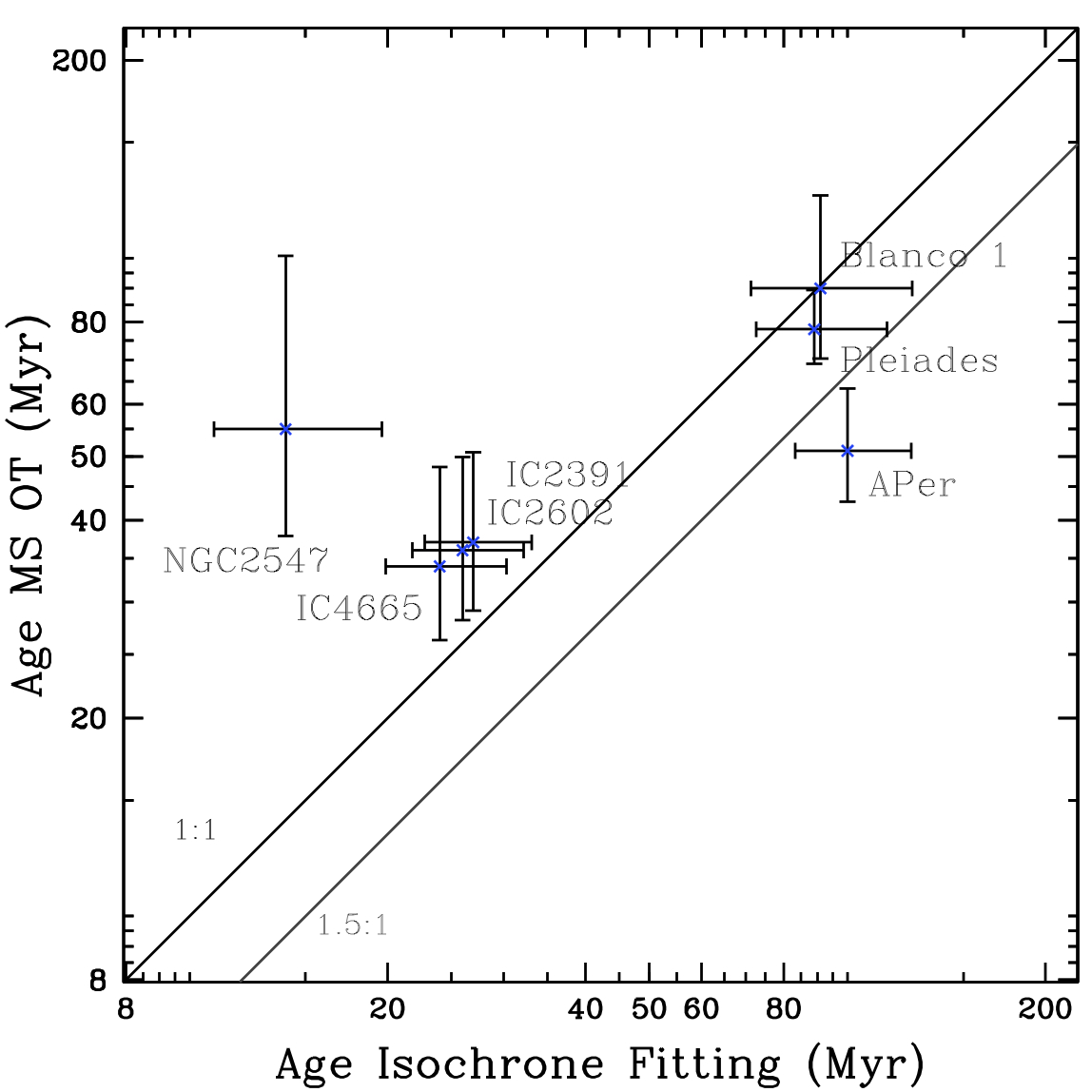} 
\caption{\label{CompAgeScales} 
%, 
Comparison of different age scales: values derived from Upper Main-Sequence Turn
OFF (nuclear, using the more massive members) versus low-mass isochrone fitting (contracting
age in the PMS). The two lines correspond to the relation 1:1 and to an increase in a 50\% between both methodologies to derive ages. Figure taken from \citet{Barrado2011-LithiumAges}.
}
\end{figure}
%%%%%%%%%%%%%%%%%%%%%%%%%%%%%%%%%%%%%%%

%%%%%%%%%%%%%%%%%%%%%%%%%%%%%%%%%%%%%%%%%%%%%%%%%%%%%%%%%%%%%%%%%%%%%%%%%%%%%%%%%%%%%%%%%%%%%  SECTION
%%%%%%%%%%%%%%%%%%%%%%%%%%%%%%%%%%%%%%%%%%%%%%%%%%%%%%%%%%%%%%%%%%%%%%%%%%%%%%%%%%%%%%%%%%%%%
%%%%%%%%%%%%%%%%%%%%%%%%%%%%%%%%%%%%%%%%%%%%%%%%%%%%%%%%%%%%%%%%%%%%%%%%%%%%%%%%%%%%%%%%%%%%%
%
\section{Age scale: anchors and epistemology\label{sec:anchors}}
%
%%%%%%%%%%%%%%%%%%%%%%%%%%%%%%%%%%%%%%%%%%%%%%%%%%%%%%%%%%%%%%%%%%%%%%%%%%%%%%%%%%%%%%%%%%%%%
%%%%%%%%%%%%%%%%%%%%%%%%%%%%%%%%%%%%%%%%%%%%%%%%%%%%%%%%%%%%%%%%%%%%%%%%%%%%%%%%%%%%%%%%%%%%%
%%%%%%%%%%%%%%%%%%%%%%%%%%%%%%%%%%%%%%%%%%%%%%%%%%%%%%%%%%%%%%%%%%%%%%%%%%%%%%%%%%%%%%%%%%%%%

%%%%%%%%%%%%%%%%%%%%%%%%%%%%%%%%%%%%%%%%%%%%%%%%%%%%%%%%%%%%%%%%%%%%%%%%%%%%%%%%%%%%%%%%%%%%%
%
\subsection{Some initial thoughts and a historical perspective\label{subsec:initialthought}}
%
%%%%%%%%%%%%%%%%%%%%%%%%%%%%%%%%%%%%%%%%%%%%%%%%%%%%%%%%%%%%%%%%%%%%%%%%%%%%%%%%%%%%%%%%%%%%%

Epistemology is the discipline that study the way knowledge is achieved in science, 
especially with reference to its limits and validity. Therefore, it has to  be kept in mind,
 specially when dealing with a subject as elusive as stellar ages.

Science is a historical process as well as a mental construction.
We build upon knowledge acquired previously, it is not {\it ex novo}.
 Sometimes everything falls apart when we move a piece, like in the game called ``jenga'';
sometimes   a reshuffle is needed in order to get it right. 
But we have to keep intact the whole structure,
although it is true that sometimes we do not have to do so and a true revolution completely
modifies our view of the universe.
Moreover, we have already said that
science is a immense puzzle, but in the case of the stellar ages we are still missing many pieces.

In this context, 
what do we mean when we say that the temporal scale of disk dissipation is 10 Myr or that giant planets are formed after 
this time, to mention two examples?  Before discussing this subject, it might be illuminating to tell a two hundred year old story.

After the French revolution in 1789, the French Directory or governing body  asked the Academy of Science to provide a standard metrology, the meter. Standardization had been in the air and everybody recognized that the very large amount of scales, even within a country, was a real problem. So two young scientists, P. F. M\'echain and J.-B. Delambre, were sent to  measure by triangulation the meridian from Dunkirk (Dunkerque) to Barcelona and from this value derived a rational standard. This task lasted several years and it was carried out during internal strife and wars against other continental powers (but cooperation with Spanish scientists was maintained even were troops of both countries were fighting against each other not far away).
The computations were tied to the determination of the latitudes of the extremes. Several stars were used together with a very precise new instrument, the circle of repetition invented by J.-C. de Borda few years before, an extremely precise tool. Unfortunately the methodology itself was not so accurate.
M\'echain knew there was a problem (mostly in his several measurements of the latitude of Barcelona, in Spain), but was unable to understand the origin and to correct it. He struggled for years and tried not to publish his data.
So, although by definition the length of the meridian should have been 10,000,000 meters (actually, the other way around), the actual length is 10,002,290 meter. The culprit, actually, falls on the Directory, who rushed the publication of the preliminary results even before finishing the acquisition of the measurements (political meddling with science is never good).
In any case, M\'echain was not aware of --nobody at that time was-- the difference between 
accuracy and precision. The definition from the Webster dictionary reads: 
accuracy, {\it ``degree of conformity of a measure to a standard or a true value''};
precision, {\it ``the degree of refinement with which an operation is performed or a measurement stated''}.
So, we might be very precise, but without accuracy the analysis might be done to no avail.
Thus we should achieve true accuracy,  this is the challenge, and be aware where the limits are.

%%%%%%%%%%%%%%%%%%%%%%%%%%%%%%%%%%%%%%%   FIGURE 
\begin{figure}
\center
\includegraphics[width=1.0\textwidth,scale=1.0]{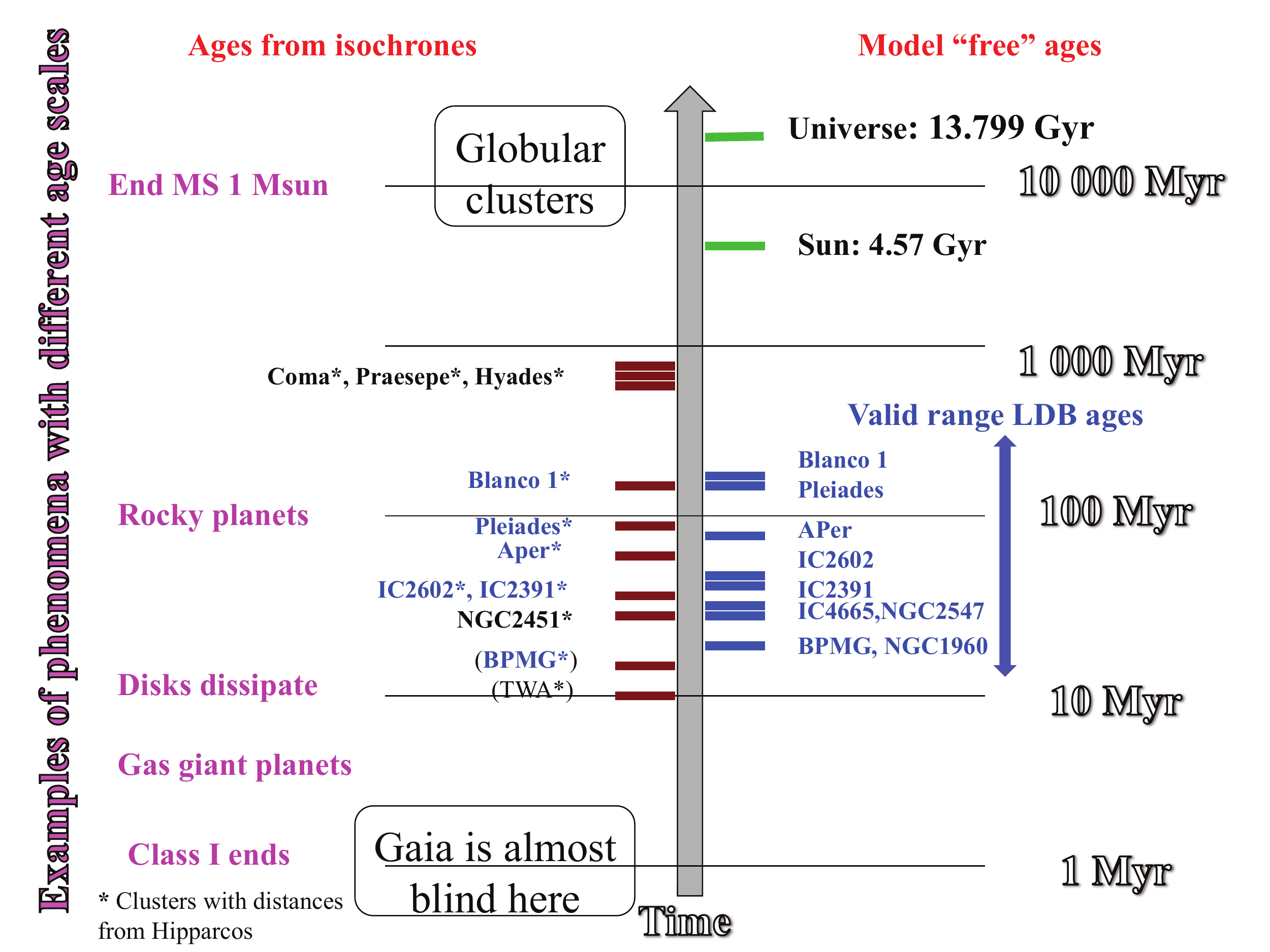} 
\caption{\label{Barrado2015-AgeAnchors} 
%, 
A comparison between different age scales: Lithium Depletion Boundary scale (blue) versus isochrones (brown). There are significant disagreements between different age scales, up to 50\%. Our true and only stellar anchors are the age of the Universe and the Sun. The time scale for several astrophysical phenomena such as the formation of rocky and gas planets or the dissipation of the protoplanetary disks are also displayed (left, in magenta). Thus, the age estimate for much younger associations can be affected by very strong biases.}
\end{figure}
%%%%%%%%%%%%%%%%%%%%%%%%%%%%%%%%%%%%%%%

%%%%%%%%%%%%%%%%%%%%%%%%%%%%%%%%%%%%%%%%%%%%%%%%%%%%%%%%%%%%%%%%%%%%%%%%%%%%%%%%%%%%%%%%%%%%%
%
\subsection{Time scale in Astrophysics: What do we really know?\label{subsec:anchors}}
%
%%%%%%%%%%%%%%%%%%%%%%%%%%%%%%%%%%%%%%%%%%%%%%%%%%%%%%%%%%%%%%%%%%%%%%%%%%%%%%%%%%%%%%%%%%%%%

In order to define an age scale (or several), it is very important to keep in mind that we cannot make experiments
except in very few situations and we rely on observations and theory. Thus, observational phenomena with very well
 defined ages, our anchors, are extremely important. As a matter of fact, there is a scarcity of them and, moreover, 
several results do not agree with each other completely.
 One example is the different age scales for open clusters, since there are several ways to estimate this  parameter, as we will see. Among them, the so called upper main sequence turn off (UMSTO), a nuclear age whose observational evidence is the departure of the massive  members of a cluster once the central hydrogen has been exhausted, has been widely used. Another one valid for young associations is the isochrone fitting for low-mass stars.  However, when these two values for several clusters are compared (see \citealt{Lyra2006-AgeDifference-NuclearContracting} or Fig.\ref{CompAgeScales} extracted from \citealt{Barrado2011-LithiumAges}), there are significant deviations from the one-to-one relationship. In both cases, the derived age depends on a comparison with stellar models. Obviously, some parameter has not been taken into account in any or the other method  (or in both). Thus, it is paramount to try to identify anchors which do not depend on theory or, when this is unavoidable, the theory is very well understood and errors and biases are minimized and fully characterized.

There are two true anchors for the astrophysical age scale. The first one it the total age of the universe. 
The current estimate is  13.799$\pm$0.021 Gyr and it has been derived based on   the
Lambda cold dark matter concordance model ( $\Lambda$CDM, \citealt{PlanckCollaboration2015_CosmologicalParameters}). Thus, it is not a direct
observable (we shall see that there is a disagreement with the age derived for the globular clusters and some
 current stellar evolutionary models). 

The second firmly established anchor is the age of the Solar System.
The Sun is about  4.57 Gyr old. This value comes from two different methodologies  based on estimates for the Sun as a star and for the  the remnants of the formation of the Solar System: 

\begin{itemize}
% i)  computer models of stellar evolution and through nucleocosmochronology (\citet{Bonanno2002-SunAge}), 
\item Computer models of stellar evolution and asteroseismology (\citealt{Bonanno2002-SunAge}), which gives a value of 
 4.57$\pm$0.11 Gyr. Again, this method is model dependent and subject of revision of a significant number of parameters.
\item Radiometric date of the oldest Solar System material. Several techniques produce  4.5672$\pm$0.0006 Gyr (\citealt{Amelin2002-AgeMeteorites}) or 4.5695$\pm$0.0002 (\citealt{Baker2005-AgeSun_Nebula}, with an amazingly small errors).
See also \citet{Bahcall1995-SunAge_SolarNebula}.
\end{itemize}

In addition, the uranium and thorium decays have been used to estimate the age of the oldest stars in the Galaxy, providing a minimum value of 12.5$\pm$3 Gyr (\citealt{Cayrel2001-AgeUraniumDecay}), not very useful but a promising technique. It is based on the atmospheric abundances of these two heavy elements, since $^{238}$U and $^{232}$Th have a half-life times of 4.5 Gyr and  14 Gyr, respectively.
In any case, radioactive ages can be considered model-free, but unfortunately their applicability is very reduced. However, they provide the most important clues when deriving age scales. However, this is restricted to population II, with low metallicity, since the uranium and thorium lines are weak and can be hidden by other spectral features.

Figure \ref{Barrado2015-AgeAnchors} displays these main age anchors and compares them with characteristic times of several  phenomena such as the formation of giant gas planets ($\sim$5 Myr), the dissipation of protoplanetary disks ($\sim$10 Myr), the formation of rocky planets ($\sim$100 Myr) or the life-time of a 1 $M_\odot$ star in the main sequence (10 Gyr). In addition, age estimates of several well known clusters have been included. On the right-hand side estimates from the Lithium Depletion Boundary (LDB, see subsection \ref{LDB}), whereas the left-hand side includes values derived from isochrone fitting (both model dependent, but the modeling behind the LDB ages is simpler). As can be  seen, a significant fraction of the most interesting phenomena occur outside the age range delimited by our anchors. This is one of the main problems of age dating: how to validate our methods and link them to a firmly established foundation.

Certainly, clusters have been used as benchmarks in order to calibrate different scales (see \citealt{Soderblom2010_AgeReview}). They provide, in principle, homogeneous samples of stars of different masses and the same chemical composition and age. However, other parameters, such as rotation, binarity, activity and magnetic fields, some of them interrelated, might play a significant role, as we will see.

One very important element, sometimes hidden, is the distance, or to be more systematic, the ``distance scales''.
They have been derived based on three different types of candles, from primary (trigonometric distances such as the Hipparcos distances used on Fig. \ref{Barrado2015-AgeAnchors}) to tertiary candles, in a ``distance ladder'', where in principle the error propagation is controlled, although the error budget is always increasing. In the same manner, a ``age ladder'' has been advocated, as discussed in \citet{Renzini1992-AgeLadder}, by extending ages in the solar neighborhood into distant phenomena such as globular clusters and other galaxies. As this author has shown, it is significantly more difficult to estimate ages than distances, among other reasons because observations and theory are intertwined. As he pointed out: ``{\it often losing track of how errors propagate and pile up''}. Thus, real errors, and not just crude estimates, have to be taken into account, and this is a task that sometimes is forgotten or just ``swept below the carpet''.

%%%%%%%%%%%%%%%%%%%%%%%%%%%%%%%%%%%%%%%   FIGURE 
\begin{figure}
\center
\includegraphics[width=1.0\textwidth,scale=1.0]{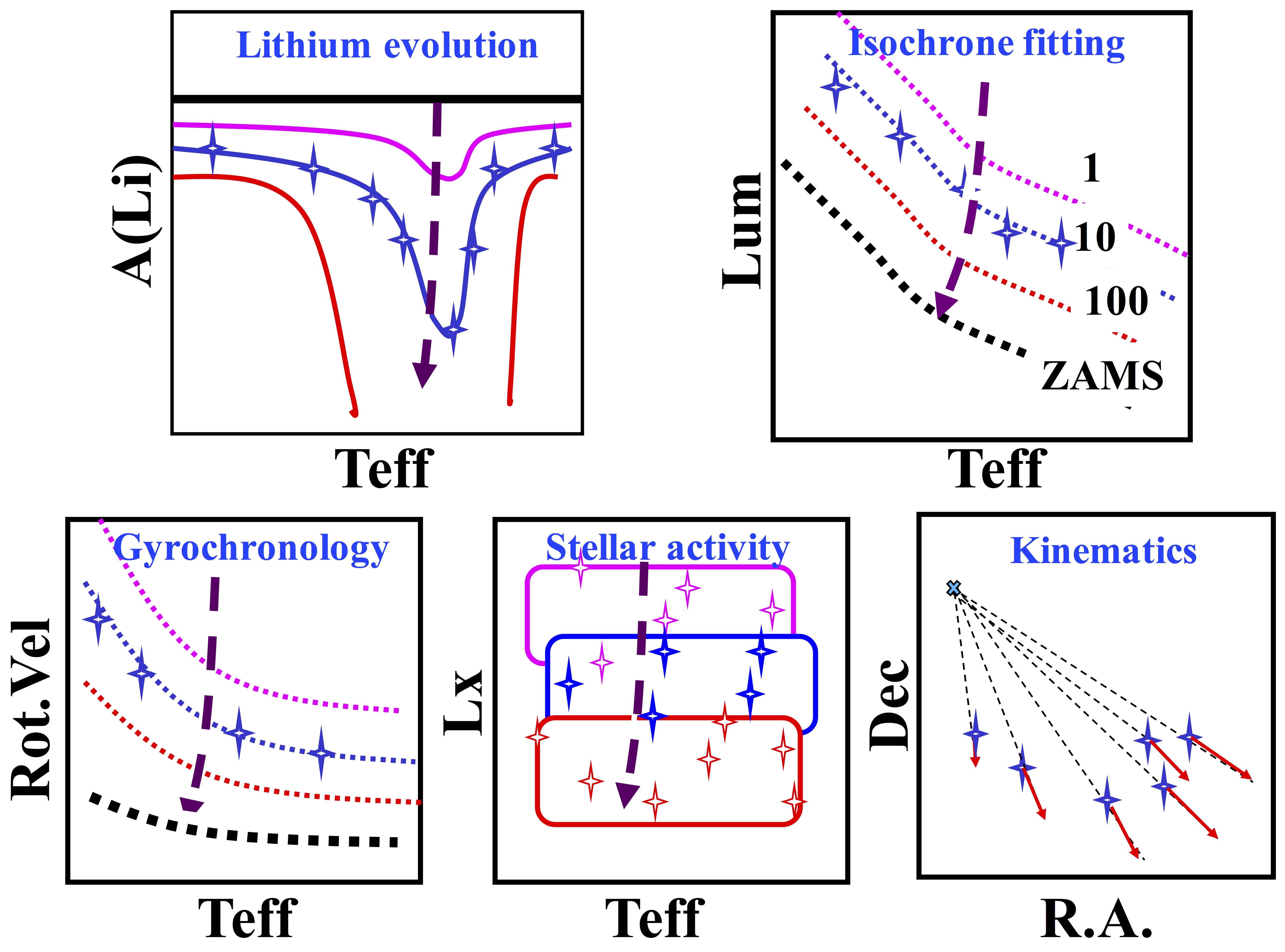} 
\caption{\label{Age_Method_cartoon} 
%, 
A schematic collection of techniques commonly used to estimate stellar ages: lithium abundance, isochrone fitting, evolution of rotation, coronal activity and kinematic groups.
}
\end{figure}
%%%%%%%%%%%%%%%%%%%%%%%%%%%%%%%%%%%%%%%

%%%%%%%%%%%%%%%%%%%%%%%%%%%%%%%%%%%%%%%%%%%%%%%%%%%%%%%%%%%%%%%%%%%%%%%%%%%%%%%%%%%%%%%%%%%%%
%
\subsection{Age scale: classifications\label{subsec:classification}}
%
%%%%%%%%%%%%%%%%%%%%%%%%%%%%%%%%%%%%%%%%%%%%%%%%%%%%%%%%%%%%%%%%%%%%%%%%%%%%%%%%%%%%%%%%%%%%%

Several proposals have been put forward when cataloguing the age-dating methods. We will revisit some of them here.

\subsubsection{A empirical classification\label{subsubsec:mermilliod}}
%
%%%%%%%%%%%%%%%%%%%%%%%%%%%%%%%%%%%%%%%%%%%%%%%%%%%%%%%%%%%%%%%%%%%%%%%%%%%%%%%%%%%%%%%%%%%%%

Jean-Claude Mermilliod (see \citealt{Mermilliod2000.4}) provided one of the first comprehensive classifications of the methods used for age-dating for clusters:

\begin{itemize}
\item The turn-off colours or earliest spectral types.
\item Morphological parameters.
\item Isochrone fitting.
\item Synthetic Color-Magnitude Diagrams. 
\item Pre-main sequence stars and Turn-On point. 
\item Lithium Depletion Boundary for very low-mass stars and brown dwarfs (BD).
\end{itemize}

We refer the reader to the quoted paper for a detailed explanation.

Figure \ref{Age_Method_cartoon} displays a cartoon with several of these methods: lithium evolution, the location in a HR diagram as the star settles in the main sequence, the change in the rotation rate as angular momentum is lost, the decrease of coronal activity due to the same reason and coevality in moving groups. They will be described in some detail in the appropriate section.

As  \citet{Mermilliod2000.4} pointed out, {\it ``all methods are essentially based on ages given by models: evolutionary models for the upper main sequence, contraction models for pre-main-sequence stars and models of fully convective objects for very-low-mass stars and brown dwarfs.''}. Thus, any age scale based on these techniques is, essentially, tied to the models used  by it and this fact should be clearly stated with the derived age (i.e., cluster ``$W$'' has an age of ``$X$'' Myr by using the technique ``$Y$'' and models ``$Z$'').

\subsubsection{A formal classification\label{subsubsec:soderblo}}
%
%%%%%%%%%%%%%%%%%%%%%%%%%%%%%%%%%%%%%%%%%%%%%%%%%%%%%%%%%%%%%%%%%%%%%%%%%%%%%%%%%%%%%%%%%%%%%

David  Soderblom (see his review in \citealt{Soderblom2010_AgeReview} and also a revision in \citealt{Soderblom2014-Ages}) has formalized this previous classification, following this scheme:

\begin{itemize}
\item Fundamental: nucleocosmochronometry.
\item Semi-fundamental: lithium depletion boundary ages (mid-M spectral type), kinematics or expansion ages.
\item Model dependent: isochrone fitting for Pre- and Main-Sequence stars, asteroseismology.
\item Empirical: gyrochronology (rotation), stellar activity, lithium depletion (FGK spectral types), photometric variability, accretion.
\end{itemize}

In the case of clusters,  the role of eclipsing binaries is discussed (accurate radii and masses, so a detailed 
comparison with models can be performed) as well as white dwarfs (WD), since the cooling age is essentially the thermal evolution of a black-body.

\subsubsection{The role of errors: the practical side\label{subsubsec:practical}}
%
%%%%%%%%%%%%%%%%%%%%%%%%%%%%%%%%%%%%%%%%%%%%%%%%%%%%%%%%%%%%%%%%%%%%%%%%%%%%%%%%%%%%%%%%%%%%%

Here,  a somewhat different classification is advocated, which is dominated by a practical approach, mirroring in a sense the ``distance ladder''. Thus, a difference between basic or  primary age indicators with those relying of the first class is established:

Primary indicators, data and models are everything needed:

\begin{itemize}
\item Nucleocosmochronometry.
\item Upper main sequence fitting.
\item White dwarf cooling.
\item Isochrone fitting.
\item Eclipsing binaries and related methods.
\item Spectral features (gravity).
\item Lithium abundance, including the LDB technique.
\item Asteroseismology.
\item  Kinematics (I): movement of components across the Galaxy.
\end{itemize}

Secondary indicators. Since they depend on the primary indicators, they have additional uncertainties:

\begin{itemize}
\item  Stellar activity (X-rays, H$\alpha$ and other activity indicators, photometric variability).
\item  Gyrochronology (rotation).
\item  The relative ratio of Class 0 / Class I / Class II / Class III (see  \citealt{Lada1987-YSO-SFR} and \citealt{Adams1987-YSO-SpectralEvolution} for definitions) in a stellar association or the disk fraction.
\item  Spectroscopic indicators for accretion.
\item  Kinematics (II, see other way above): physical association to another star or to a moving group.
\end{itemize}

As we shall see, not all these methods are equal (regarding how reliable they are) and we will propose several levels in order to achieve a preliminary although comprehensive and coherent age scale system.

Some of these methods can be applied to individual stars. Other, by definition, only to groups, so they can be considered statistical indicators. Note, in any case, the dependence with models and {\it distance} for most of these methods. In the following sections we will see the advantages and shortcomings of these methods (\citealt{Soderblom2010_AgeReview})  and will try to explicitly enunciate the dependences with other parameters. But before doing so, we have to provide the proper set-up, an elementary description of one of our main tool: stellar evolution and models.

%%%%%%%%%%%%%%%%%%%%%%%%%%%%%%%%%%%%%%%   FIGURE 
\begin{figure}
\center
\includegraphics[width=1.0\textwidth,scale=1.0]{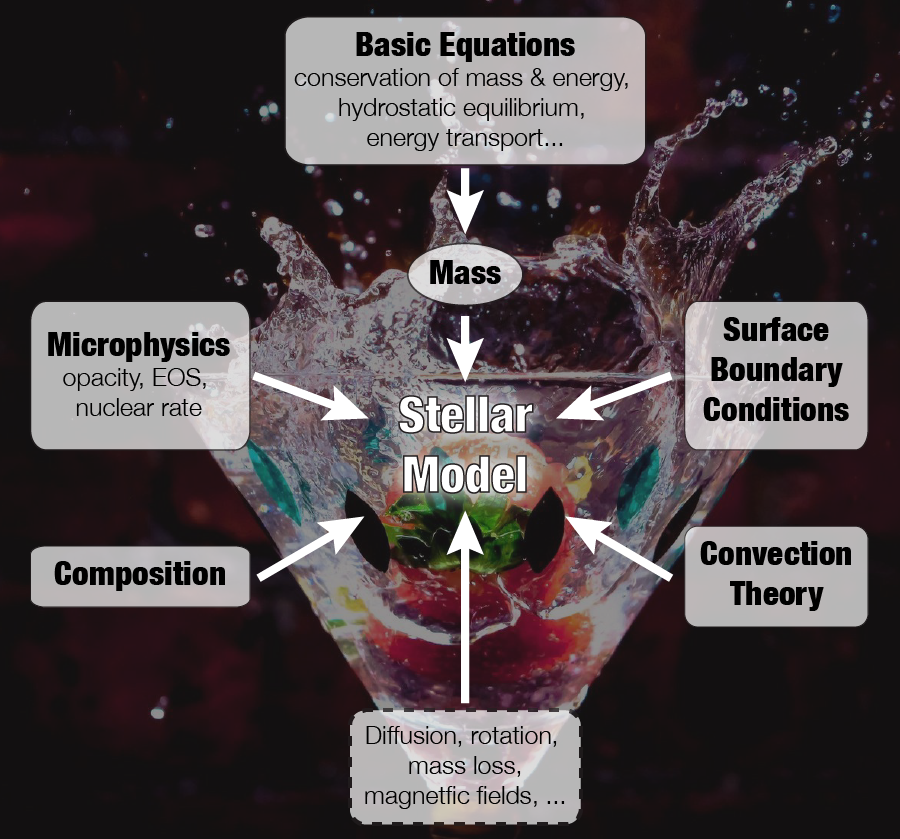} 
\caption{\label{CocktailChaboyer2001StellarModels} 
%, 
Stellar models and evolution as a ``perfect'' cocktail. All the ingredients have both uncertainties and a specific parameter space. These degrees of freedom  normally lead to significantly different physical properties and very large uncertainties when applying the models to real data. After \citet{Chaboyer2001-AgeGC}.
}
\end{figure}
%%%%%%%%%%%%%%%%%%%%%%%%%%%%%%%%%%%%%%%

%%%%%%%%%%%%%%%%%%%%%%%%%%%%%%%%%%%%%%%   FIGURE 
\begin{figure}
\center
\includegraphics[width=1.0\textwidth,scale=1.0]{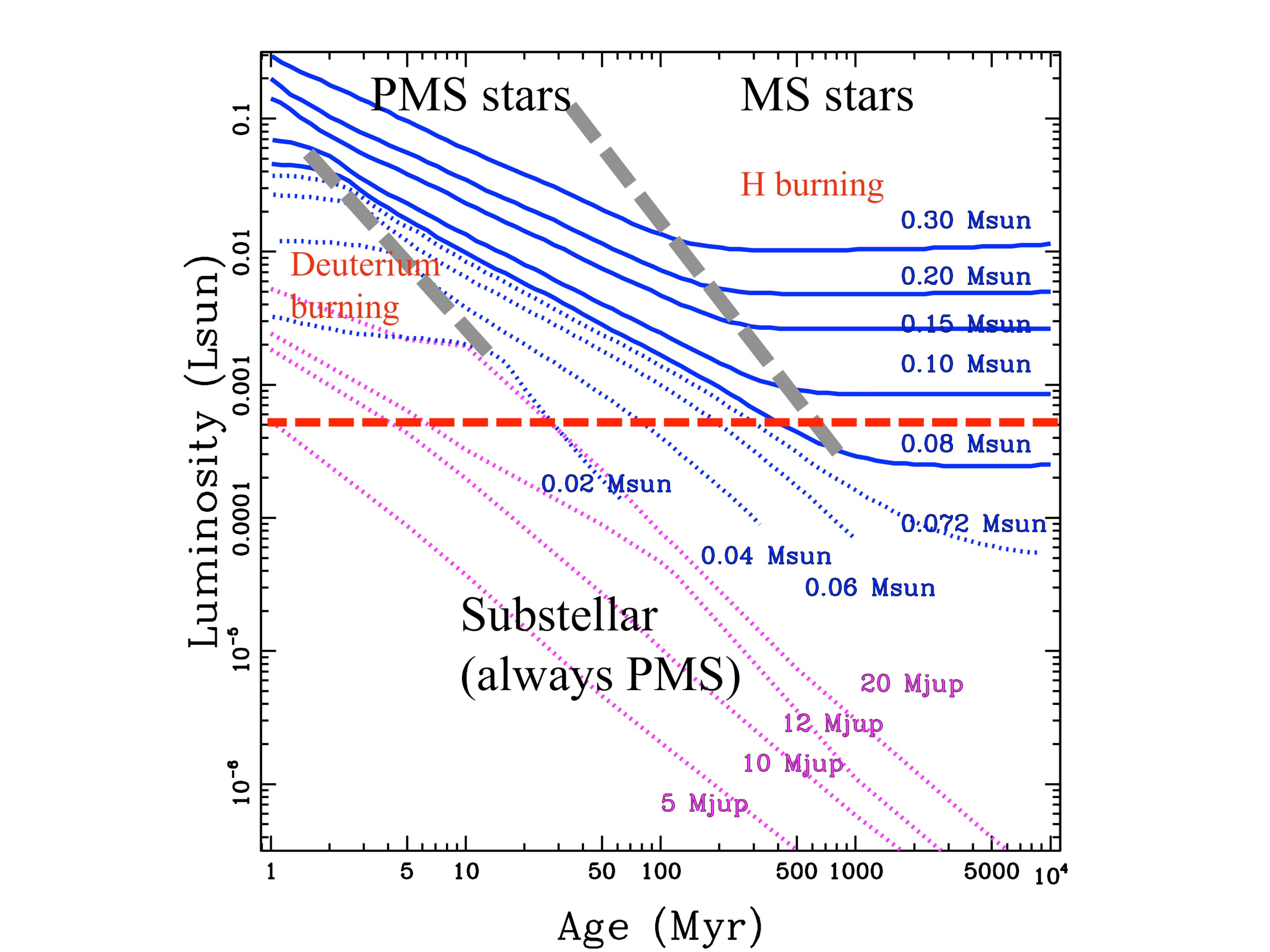} 
\caption{\label{EvolutionLumAgeModelsLyon} 
Luminosity as a function of age, for different evolutionary tracks from
the Lyon group: blue for NextGen and magenta for COND models (\citealt{Baraffe1998.1};
\citealt{Chabrier2000.1}). Solid and dotted lines represent stars and substellar objects, respectively. The
age ranges when deuterium and hydrogen burning happens are easily identified as plateaus in
each track. 
%, 
}
\end{figure}
%%%%%%%%%%%%%%%%%%%%%%%%%%%%%%%%%%%%%%%

\subsubsection{Stellar evolution and models\label{subsubsec:models}}
%
%%%%%%%%%%%%%%%%%%%%%%%%%%%%%%%%%%%%%%%%%%%%%%%%%%%%%%%%%%%%%%%%%%%%%%%%%%%%%%%%%%%%%%%%%%%%%

It cannot be emphasized enough: except in few exceptions, our age estimates (and a significant fraction of what we know about stars) depend on models.
Figure \ref{CocktailChaboyer2001StellarModels}, extracted from \citet{Chaboyer2001-AgeGC}, clearly illustrates the
cocktail we call a stellar model. We refer the reader to the very wide literature dealing with this problem, starting with the quoted paper but also several chapters in volumes in these series.

Regarding the stellar evolution and the underlying physics, \citet{Lebreton2014-Model-Uncertainties} discussed the equations involved in stellar structure and evolution, including general factors  and specific parameters to every star. Among the first ones can be listed: the equation of state, the opacities, microscopic diffusion (i.e., transport of material inside the stars)  and the nuclear reaction rates; whereas the individual ingredients are the
chemical composition, the rotation, binarity and the magnetic fields (usually ignored). This cocktail translates into several phenomena which affect the age determination: rotation and overshooting (i.e., macro scale mixing pushed farther than the convection layer,
for a very recent attempt to calibrate it for stars more massive than the Sun, see \citealt{Deheuvels2016-LowMassOveshooting-Kepler}),
 convection for intermediate  and low-mass stars and atmospheric effects (the colors from the 
observational side, \citealt{Stauffer2003.1}) for the low-mass end. Moreover, the chemical composition has a significant effect on the age estimate not only because of the helium content or the overall metallicity [Fe/H], but anomalies in specific abundances might play a role too, although this problem might have a reduced effect. A recent work by \citet{Bovy2016-ChemicalHomogeneity-Clusters},  dealing with the chemical homogeneity of open clusters, illustrates this point.

Thus, as a brief summary,  several relevant issues can be pointed out. First, the essential role of the stellar mass, the most basic parameter we are dealing with but unknown in most cases. Except in few cases, we can have 
and idea based on the color or spectral type (correct up to 10\%, \citealt{Soderblom2010_AgeReview}), and this estimate depends on age, the parameter we would like to determine, in a tautological or circular argument (in a sense, a loop within a loop). Second, chemical composition, convection and the microphysics are also very important factors which, to a large extent, are not very well determined. One common situation, for instance, is to assume solar metallicity when a detailed chemical analysis is not available, but both the helium content (the fraction Y used in the models) or the overall composition of more massive elements (Z) does affect the nuclear reaction rates and therefore the stellar lifetime, just to mention a couple of effects.

There is a significant number of evolutionary tracks and interior models in the literature, some of them customized of specific problems and others computed for a more general use. The properties we are interested in (for age-dating)
 are those which change fast enough with age and, hopefully, with mass. They would be perfect if they do not depend on any other parameter, such as metallicity, but in practice this does not happen. What seems to be true is that models for the main sequence make a much better job than models for the pre-main sequence. In fact, for this last case, the situation gets worse for lower masses and/or for younger ages (\citealt{Hillenbrand2008-CS14-AgeSpread}).

Figure \ref{EvolutionLumAgeModelsLyon} displays  models by the Lyon group (\citealt{Baraffe1998.1};  \citealt{Chabrier2000.1}), a set  which has been widely used to study low-mass stars and brown dwarfs. This last category corresponds to objects whose mass is so low that the internal temperature never reaches the threshold to burn hydrogen and to release energy, although other fusion of light elements, such as deuterium or lithium, can happen at different moments. Thus, it can be said that BDs remain in the pre-main sequence forever and they cool down in a steady manner. However, {\it bona fide} stars do reach this critical point (the arrival to the MS) at some moment, which depends primarily on the total mass. This property  (the contraction toward the zero age main sequence) can be used to estimate the stellar age. Once the hydrogen burning has started, the luminosity (and the temperature) is stabilized and the star remains on the MS for most of its total lifetime. A good estimate is $\tau_{MS}$=$10^{10}$$\times$$Mass(M_\odot)^{-2.5}$ (for instance, see \citealt{Hansen1994-EvolutionInteriors}).

For more massive stars an option, among others, is the models by the Geneva group. Figure \ref{Mermilliod1981-CMD-OpenClustersIsochrones}, taken from  \citet{Mermilliod1981.2}, displays a set of isochrones which represent very well known clusters of very different ages, up to the Hyades (about 600 Myr). The original caption is: 
{\it ``Composite HR diagram presenting the sequences deduced from 14 pairs of composite
diagrams ... The age groups are designated
by the name of the most representative cluster. The darkened
areas show the positions of the red giant concentrations. Triangles stand for Cepheids 
and dots for non-Cepheid stars in the Hertzsprung gap. The dashed lines have been adapted
from models by Maeder.''}
Thus, the departure of a massive member of an association from the main sequence is complementary to the previous case, the settling of the low-mass counterparts onto the MS. Although the method is plagued with problems and caveats, one very important fact is that {\it relative} ages can be easily estimated (\citealt{Mermilliod2000.4}).

In any case, to conclude, different sets of models, since they do not include the same ingredients,  produce different answers, both in intrinsic  properties (luminosity, temperature) or the derived age. But even if they would give exactly the same answer, a practical problem arises: how to convert parameters computed by these models into observable quantities.

\subsubsection{From theory to observations and vice versa\label{subsubsec:theryobservations}}
%
%%%%%%%%%%%%%%%%%%%%%%%%%%%%%%%%%%%%%%%%%%%%%%%%%%%%%%%%%%%%%%%%%%%%%%%%%%%%%%%%%%%%%%%%%%%%%

One crucial step, already mentioned here and in the specialized literature, is how to compare observables with 
values computed with the theory: i.e., the conversion from photometric colors, magnitudes (or the information derived from spectroscopy) to luminosities, chemical abundances, masses and  temperatures.  Essentially, what is needed are empirical conversions, 
based on the Sun, very well known clusters and different types of binaries with accurate parameters. 
And, again, models play a fundamental role and they are far from perfect.
 Although extraordinary improvements have been achieved on different fronts, there is still a long way to go. For a detailed discussion of these issues, see \citet{Lebreton2014-Model-Uncertainties, Lebreton2014-Asteroseismology} and 
\citet{Cassisi2014-ModelUncertainties} in these series.
Some intrinsic problems are discussed in
\citet{Terndrum2008-CS14-CalibratingIsochrones}, who affirmed: 
{\it ``I show that models which work near the solar abundance currently fail at lower metallicity. I argue that new parallax surveys aid model calibration only if we also have highly accurate temperatures and metallicities over wide swaths of the H-R diagram''}.
Data, data and more data, and analysis from a holistic perspective, are needed.

Another important aspect, usually forgotten, is that models normally are not validated at young ages. This fact has been illustrated in Figure \ref{Barrado2015-AgeAnchors} and has been discussed, for instance, in 
\citet{Stassun2008-CS14-Constrains} for pre-main sequence low-mass stars and brown dwarfs.
Moreover,  \citet{Naylor2009.1} have derived age estimates significantly older than the canonical values for several young associations.

Several works have tried to asses how accurate theoretical isochrones are. 
For instance, \citet{Hillenbrand2004-IsochroneMasses} have compared 
measured masses  (148 stars)  with the values derived by using several sets of models.
For the  main sequence, all models seem to reproduce the data for $M>1.2 M_\odot$, some sets 
are OK down to 0.5 $M_\odot$ and all fail below 0.5 $M_\odot$, producing differences  between 5 and 20\%.
In the case of  pre-main sequence (PMS), the results are good enough for $M>1.2 M_\odot$, but for stars with masses in the
range  1.2-0.3  $M_\odot$ the differences are in the range 10-30\%.
Conversely, \citet{Hillenbrand2008-CS14-AgeSpread} confronted the ages of several young stellar associations and  concluded that the observed spread in HR diagrams are not due to age spread (at least this is not the most significant factor). Similar conclusion has been reached by  \citet{Preibisch2012-Age-PMS-HRD-CMD} or
 \citet{Jeffries2012-AgeSpread-SFR}.
 But in any case it plays a significant role when trying to estimate accurate ages.

But even if our theoretical models would be perfect when deriving masses, another essential problem is present: confronting theory with observations for other properties.
When comparing theoretical data with observational properties, such as the plane effective temperatures and Luminosities ($T_\mathrm{eff},L_\mathrm{bol}$) with
 magnitudes and colors (mag, colors), either bolometric corrections (BC) and/or a specific temperature scale are used.
Some of these problems could be avoided or mitigated using multi-wavelength photometry and the complete Spectral Energy Distribution (SED, see \citealt{Bayo2008-VOSA}).
The basic problem here is the modeling of stellar atmospheres and the reproduction of colors for all masses  and metallicities.
 Thus, the results are necessarily linked to these transformations. 
 Coming back to   \citet{Terndrum2008-CS14-CalibratingIsochrones}: {\it ``One problem for sure is that the color-$T_\mathrm{eff}$ relations are derived using large
samples of local stars. An inspection of the original sources and the spectroscopic
sample in Figure 5} --in that article-- {\it shows a lack of cool, metal-poor stars in the determination
of the color-$T_\mathrm{eff}$ relations.''}. Thus, the samples we have used to define these relations are also a hidden factor which might be affecting the final results because of selection effects.

\subsubsection{Coevality,  distances and other effects: stars are individuals\label{subsubsec:coevality}}
%
%%%%%%%%%%%%%%%%%%%%%%%%%%%%%%%%%%%%%%%%%%%%%%%%%%%%%%%%%%%%%%%%%%%%%%%%%%%%%%%%%%%%%%%%%%%%%

When dealing with stellar groups, several assumptions are kept, even when they are not explicitly stated:
coevality is assumed for star clusters and moving groups, as well as the same chemical composition.
What lies at the background is whether the star formation is essentially instantaneous or not.
 From the observational point of view, young stellar associations show a large range of luminosities for the same effective temperature in the Hertzsprung-Russell diagram. This fact remains for globular and open clusters (the case of  M15  and the Pleiades in Figures \ref{Krauss2003_Globular_CMD} and \ref{Pleiades_HRD_LiAbyss}). This width in L$_{bol}$ could be translated onto a spread in age $\delta$$\tau$. The meaning of $\delta$$\tau$ is not clear. Is it related to a true diversity in ages (a star formation lasting several Myr), it is due to the size of the star forming region and/or the original molecular cloud, does it come from errors and biases of unknown origin?
For an overview, see  \citet{Hillenbrand2008-CS14-AgeSpread}.

One aspect we all tend to forget is that stars are individuals. First, each of them is born with specific initial conditions and has a distinct history (which might be very different for the first 10 Myr). They show specific characteristics such as photospheric spots and other activity related phenomena, rotation, accretion during the early phases (including peaks in the accretion rates), circumstellar disks,  reddening,   magnetic fields, metallicity, individual distances (a cluster has a size and a shape), spatial segregation (differences in the properties depending on the location inside the association),  the potential effect of planets and other nearby stellar companions, and so on. On top of this, the stellar history
 or all that has happened to the star before it is observed that can modify the subsequent evolution from its initial conditions
(including previous interactions with other stars, specially in compact star forming regions or SFRs for short). 

Certainly, accurate ages and complete censuses (at least the removal of interlopers based on accurate  and precise  proper motion), will be very helpful and here the role of the Gaia satellite will be paramount. Its hopefully extraordinary archive will be exploited for many years to come and we can only speculate what wonders it will contain. After all, Nature has a ``tendency'' to surprise us.

%%%%%%%%%%%%%%%%%%%%%%%%%%%%%%%%%%%%%%%   FIGURE 
\begin{figure}
\center
\includegraphics[width=1.0\textwidth,scale=1.0]{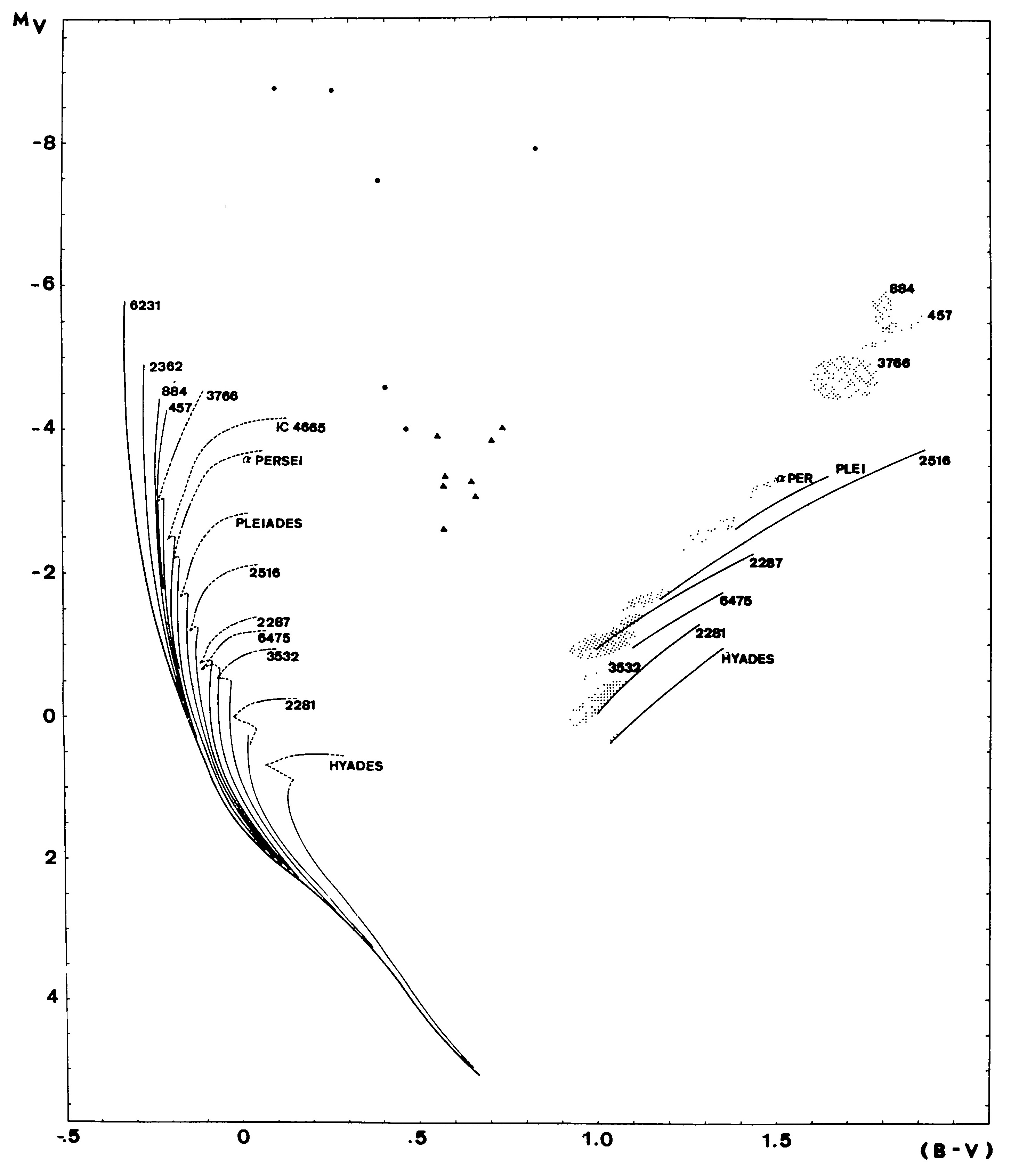} 
\caption{\label{Mermilliod1981-CMD-OpenClustersIsochrones} 
%, 
HR diagram with isochrones for nearby clusters.
Figure selected  \citet{Mermilliod1981.2}, see text for an explanation.
}
\end{figure}
%%%%%%%%%%%%%%%%%%%%%%%%%%%%%%%%%%%%%%%

%%%%%%%%%%%%%%%%%%%%%%%%%%%%%%%%%%%%%%%%%%%%%%%%%%%%%%%%%%%%%%%%%%%%%%%%%%%%%%%%%%%%%%%%%%%%%  SECTION
%%%%%%%%%%%%%%%%%%%%%%%%%%%%%%%%%%%%%%%%%%%%%%%%%%%%%%%%%%%%%%%%%%%%%%%%%%%%%%%%%%%%%%%%%%%%%
%%%%%%%%%%%%%%%%%%%%%%%%%%%%%%%%%%%%%%%%%%%%%%%%%%%%%%%%%%%%%%%%%%%%%%%%%%%%%%%%%%%%%%%%%%%%%
%
\section{Methods to estimate the stellar age\label{sec:methods}}
%
%%%%%%%%%%%%%%%%%%%%%%%%%%%%%%%%%%%%%%%%%%%%%%%%%%%%%%%%%%%%%%%%%%%%%%%%%%%%%%%%%%%%%%%%%%%%%
%%%%%%%%%%%%%%%%%%%%%%%%%%%%%%%%%%%%%%%%%%%%%%%%%%%%%%%%%%%%%%%%%%%%%%%%%%%%%%%%%%%%%%%%%%%%%
%%%%%%%%%%%%%%%%%%%%%%%%%%%%%%%%%%%%%%%%%%%%%%%%%%%%%%%%%%%%%%%%%%%%%%%%%%%%%%%%%%%%%%%%%%%%%

Before starting the discussion regarding the methods to estimate the stellar age and, in fact, before carrying
out any analysis, the first step is to verify whether the star belongs to a binary or multiple system and, in case of binary, the type (wide physical binaries, spectroscopic, eclipsing), since the information that can derived is very different and in some cases could be very useful.

The simplest case is isolation. A single star can reveal its age in an approximately way if the distance is known (the magnitude gives the luminosity and the color the mass/$T_\mathrm{eff}$, assuming a given value of the metallicity). If a high quality spectrum is available, the surface gravity can be estimated, which is also related to age. Finally, pulsations, when present, can provide a detailed insight of the internal structure and its age, although the story is not so simple.

On the other hand, the light curve (LC) of the eclipsing binaries (EB) gives the stellar radii (or ratio), spectroscopic binaries the masses (affected by the inclination of the orbit with respect to the line of sight) and the astrometric data of wide binaries give the absolute masses of both components.

The different versions of multiplicity (membership to star forming regions, open and globular clusters, loose associations and moving groups) can be used to derive a common age for the whole group. See the chapters by Estelle Moraux and Corinne Charbonnel in this volume. However, binarity and other hierarchical systems are not normally considered and if this factor is not taken into account the final age estimated can be biased to a considerable amount.

%%%%%%%%%%%%%%%%%%%%%%%%%%%%%%%%%%%%%%%%%%%%%%%%%%%%%%%%%%%%%%%%%%%%%%%%%%%%%%%%%%%%%%%%%%%%%
%
\subsection{Isochrone fitting\label{subsec:isochrone}}
%
%%%%%%%%%%%%%%%%%%%%%%%%%%%%%%%%%%%%%%%%%%%%%%%%%%%%%%%%%%%%%%%%%%%%%%%%%%%%%%%%%%%%%%%%%%%%%

\subsubsection{Open clusters: turn off ages\label{openclusterTO}}
%
%%%%%%%%%%%%%%%%%%%%%%%%%%%%%%%%%%%%%%%%%%%%%%%%%%%%%%%%%%%%%%%%%%%%%%%%%%%%%%%%%%%%%%%%%%%%%

Few years after it was recognized the nuclear burning was the energy sources of the stars (\citealt{Bethe1939-EnergyStars}), \citet{Sandage1952-Models-GravContraction} identified how the evolution off the main sequence and the position in the HR diagram could be used to estimate the age of giants (and globular clusters, see below).

As stated before, Fig.  \ref{Mermilliod1981-CMD-OpenClustersIsochrones}, taken from  \citet{Mermilliod1981.2}, provides an excellent example of isochrone fitting for several clusters and a {\it relative age scale} for them (a sorting in evolutionary status). These fittings are carried out using the most massive members in the clusters, either once they have become giants or when they are in the process of evolving off the main sequence, once they have exhausted the hydrogen in the core and when the stellar evolution is fast. 

There is a significant disadvantage of  this technique: the Initial Mass Function (\citealt{Salpeter1955-LF}, see also \citealt{Barrado2001-IMF-M35}, \citealt{Barrado2005-C69-IMF} or \citealt{Bayo2011-IMF-C69}). The statistics make more difficult to fit isochrones at the top of the mass spectrum because the scarcity of cluster members when compared with the bottom of the main sequence. On the other hand, they are much brighter and allow an age determination for much further away clusters.

Another important issue has to be taken into account: the interplay between distance, interstellar reddening and age. All these parameters have to be derived simultaneously and, as a matter of fact, they are intertwined (a increased in the reddening would modify the distance and/or the age estimate and vice versa). Additional data, such as distances from parallax or reddening from spectral types and colors, are extremely useful to avoid these degeneracies.

In any event, the use of more sophisticated fitting techniques, such as Bayesian estimation, can help to dilute the effects of some of these shortcomings (see, for instance, \citealt{Jorgensen2005-AgeIsochroneBayesian}).

%%%%%%%%%%%%%%%%%%%%%%%%%%%%%%%%%%%%%%%   FIGURE 
\begin{figure}
\center
\includegraphics[width=1.0\textwidth,scale=1.0]{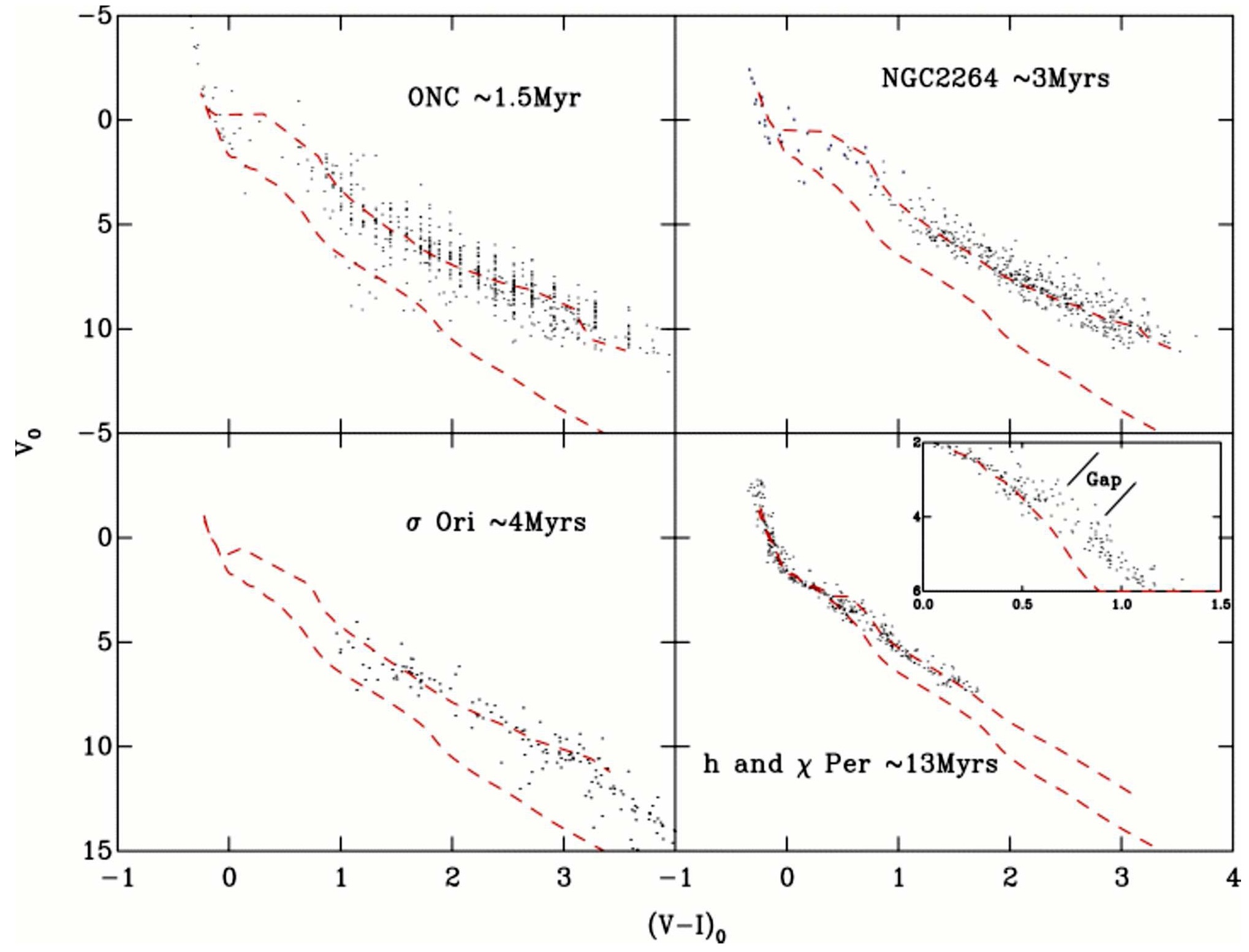} 
\caption{\label{Naylor2009-AgesPMS-Methods} 
%, 
Color-Magnitude Diagrams for several young and very young stellar associations. Figure from \citet{Naylor2009.2}. 
}
\end{figure}
%%%%%%%%%%%%%%%%%%%%%%%%%%%%%%%%%%%%%%%

\subsubsection{Open clusters: PMS ages\label{openclusterPMS}}
%
%%%%%%%%%%%%%%%%%%%%%%%%%%%%%%%%%%%%%%%%%%%%%%%%%%%%%%%%%%%%%%%%%%%%%%%%%%%%%%%%%%%%%%%%%%%%%

Pre-main sequence isochrones and tracks present additional problems based on the lack of good anchors (Figure \ref{Barrado2015-AgeAnchors}). 
Few  spectroscopic and eclipsing binaries, specially low-mass members, are being analysed (see below, subsection \ref{eclipsingbinaries}) and this fact will have an extraordinary impact of the models and the improvement in the accuracy of the age determination. In any case, the bottom of the cluster sequence has the advantage of sheer numbers: there are tens or hundreds of objects so any fit could, in principle, be very sound. Unfortunately Nature is never easy and several phenomena modify the observational properties of the low-mass objects and these second-order effects do have a strong impact.

In any case, there are several methods which can be applied. Figure \ref{Naylor2009-AgesPMS-Methods} 
illustrates new proposals for  several  young associations (\citealt{Naylor2009.2}).
The original explanation of the figure says: {\it ``The CMDs for a selection of young groups in absolute magnitude and intrinsic colour. In each case the lower red dotted line is the position of the MS, the upper an appropriate \citet{Siess2000.1} isochrone.''}
Note that, as we have already shown and will be discussed later, one of these methods
 {\it ``... suggests that there is a factor two difference between these “nuclear” ages, and more conventional pre-main-sequence contraction ages.''}, as that article concluded. 

Coming back to the issue of individuality or second order effects, such as accretion activity, magnetic fields, and so on, early stellar evolution during the Pre-Main  Sequence phase can be modified to a significant extent. For instance, \citet{Somers2015-Spots-OnMassAge-PMS} have estimated that  large surface coverage by stellar spots can increase the stellar radii up to 10\%, modifying the location in the HR diagram (thus, the age estimate).  Moreover, \citet{Jeffries2012-AgeSpread-SFR} concluded that 
 ``{\it the traditional HR diagram is a
poor tool for estimating the ages of young ($<$20Myr) PMS stars and also perhaps
for estimating age-dependent masses}''. An additional reason to be cautious.

Semi-empirical isochrones, which essentially rely on data of well known associations, can be used in order to avoid or ameliorate some of the problems discussed above. This approach has been followed, for instance, by 
\citet{Bell2015-IsochroneAge-MG},
in order to provide a sorting scale for young moving groups. These loose associations (subsection \ref{subsec:kinematics})  are very important, since they provide samples of well characterized nearby low-mass stars, specifically for an age range between very young associations and open clusters. Moreover, they  can be used, among other things, to search for planets (primarily by direct imaging), a very active research field nowadays.

Young stellar associations offer an alternative with a lot of potential, the ``H feature'' or radiative-convective  gap
(R-C gap, see, for instance, \citealt{Piskunov1996-LF-RCgap} or \citealt{Mayne2007-EmpiricalIsochrones-Age-RCgap}). This feature in a cluster isochrone (or, better, in the luminosity function, LF) appears when a radiative core is developed as a star begins the hydrogen fusion (i.e, it corresponds to the separation between to the MS and PMS in a cluster sequence). The location of the R-C gap depends on the age (both the $T_\mathrm{eff}$ and the $L_\mathrm{bol}$ in a HRD), but at $\sim$10 Myr and older ages it cannot be distinguished any longer.

%%%%%%%%%%%%%%%%%%%%%%%%%%%%%%%%%%%%%%%   FIGURE 
\begin{figure}
\center
\includegraphics[width=1.0\textwidth,scale=1.0]{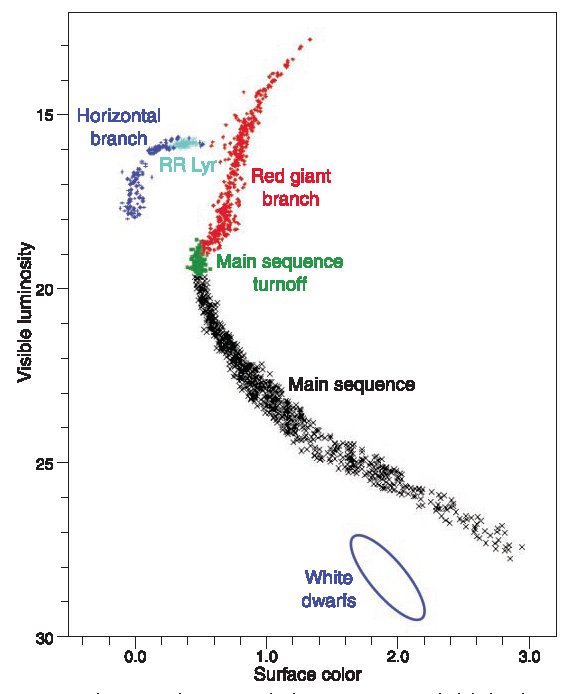} 
\caption{\label{Krauss2003_Globular_CMD} 
%, 
Different phases of the stellar evolution of a solar-like star, as seen in the M15 globular cluster.
The figure comes  from \citet{Krauss2003-Globular-AgeEstimates}, after \citet{Durrell1993-GC-M15}. See body-text for a detailed explanation.
}
\end{figure}
%%%%%%%%%%%%%%%%%%%%%%%%%%%%%%%%%%%%%%%

%%%%%%%%%%%%%%%%%%%%%%%%%%%%%%%%%%%%%%%   FIGURE 
\begin{figure}
\center
\includegraphics[width=1.0\textwidth,scale=1.0]{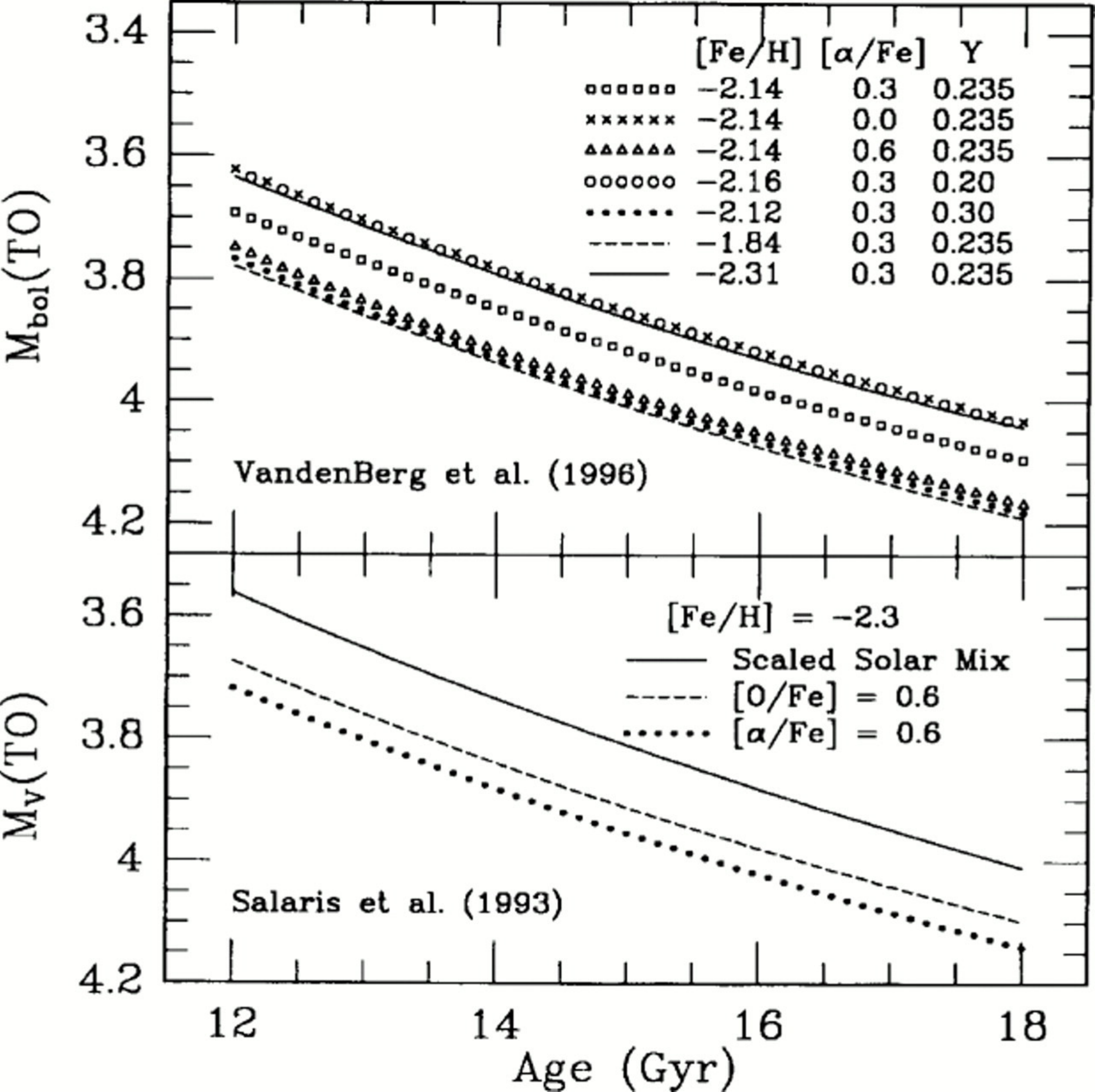} 
\caption{\label{VandenBerg1996_MvAge_Composition} 
%, 
Stellar luminosity at the Turn Off point, after,  \citet{VandenBerg1996-AgeGC}. 
See text.}
\end{figure}
%%%%%%%%%%%%%%%%%%%%%%%%%%%%%%%%%%%%%%%

\subsubsection{Globular clusters\label{globularcluster}}
%
%%%%%%%%%%%%%%%%%%%%%%%%%%%%%%%%%%%%%%%%%%%%%%%%%%%%%%%%%%%%%%%%%%%%%%%%%%%%%%%%%%%%%%%%%%%%%

The Milky Way contains about 150 known globular clusters (\citealt{Harris1996-GC-Catalog}). They are very compact and contain several tens of  
thousand of members. In addition, they are among the oldest objects in the universe. 

Figure \ref{Krauss2003_Globular_CMD} represents the evolution of a solar-type star across the Hertzsprung-Russell
diagram. The diagram has been taken  from \citet{Krauss2003-Globular-AgeEstimates}. The original caption reads:
{\it ``A schematic color-magnitude diagram for a typical globular cluster
... showing the location of the principal stellar evolutionary sequences.
This diagram plots the visible luminosity of the star (measured in magnitudes)
as a function of the surface color of the star (measured in B-V
magnitude). Hydrogen-burning stars on the main sequence eventually exhaust
the hydrogen in their cores (main sequence turnoff ). After this, stars
generate energy through hydrogen fusion in a shell surrounding an inert
hydrogen core. The surface of the star expands and cools (red giant branch).
Eventually the helium core becomes so hot and dense that the star ignites
helium fusion in its core (horizontal branch). A subclass is unstable to radial
pulsations (RR Lyrae). When a typical globular cluster star exhausts its
supply of helium, and fusion processes cease, it evolves to become a white
dwarf.''}. The figure clearly shows why giants are perfect in order to determine the association age,
 since they are tens or hundreds of times brighter than MS of the same color. On the other hand, globular clusters
contain at least several thousands members and even despite the fact that the IMF goes against more massive members,
 in number, there are still a significant amount of them, so a good fit can be achieved (compare this HRD with 
Fig.  \ref{Mermilliod1981-CMD-OpenClustersIsochrones}).

The most significant feature is the turn off (TO) and its relation to the cluster sequence.
Almost half a century ago \citet{Sandage1970-GC-Age} explicitly formulated how the age of a globular cluster can be estimated based on the position of the main sequence turn off, the point when the star members ceased to burn hydrogen in the central part of the nucleus (see Charbonnel, this volume):

\begin{equation}
Log (\tau/10^9 yr) \simeq -0.41 + 0.37 \times M_V^{TO} - 0.43 \times Y - 0.13 \times \mathrm{[Fe/H]}
\end{equation}

As \citet{Renzini1992-AgeLadder} pointed out, the main source of errors comes from a "hidden" factor, the distance. It can amount to 22\% for a error in the distance modulus of 0.25 mag (9\% for 0.1 mag). Helium abundance is, in principle, well constrained and might only add a 2\% of uncertainty on the age estimate. The metallicity, however, is another matter. A $\Delta$[Fe/H]=0.3 dex --a poor value nowadays-- translate into another $\Delta$$\tau$=9\%. Therefore, both Gaia and its very precise distances and systematic spectroscopic surveys (large number of samples combined with high-resolution, high signal-to-noise ratio spectra) might eventually change the game. Additional, more sophisticated techniques such as a  Bayesian fittings to derive ages in GCs are described in
\citet{VallsGabaud2014-EES2013-AgesBayesian}.

In addition, the overall chemical composition (different fraction X and Y of hydrogen and helium, but also the detailed
split for Z, the heavier elements) do modify to a very large extent the age estimate. This is clearly shown in   
Fig. \ref{VandenBerg1996_MvAge_Composition}  from  \citet{VandenBerg1996-AgeGC}. The original caption reads:
{\it ``Turnoff luminosity vs age relations from the indicated investigations for the particular choice of Y=0.20 and Z=0.0001 for the mass-fraction abundances of helium and the heavier elements, respectively. The Mbol(TO) values were calculated on the assumption that the solar value is 4.72 mag.''}
 Differences can be up to 15\%.

The age of a GC can be determined by additional techniques (\citealt{VandenBerg1996-AgeGC}; \citealt{Chaboyer2001-AgeGC}).  Apart of the standard isochrone fitting, most of them essentially depend on differences in color or magnitude.

\begin{itemize}
\item Isochrone fitting.
\item Relative MS-fitting Method.- This is a version of first, using age-insensitive regions (the location of the MS and RGB).
\item $\Delta$Color.- The color of the MSTO is a strong function of age, while the color of the RGB is relatively insensitive to age. Thus, $\Delta$Color (MSTO - RGB) is sensitive to the age of a globular cluster (\citealt{Sarajedini1990-GC-Age}; \citealt{VandenBerg1990-RelativeAgeClusters}).
\item  $\Delta$Magnitude.- It uses the difference in magnitude between the MSTO (or the SGB) and the HB as an age diagnostic (e.g., \citealt{Renzini1991-Magallanic}).
\item The Horizontal Branch Morphology Method (\citealt{Faulkner1966-HB}; \citealt{Lee1994-GC-Age_HBmorphology}; \citealt{Jimenez1998-GC-Age}).
\item The Luminosity Function Method  (\citealt{Jimenez1996-GC-Age}; \citealt{Jimenez1998-GC-Age}).
\item Comparison of the Spectral Energy Distribution with theoretical synthesis models (extragalactic method, starburst, etc. See, for instance, \citealt{Wang2010-M31-GC-Age}).
\end{itemize}

Note, however, that \citet{VandenBerg1990-AgeGC} have concluded that age is not the reason of the diversity of the HB morphology for cluster of the same metallicity. For the impact of different chemical compositions on the age dating technique, an interesting discussion is presented in \citet{MarinFranch2009-GC-AbundanceAge}.

The Luminosity Function Method proposed by \citet{Jimenez1996-GC-Age} --see additional details in \citet{Jimenez1998-GC-Age}-- takes into account the whole cluster sequence. It is divided into five different sectors, each of them providing a different parameter: the more massive red giants provide a normalization factor; the distance modulus comes from the red giants and subgiants; the more massive MS for a completeness check, the other MS provide the IMF, and the age is derived from the subgiants coming off the MS.

An additional point that needs to be remarked is the fact that the derived ages for a significant number of clusters are older that the very precise value establish by the $\Lambda$CDM concordance model (13.799$\pm$0.021 Gyr estimated by \citealt{PlanckCollaboration2015_CosmologicalParameters}).
 \citet{Krauss2003-Globular-AgeEstimates}, using Monte Carlo simulations, estimated that about one third might be older.
More recently,   \citet{Wang2010-CosmicAge} re-estimated the age of nine very old GC in M31 (thus, the distance is not an issue) and another located at a distance of z=1. Six of them, including the last one, present a conflict with the 
cosmological age. Thus, it seems that some problem still remains.

%%%%%%%%%%%%%%%%%%%%%%%%%%%%%%%%%%%%%%%   FIGURE 
\begin{figure}
\center
\includegraphics[width=1.0\textwidth,scale=1.0]{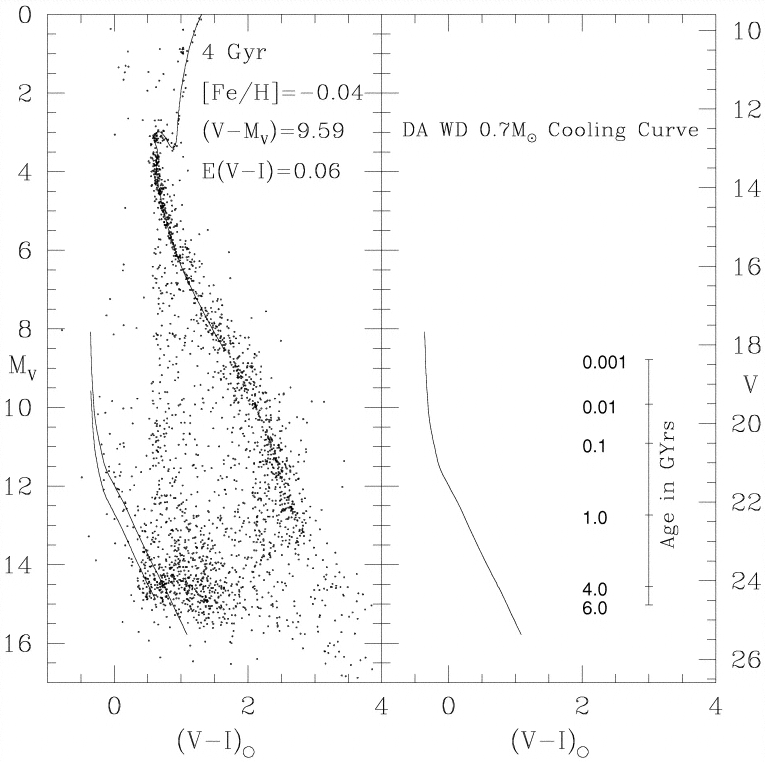} 
\caption{\label{Richer1998M67CoolingAgeWD} 
%, 
Color-Magnitude diagram for the $\sim$4 Gyr cluster M67, including the WD cooling sequence (figure from \citealt{Richer1998-M67-WD}).
}
\end{figure}
%%%%%%%%%%%%%%%%%%%%%%%%%%%%%%%%%%%%%%%

\subsubsection{White dwarfs\label{whitedwarfs}}
%
%%%%%%%%%%%%%%%%%%%%%%%%%%%%%%%%%%%%%%%%%%%%%%%%%%%%%%%%%%%%%%%%%%%%%%%%%%%%%%%%%%%%%%%%%%%%%

White dwarfs correspond to the last phase of a solar-like or a low-mass star, after they have exhausted all available 
energy sources based on fusion. All that is left is a slow cooling process, when they  release their thermal energy. Thus, they become redder and fainter with age. This fact is very well illustrated for one of the best known old open clusters, M67.
Figure \ref{Richer1998M67CoolingAgeWD}, taken from \citealt{Richer1998-M67-WD}), has a original caption stating: 
{\it ``Left.- M67 MV, (V-I) CMD for the entire cluster is shown. Only objects passing a shape test, indicating that they were likely to be stars are included in these diagrams except that theoretical cooling curves for 0.7 (upper) and 1.0 $M_\odot$ DA WDs are included. Also shown is an isochrone for 4 Gyr for a metallicity of [Fe/H] = -0.04, which is appropriate to M67. Right.- Re-plot of the 0.7 
$M_\odot$ cooling curve, indicating along it cooling times to various magnitudes.''}

There are several draw backs for this technique: the mass of the progenitor is not very well constrained, the properties depend on H and He composition, the WD cooling track does not provide information regarding the time elapsed in the previous phases (as a giant and a MS star), WD are very faint and, in order to apply the method to an association, numbers are needed (a significant population, since, again, the IMF is ``conspiring'' against us). In addition, confusion by interlopers can be a real headache ({\it bona fide} membership always is). In any case, by definition, ages from WD cannot be absolute and are model dependent.

Different sets of models, to name a few,  can be found in \citet{Bergeron1995-WD-models}, \citet{Hansen1999-WD-Models}
or \citet{Salaris2000-WD-CoolingAges}.
For a use of WD to date a multiple system, and interesting exercise has been carried out recently by 
\citet{Huelamo2009-WD-EB}, as it will described in the next subsection (\ref{eclipsingbinaries}). In addition, the WD mass function has been used to estimate de ages of the galactic population  (\citealt{Kepler2007-WD-Age-DiskGalaxy}).

%%%%%%%%%%%%%%%%%%%%%%%%%%%%%%%%%%%%%%%   FIGURE 
\begin{figure}
\center
\includegraphics[width=1.0\textwidth,scale=1.0]{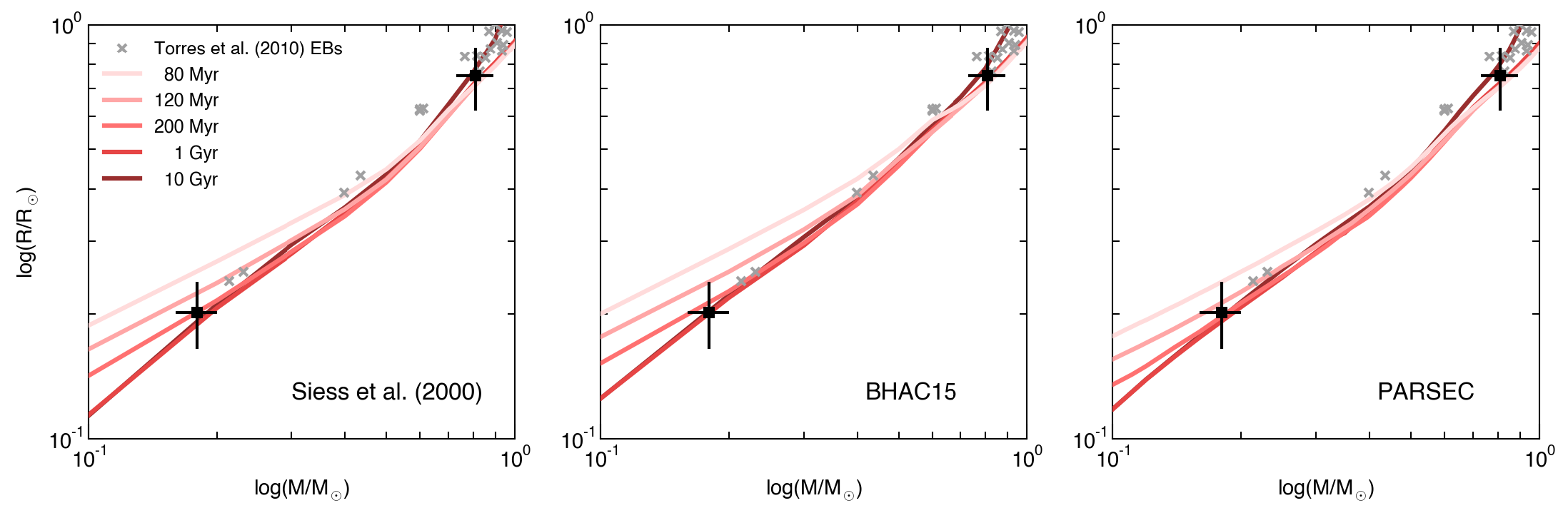} 
\caption{\label{David2015_HII2407_RadMass} 
 Figure taken from  \citet{David2015-EB-Pleiades-K2}. The new eclipsing binary HII2407, a  {\it bona fide} member of 
the Pleiades cluster.  The different panels display comparisons with several sets of isochrones.
% \citet{}
}
\end{figure}
%%%%%%%%%%%%%%%%%%%%%%%%%%%%%%%%%%%%%%%

\subsubsection{Sizes: eclipsing binaries\label{eclipsingbinaries}}
%
%%%%%%%%%%%%%%%%%%%%%%%%%%%%%%%%%%%%%%%%%%%%%%%%%%%%%%%%%%%%%%%%%%%%%%%%%%%%%%%%%%%%%%%%%%%%%

Coevality is a key property when deriving ages. In binaries (or multiple systems), some components are easier to handle  than others because their mass or evolutionary stage, so if both were born at the same time the age estimate derived for one can be  transferred to the other/s. In addition, coevality  is essential to calibrate theoretical isochrones, not only by using stellar associations, but also in binaries. Eclipsing binaries play an essential role.

There has been a very large amount of effort invested in the search and characterization of eclipsing binaries in different evolutionary stages 
(\citealt{Popper1980-EB-Masses}; \citealt{Strassmeier1993-CABS} \citealt{Barrado1994-Status-RSCVn}; \citealt{Malkov2006-EclipsingVariables-EB}). Recently a significant number of both low-mass and PMS EB have been incorporated into this essential database (\citealt{Shkolnik2008-QuadrupleSystem-EB}; \citealt{Stassun2014-PMS-EB}; \citealt{David2016-EB-UpperSco-K2}).

One very recent result comes from  \citet{David2015-EB-Pleiades-K2}, as shown in Fig. \ref{David2015_HII2407_RadMass}. The data come from the Kepler K2 mission (\citealt{Howell2014-Kepler-K2}). The EB belongs to the Pleiades cluster (130 Myr). As can be seen, even in this optimal situation the EB does not provide a strong constraint for the age. Another comparison can be found in \citet{Huelamo2009-WD-EB}, in this case with a triplet which includes a EB and a  visual WD. Despite the fact the spectroscopic properties of the WD can be established with a certain degree of precision, the overall properties of the system remain elusive.

In general, it can be concluded that masses and radii do not fit in the isochrones, even for objects with the same age and metallicity, such as eclipsing binaries in open clusters, the optimal case. So, this situation shows our real limits to our ability to estimate the stellar ages (and other properties).

%%%%%%%%%%%%%%%%%%%%%%%%%%%%%%%%%%%%%%%   FIGURE 
\begin{figure}
\center
\includegraphics[width=1.0\textwidth,scale=1.0]{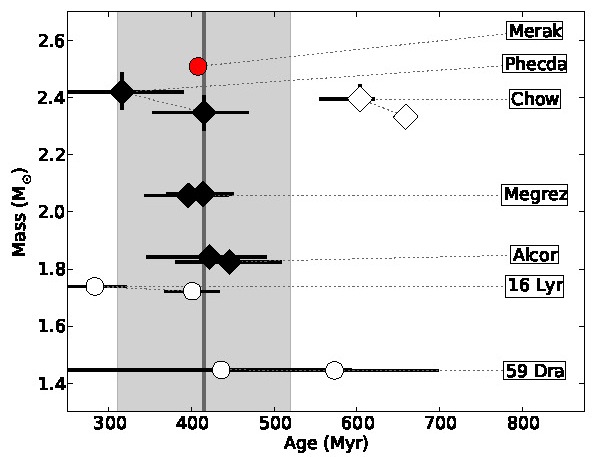} 
\caption{\label{Jones2015_A_Age_UMaMG} 
%, 
Panel c from figure 10 in \citet{Jones2015-UMaG-Interferometry}.
}
\end{figure}
%%%%%%%%%%%%%%%%%%%%%%%%%%%%%%%%%%%%%%%

\subsubsection{Sizes: interferometry\label{interferometry}}
%
%%%%%%%%%%%%%%%%%%%%%%%%%%%%%%%%%%%%%%%%%%%%%%%%%%%%%%%%%%%%%%%%%%%%%%%%%%%%%%%%%%%%%%%%%%%%%

Figure \ref{Jones2015_A_Age_UMaMG}, taken from \citet{Jones2015-UMaG-Interferometry}, shows the results of an interferometry survey in seven members of a coeval moving group, Ursa Majoris (UMaG, \citealt{Roman1949-UMaG}). A review regarding this technique can be found in \citet{Cunha2007-AsteroseismologyInterferometry-Review}. The canonical age estimate for this comoving stream is  $\sim$300 Myr (\citealt{Soderblom1993-Dynamics-UMaG} and references therein). A more recent estimate, by adding a significant number of new candidate members evolving off the MS or in the giant branch produces 500$\pm$100 Myr, significantly older (\citealt{King2003-Dinamics-Activity-UMaG}).
The original caption of this figure reads:
{\it ``Distribution of stellar masses versus age for 7 stars in the Ursa Major moving group as determined 
using the vZ gravity darkening law (10a), ELR law (10b), and both (10c) with the model described in Section 4.1.}
--their section in the original article-- {\it The circles are slowly rotating stars (Ve $<$ 170 km s$^{-1}$) and the diamonds are rapidly rotating
 (Ve $>$ 170 km s$^{-1}$). The black points are nucleus members and the white points are stream
members. The red point shows the mass and age of the nucleus member, Merak, that was previously observed by
\citet{Boyajian2012-RadiusTemp-AFG} and is discussed here in Section 4.3.}
--again, their section-- {\it In some cases, the size of the statistical
 error bar is smaller than the size of the symbol. The dark vertical lines represent the median in the ages, 
the shaded regions represent the gaper scale (the standard deviation equivalent discussed
in Section 5.4). The dotted lines in 10c connect the age and mass estimates from the two different laws.''}
This works concludes that the MG age is 414$\pm$28 Myr. It makes use of an evolutionary code  \citep[MESA,][]{Paxton11}, but also rejects some outliers. Note, however, that this error is significantly smaller than the grey area in the figure. We refer the readers to the paper for additional information regarding the details.

In any case, the technique has severe limitations, since essentially only can be applied to stars large enough or nearby enough (or both) to be able to derive a measurement (a radius) with a reduced error-bar and, in any case, other stellar parameters, such as the mass, should be derived with other techniques.

%%%%%%%%%%%%%%%%%%%%%%%%%%%%%%%%%%%%%%%%%%%%%%%%%%%%%%%%%%%%%%%%%%%%%%%%%%%%%%%%%%%%%%%%%%%%%
%
\subsection{Lithium, a ``perfect'' element as a stellar tracker\label{subsec:lithium}}
%
%%%%%%%%%%%%%%%%%%%%%%%%%%%%%%%%%%%%%%%%%%%%%%%%%%%%%%%%%%%%%%%%%%%%%%%%%%%%%%%%%%%%%%%%%%%%%

%%%%%%%%%%%%%%%%%%%%%%%%%%%%%%%%%%%%%%%   FIGURE 
\begin{figure}
\center
\includegraphics[width=1.0\textwidth,scale=1.0]{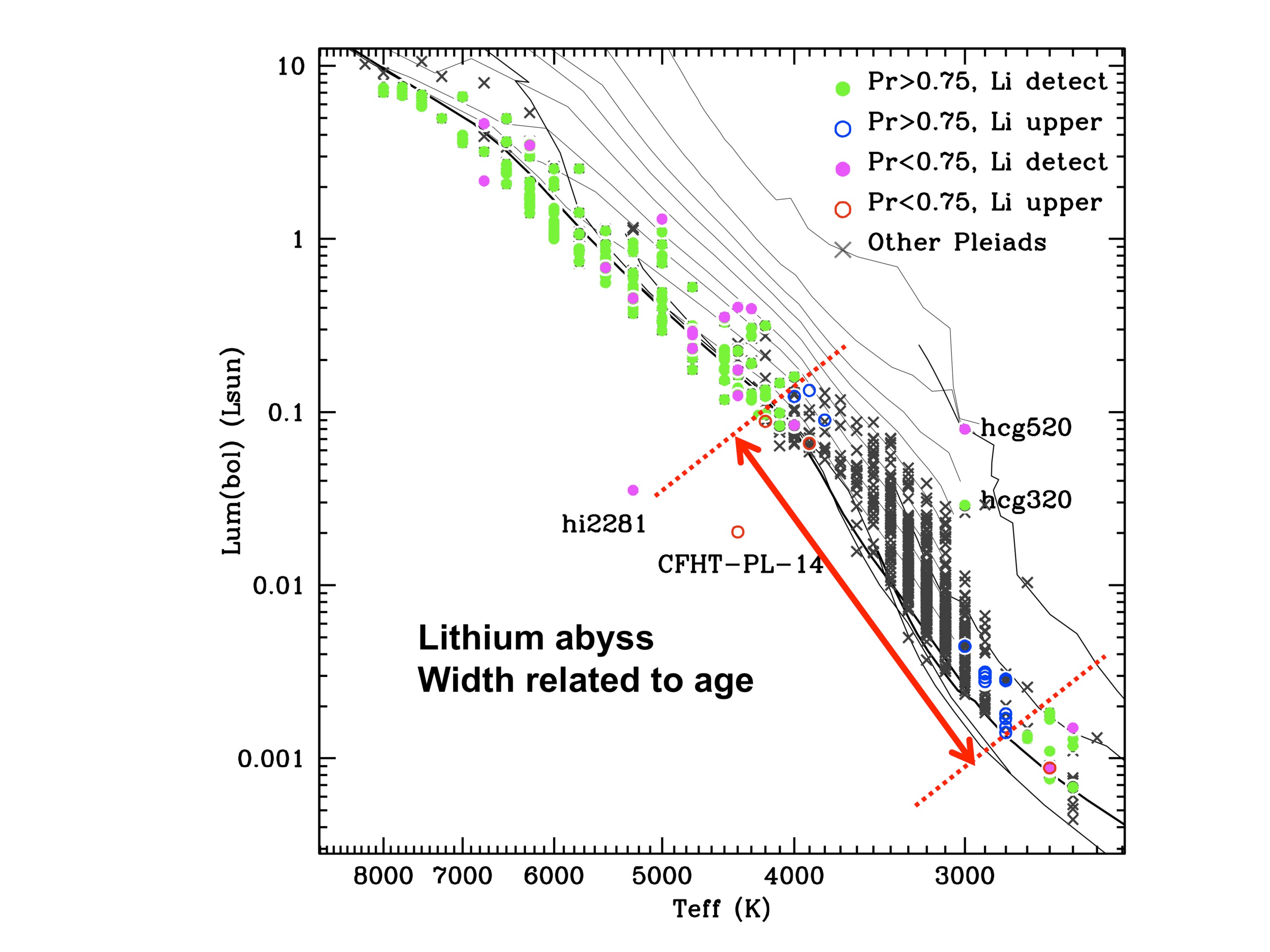} 
\caption{\label{Pleiades_HRD_LiAbyss} 
%, 
Herzprung-Russell diagrams displaying Pleiades data. 
The isochrones correspond to \citet{Siess2000.1} --1, 3, 5, 7, 10, 15, 20, 30, 50, 125, Myr and 5 Gyr--
and BT-Settl by  the Lyon group (\citealp{Allard2012.1}) --1, 20, 120 Myr and 10 Gyr--.
The 120/125 Myr isochrones are high-lighted.
 We have distinguished four cases:
 green solid circles for  lithium detection and membership probability larger than 0.75;
 blue open circles for  lithium upper limits and membership probability larger than 0.75;
 magenta solid circles for  lithium detection and membership probability less than 0.75;
 red open circles for  lithium upper limits and membership probability less than 0.75.
 Membership probabilities come from  \citet{Bouy2013.1}. Details in \citet{Barrado2016-Li-Pleiades}.
}
\end{figure}
%%%%%%%%%%%%%%%%%%%%%%%%%%%%%%%%%%%%%%%

%%%%%%%%%%%%%%%%%%%%%%%%%%%%%%%%%%%%%%%   FIGURE 
\begin{figure}
\center
\includegraphics[width=1.0\textwidth,scale=1.0]{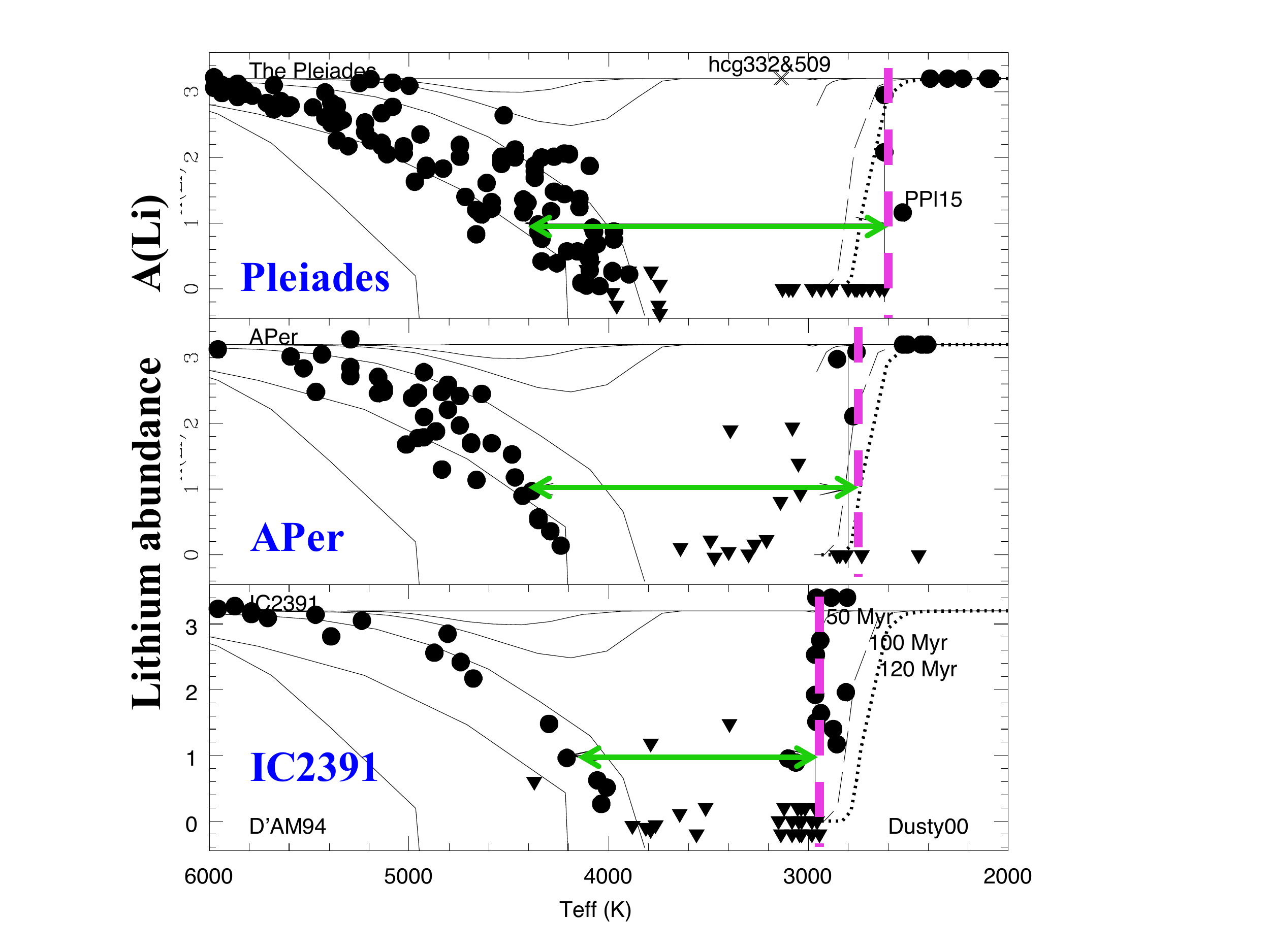} 
\caption{\label{LDB_ALi_Teff-3clusters} 
Lithium abundance versus the effective temperature
for three well known clusters: the Pleiades, Alpha Per and IC 2391
(see \citealp{Barrado2004-Li-Ha-IC2391}
and \citealp{Barrado2011-LithiumAges}). The figure also shows the ``lithium abyss'' for late-F and early-M and  depletion boundary (see body-text). Note the two mid-M dwarfs with lithium
(HCG332 and HCG509, \citealt{Oppenheimer1997.1}). 
%, 
}
\end{figure}
%%%%%%%%%%%%%%%%%%%%%%%%%%%%%%%%%%%%%%%

%%%%%%%%%%%%%%%%%%%%%%%%%%%%%%%%%%%%%%%   FIGURE 
\begin{figure}
\center
\includegraphics[width=1.0\textwidth,scale=1.0]{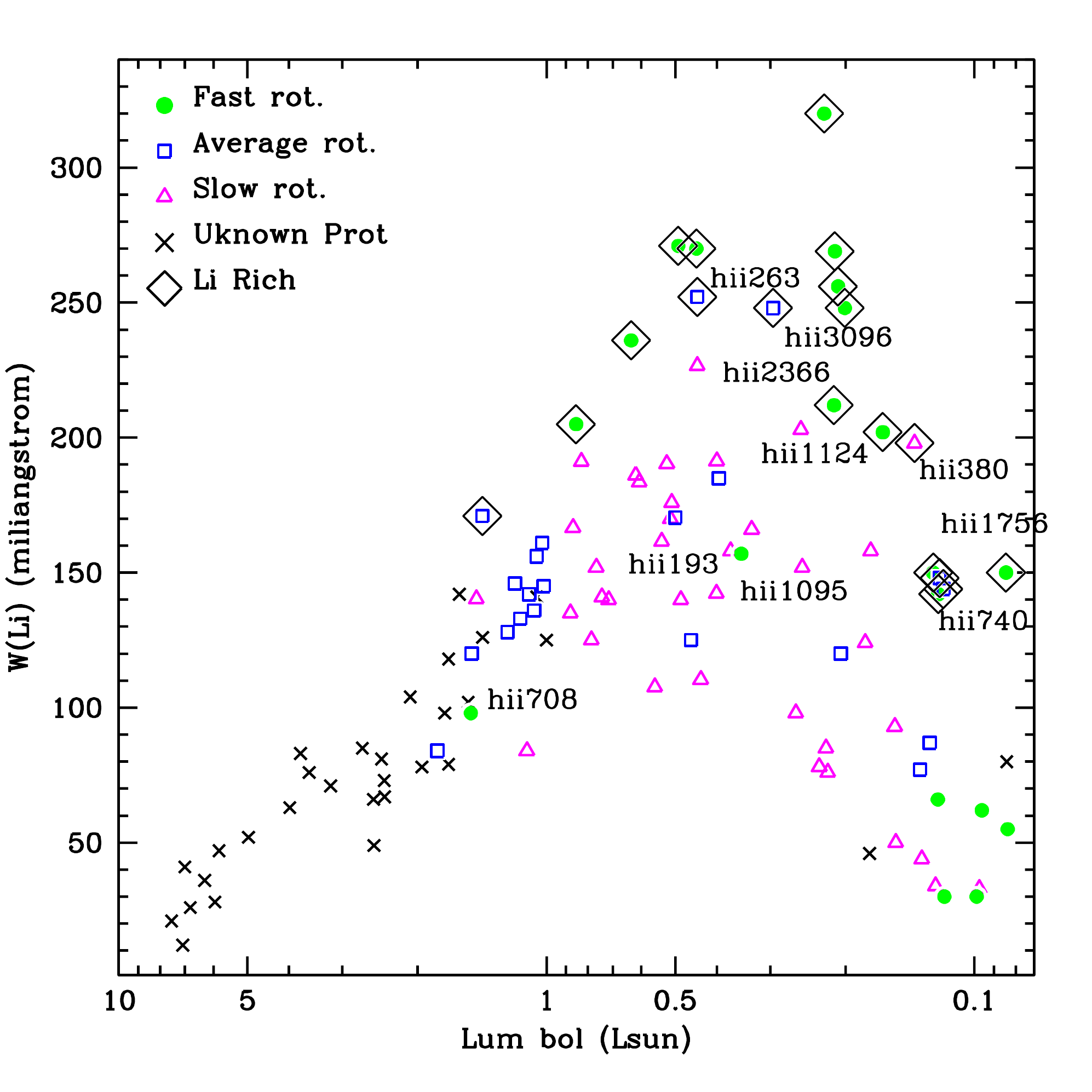} 
\caption{\label{LumBol_WLi_Pleides_Pr0p75_S_Pclass} 
%, 
 Lithium equivalent with versus the bolometric luminosity
for the high probable members (probability larger than 0.75).
For those stars with multiple values of W(Li), only one value have been selected.
 Only single stars are displayed. Lithium-rich stars are also indicates with
large, open diamonds.  Details in \citet{Barrado2016-Li-Pleiades}.}
\end{figure}
%%%%%%%%%%%%%%%%%%%%%%%%%%%%%%%%%%%%%%%

\subsubsection{Lithium evolution in F, G and K stars\label{lithiumFGK}}
%
%%%%%%%%%%%%%%%%%%%%%%%%%%%%%%%%%%%%%%%%%%%%%%%%%%%%%%%%%%%%%%%%%%%%%%%%%%%%%%%%%%%%%%%%%%%%%

The lithium abundance found in meteorites of the Solar System is in the range
A(Li)=3.1-3.2. This is, in fact, the assumed initial abundance for population I stars (for cosmological abundance and population II, see C. Charbonnel in this volume). However, lithium is a fragile element and can be destroyed easily, as happens inside the stars.

Standard models predict that the depletion happens during the pre-main sequence evolution. However, the observations show that it continues beyond the arrival to the ZAMS, so additional, non-standard mixing has to take place. 
As a matter of fact, this behaviour and dependence are clearly seen in open clusters of different ages  for stars cooler than  about 6300 K (\citealt{Boesgaard1991-LiAge-Cluster}).
Moreover,  for clusters older than the Pleiades there is a narrow effective temperature
range (6400-6900 K) which shows a large depletion of lithium abundance due to
non-standard mixing, the so 
called lithium gap, dip or chasm  (\citealp{Boesgaard1986-Li-Fdwarfs}; \citealp{Michaud1991-Li}; 
\citealp{Balachandran1995-Li-F-M67}).
In any case, the complexity of the evolution has been established by multiple studies focusing on
clusters of different ages. Seminal papers, to name a few,  are
\citet{Pilachowski1984-Li-NGC7789},  \citet{Pilachowski1986-Li-NGC7789},
\citet{Pilachowski1987-Li-F-Pleiades},  \citet{Pilachowski1988-L-NGC752-M67},  
for NGC7789, the Pleiades, NGC752, and M67;
\citet{Boesgaard1986-Li-Hyades},  \citet{Boesgaard1987-Li-Coma}, \citet{Boesgaard1987-Li-F-Hyades}, 
 \citet{Boesgaard1988-Li-Pleiade-APer},  \citet{Boesgaard1988-Li-HyadesMG-Praesepe},  
\citet{Boesgaard1989-LiBe-Hyades}, \citet{Barrado1996-Li-Hyades}
  for the Hyades, Coma,  the Pleiades and Alpha Per,  and Praesepe;
\citet{Soderblom1993-Li-Praesepe}, \citet{Soderblom1993-Li-Pleiades}, \citet{Soderblom1993-Li-UMaG} 
for Praesepe, the Pleiades and Ursa Majoris moving group.

More recently, additional observations for  clusters, generally younger,
 have been added. Again, just to provide some references: 
NGC2516 and M35, almost Pleiades twins 
(\citealt{Jeffries1998-LiRotation-NGC2516}; \citealt{Barrado2001-Li-M35}),
IC2602 and IC2391 (\citealt{Barrado1999-Li-LDB-IC2391}; \citealt{Randich2001-Li-K-IC2602}; 
\citealt{Randich2001_Li_Metal_IC2602_IC2391}; \citealt{Barrado2004-Li-Ha-IC2391}),
NGC2547 (\citealt{Jeffries2003-Li-NGC2547}),
IC4665 (\citealt{Jeffries2009-Li-IC4665}), and
Collinder 69 (\citealt{Dolan1999-Li-C69}, \citealt{Bayo2012-Li-Rotation-Activity-C69}).
In the very near future, the large scale spectroscopic survey Gaia-ESO 
(\citealt{Gilmore2012.GaiaESO}; \citealt{Randich2013.GaiaESO})
 will provide an extended database. A recent example is  the Vela OB2 association 
(\citealt{Sacco2015.VelaOB2}).

In the case of the Pleiades, a quite complete database is at hand, both in number of measurement of lithium in members
and in ancillary data such as accurate photometry (from the DANCE project, \citealt{Bouy2013.1}), rotational periods
(\citealt{Hartman2010.1}; \citealt{Rebull2016-Pleiades-K2-I, Rebull2016-Pleiades-K2-II};
\citealt{Stauffer2016-Pleiades-K2-III}) or activity (see detail in Barrado et al. 2016, submitted).
Figure \ref{Pleiades_HRD_LiAbyss} 
corresponds to an HR diagram where lithium information has been appended. The diagram clearly shows the ``lithium abyss'' between late-K and mid-M, significantly wider than the gap present in older stars at mid-F spectral type, already described above.

As a matter of fact, the equivalent width of this feature --W(Li)-- depends on the age, since late K and early M deplete lithium very fast. This is illustrated in Fig. \ref{LDB_ALi_Teff-3clusters}, where the lithium abundance and the effective temperature are displayed for three clusters (IC2391, Alpha Per and the Pleiades) with ages between 50 Myr and  125 Myr in the new LDB scale. 

The role of the rotation in the W(Li) spread observed for stars cooler than the Sun was well established by \citet{Soderblom1993-Li-Pleiades}. Using all the available data, Figure \ref{LumBol_WLi_Pleides_Pr0p75_S_Pclass} displays the lithium equivalent width versus the luminosity for probable single stars belonging to the Pleiades, where rotation has been taken into account. Thus, although the lithium content is a characteristic that depends primarily on mass and age, other factors modify its evolution and create a quite complicated picture.

In fact, several works have been published lately trying to understand the lithium content in solar-type stars from different perspectives.
\citet{Bouvier2008.1} (see also \citealt{Eggenberger2012-Li-Rotation-Disk-PMS}) 
investigates the effect of the disk life-time on the rotation and lithium:
 slow rotation would be the consequence of long-lasting star-disk interaction during the PMS and would produce a significant decoupling 
between the core and the convective envelope, with the final consequence of extra-mixing and higher lithium depletion.
On the other hand, \citet{Somers2014.1} argue that the strong magnetic field in fast rotators during the early 
PMS enlarges the radii and diminishes the temperature of the bottom of the convective envelope, provoking over-abundances. The effective 
temperature would also be affected, due to the larger spot coverage (with cooler temperatures).
These investigations assume that the lithium spread for a given mass corresponds to real abundance differences. 
However, on the other side, \citet{Soderblom1993-Li-Pleiades}, \citet{Stuik1997-Pleiades-Li-K-Activityspread}, \citet{Jeffries1999-Li-Pleiades}, 
 \citet{King2000-Lithium-Rotation}, 
\citet{Barrado2001-Li-Na-Ka-Activity-Pleiades}, \citet{King2004-Activity-Alkali},  \citet{King2010-Li-K-Scatter-Pleiades},
 all in the case of the Pleiades but with very different approaches,
 have tried to verify whether the real cause
is related to the presence of surface  inhomogeneities and their effect on the observed lithium equivalent width. Some of these
works conclude that at least partially the spread is due to atmospheric effects, others argue that most come from real differences in the
depletion rate during the PMS evolution. The debate is still open.

%%%%%%%%%%%%%%%%%%%%%%%%%%%%%%%%%%%%%%%   FIGURE 
\begin{figure}
\center
\includegraphics[width=1.0\textwidth,scale=1.0]{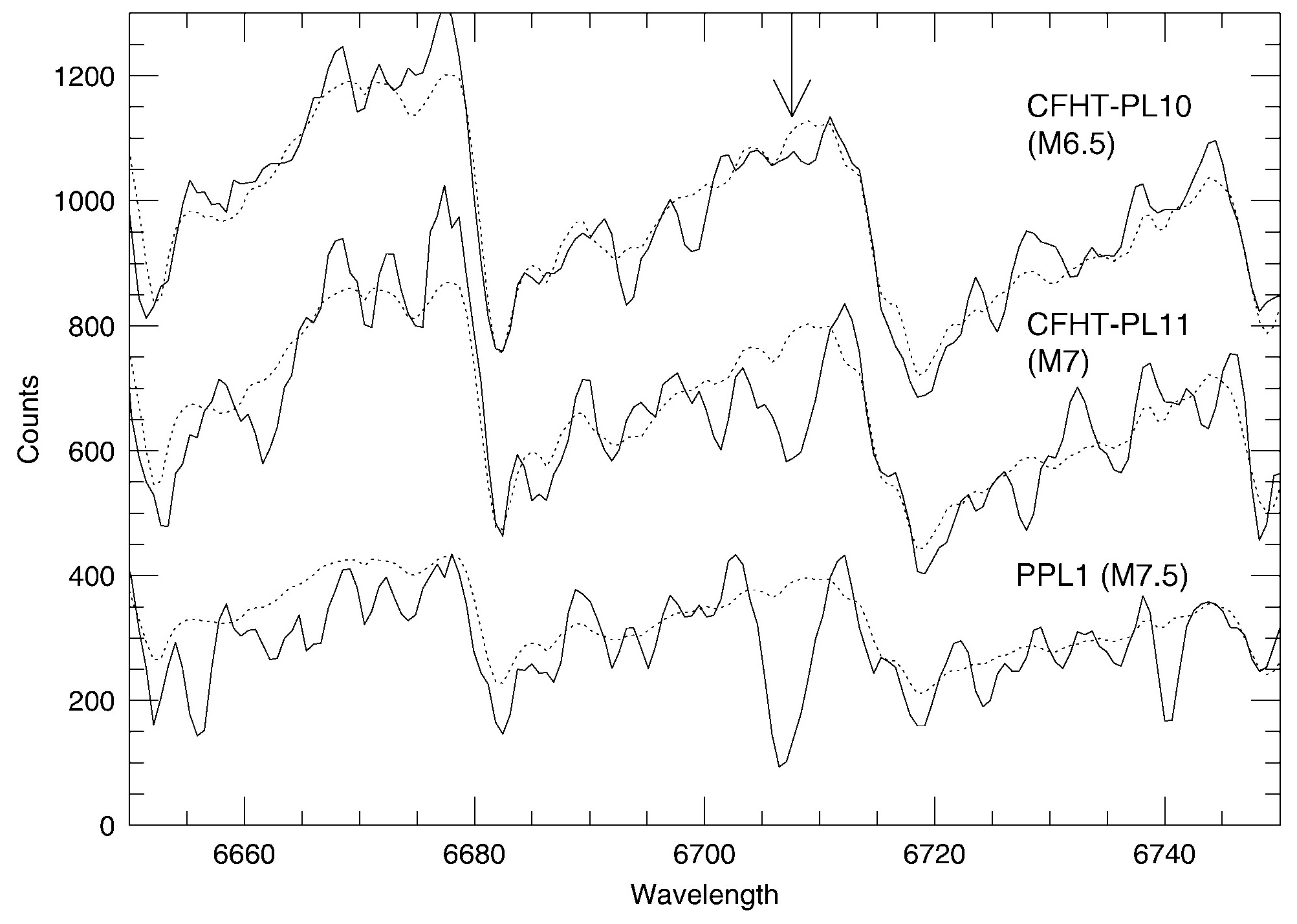} 
\caption{\label{Stauffer1998_Pleiades_LDB_specra} 
%, 
Spectra of three mid-M Pleiades members, from  \citet{Stauffer1998.2}. 
The presence of LiI 6707.8 \AA{ } feature is easily discerned for the coolest objects.
}
\end{figure}
%%%%%%%%%%%%%%%%%%%%%%%%%%%%%%%%%%%%%%%

%%%%%%%%%%%%%%%%%%%%%%%%%%%%%%%%%%%%%%%   FIGURE 
\begin{figure}
\center
\includegraphics[width=1.0\textwidth,scale=1.0]{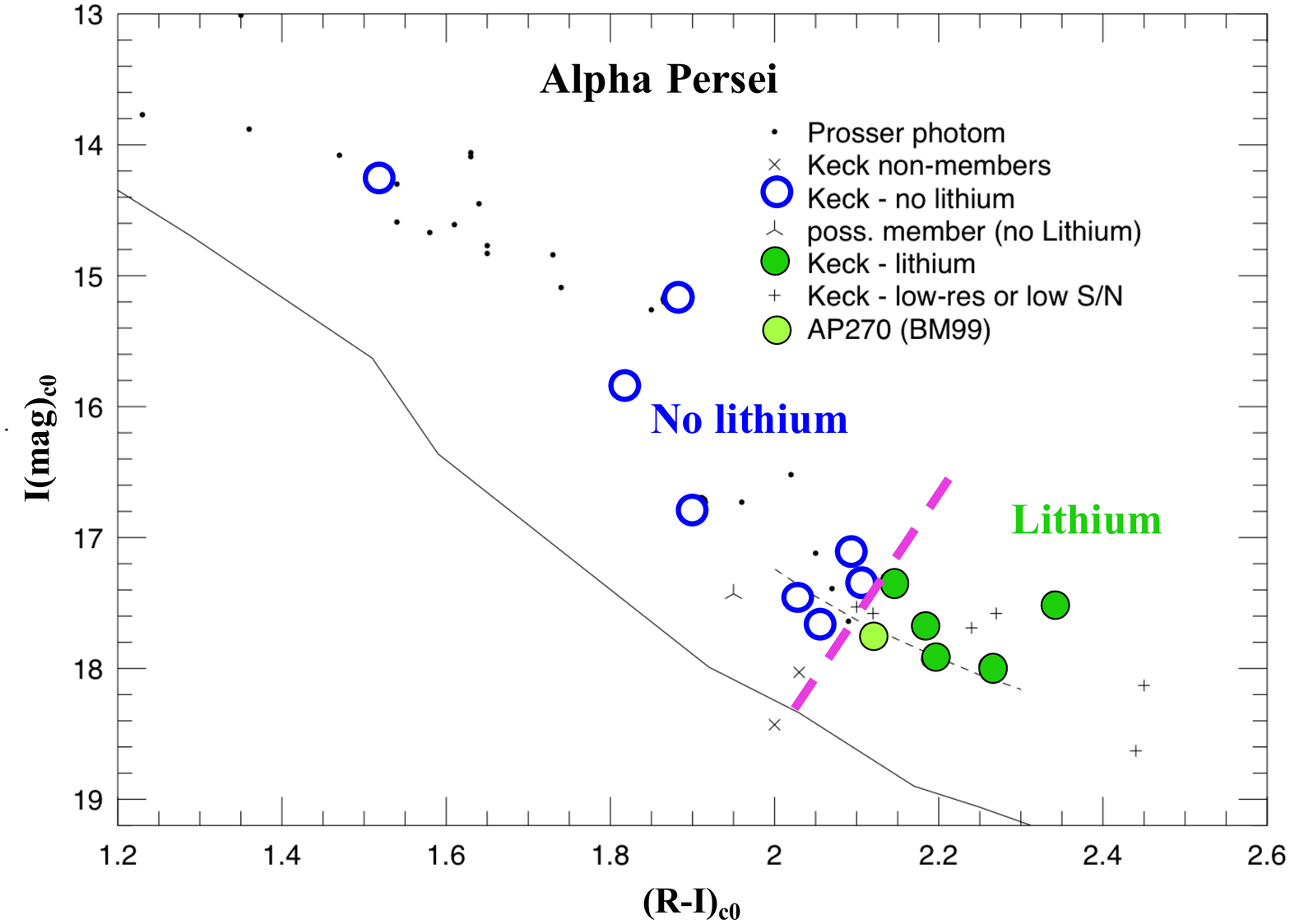} 
\caption{\label{LDB_CMD_APer} 
Color-Magnitude diagram for  low-mass members of  the Alpha Per cluster, after \citet{Stauffer1999.1}.
Note the very narrow gap between members with and without lithium, since in this particular case the sampling 
is excellent.
%, 
}
\end{figure}
%%%%%%%%%%%%%%%%%%%%%%%%%%%%%%%%%%%%%%%

%%%%%%%%%%%%%%%%%%%%%%%%%%%%%%%%%%%%%%%   FIGURE 
\begin{figure}
\center
\includegraphics[width=1.0\textwidth,scale=1.0]{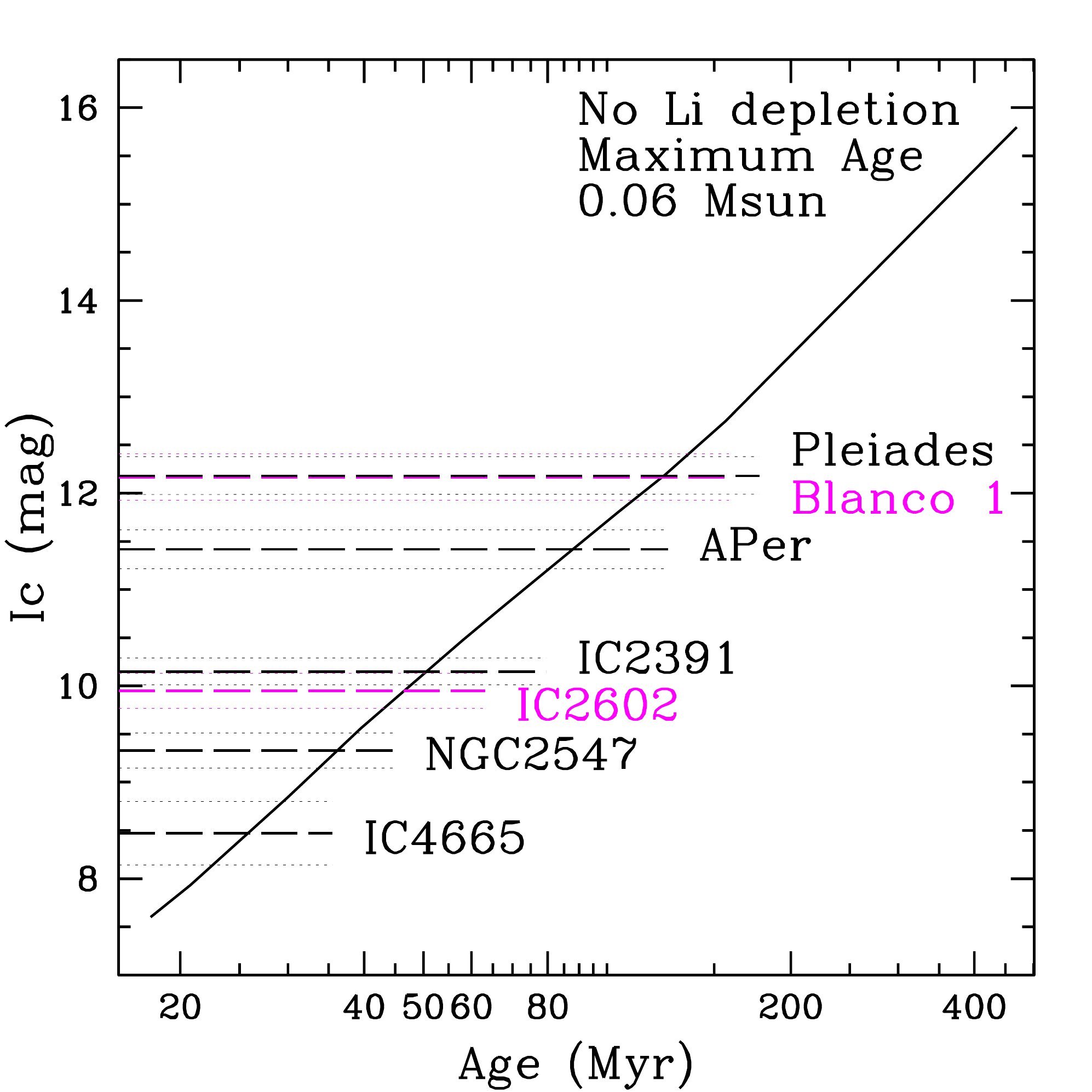} 
\caption{\label{LDB_MagAge} 
%, 
From \citet{Barrado2011-LithiumAges}.
Location of the LDB boundary and its relation with the cluster age.
We have represented the absolute $I$ magnitude for the LDB --dashed lines-- and the associated
errors --dotted lines. The wide line corresponds to the complete lithium depletion from theoretical BT-settl models from \citet{Allard2012.1}. 
}
\end{figure}
%%%%%%%%%%%%%%%%%%%%%%%%%%%%%%%%%%%%%%%

%%%%%%%%%%%%%%%%%%%%%%%%%%%%%%%%%%%%%%%   FIGURE 
\begin{figure}
\center
\includegraphics[width=1.0\textwidth,scale=1.0]{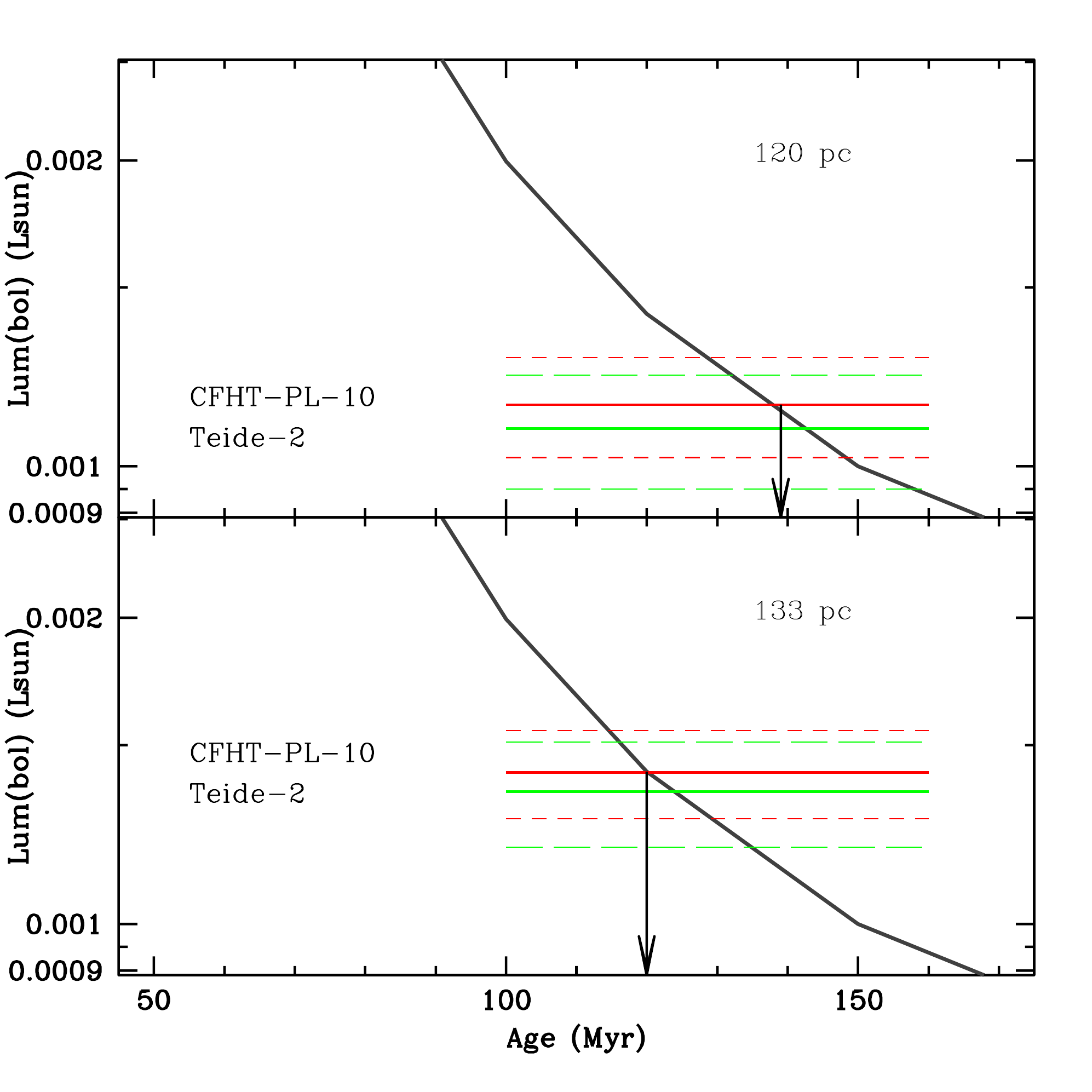} 
\caption{\label{Lumbol_Age_LDB_Pleiades} 
%, 
Bolometric luminosity  versus age. The thick line corresponds to 1\% depletion from \citet{Baraffe1998.1}.
 Red lines represent the last high probable Pleiades member, whereas the green line comes from the 
first member with lithium. Errors are included as dotted and dashed lines. 
The effect of the distance is also displayed in the panels. 
}
\end{figure}
%%%%%%%%%%%%%%%%%%%%%%%%%%%%%%%%%%%%%%%

%%%%%%%%%%%%%%%%%%%%%%%%%%%%%%%%%%%%%%%   FIGURE 
\begin{figure}
\center
\includegraphics[width=1.0\textwidth,scale=1.0]{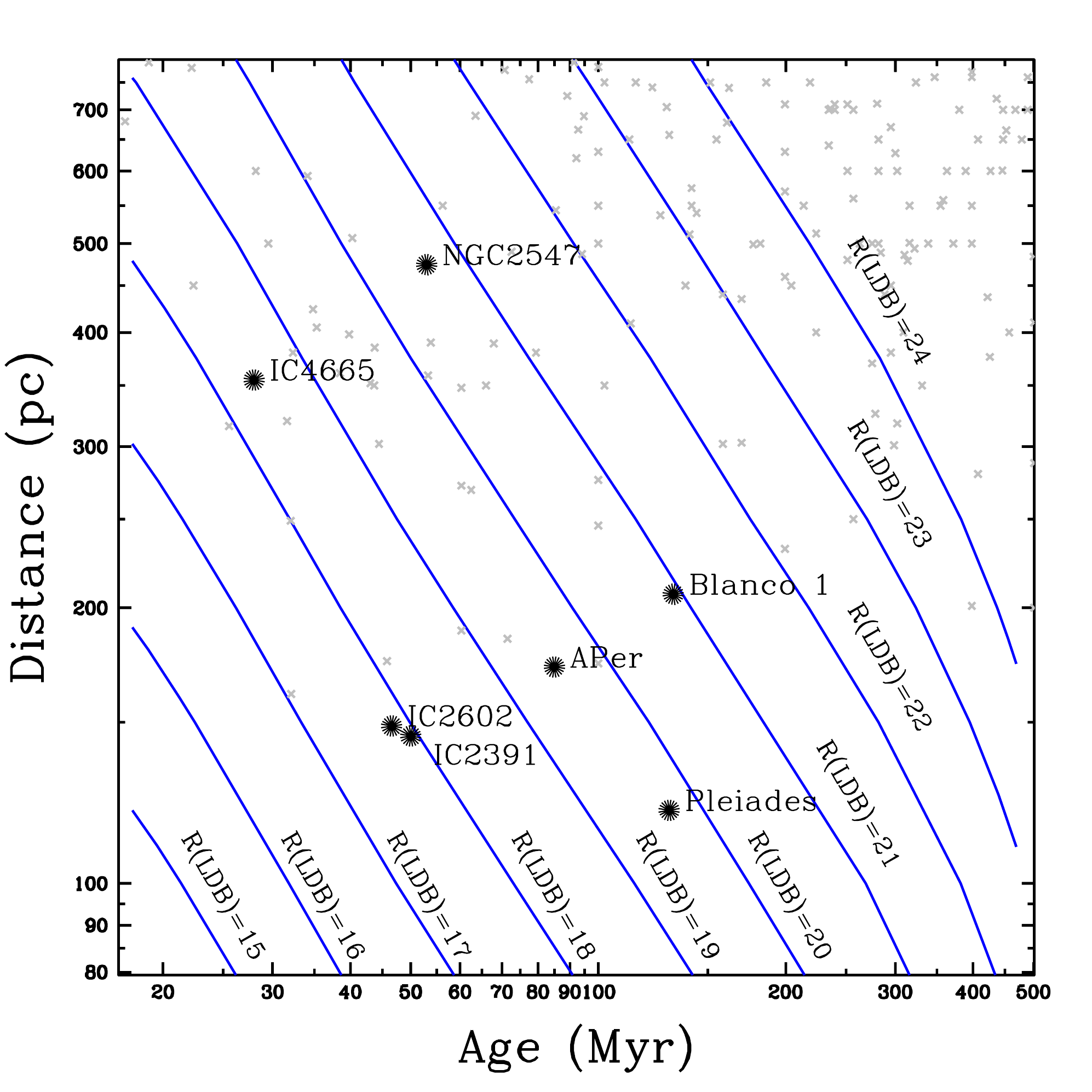} 
\caption{\label{LogDist_AgeLog_LDB_Rmag} 
%, 
After \citet{Barrado2011-LithiumAges}.
Relation between the distance and the age for the Lithium Depletion Boundary.
We have included the clusters listed in \citet{Dias2002-CatalogOpenClusters} as small grey crosses, whereas the seven
clusters with derived lithium ages appear as big symbols. The lines correspond to the apparent
magnitudes when lithium is 99\% depleted, from models from \citet{Baraffe1998.1}. Magnitudes
in the R band, where the LiI6707.8\AA{ }  doublet is located.
}
\end{figure}
%%%%%%%%%%%%%%%%%%%%%%%%%%%%%%%%%%%%%%%

%%%%%%%%%%%%%%%%%%%%%%%%%%%%%%%%%%%%%%%   FIGURE 
\begin{figure}
\center
\includegraphics[width=1.0\textwidth,scale=1.0]{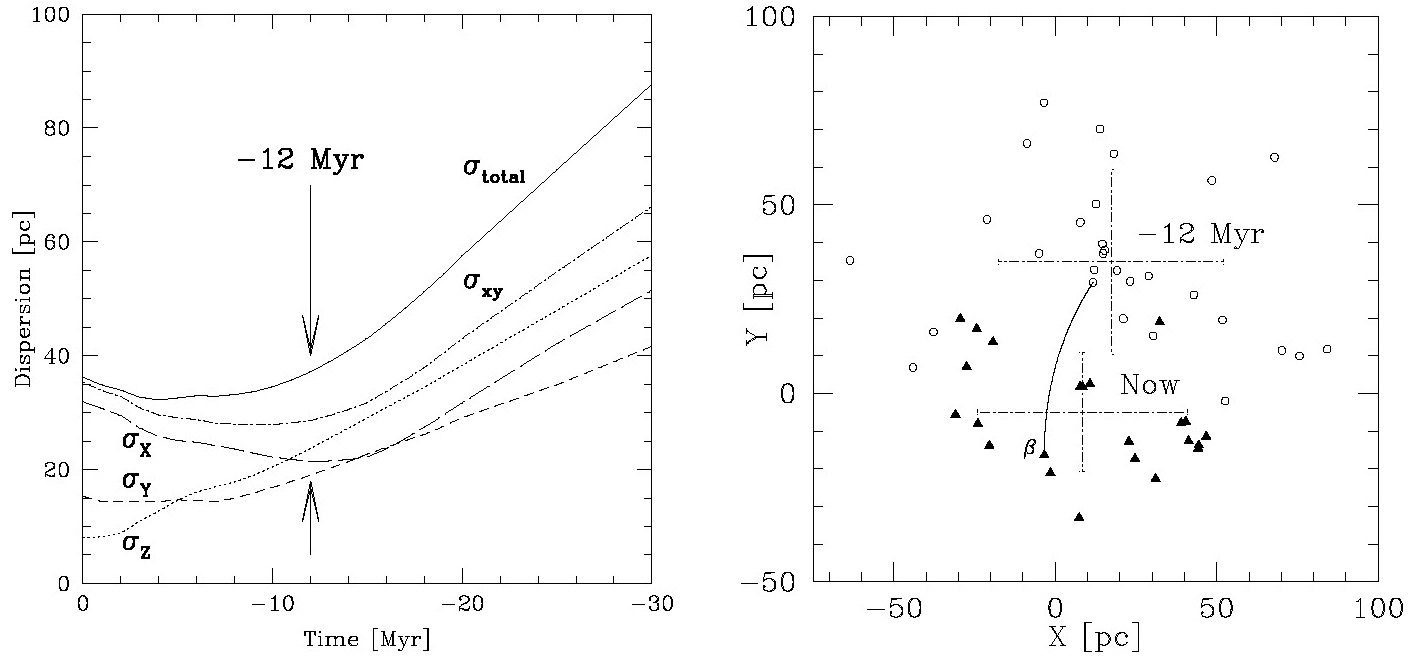} 
\caption{\label{Mamajek2014_BPMG_Age_Traceback} 
%, 
These two panels correspond to figures obtained from \citet{Mamajek2014-AgeBPMG}.
Left panel shows the dispersion in distance for different  members of the Beta Pic MG as a function of time, whereas the
right panel displays the current positions and the computed values 12 Myr ago in the (X,Y) space.
}
\end{figure}
%%%%%%%%%%%%%%%%%%%%%%%%%%%%%%%%%%%%%%%

%%%%%%%%%%%%%%%%%%%%%%%%%%%%%%%%%%%%%%%   FIGURE 
\begin{figure}
\center
\includegraphics[width=1.0\textwidth,scale=1.0]{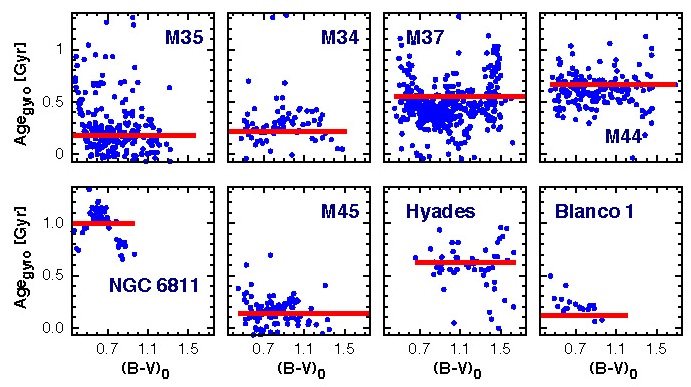} 
\caption{\label{Kovacs2015_GyroAges} 
%, 
The solid circles are  {\it computed} values using gyrochronology, whereas the horizontal blue line is the
adopted age for each cluster. Figure  from  \citet{Kovacs2015-Gyro}.
%{\it ``Predicted individual gyro-ages (dots) by using Eqs. (3)
% and (5) and the adopted isochrone cluster ages of Table 1 (lines).''}
}
\end{figure}
%%%%%%%%%%%%%%%%%%%%%%%%%%%%%%%%%%%%%%%

%%%%%%%%%%%%%%%%%%%%%%%%%%%%%%%%%%%%%%%   FIGURE 
\begin{figure}
\center
\includegraphics[width=1.0\textwidth,scale=1.0]{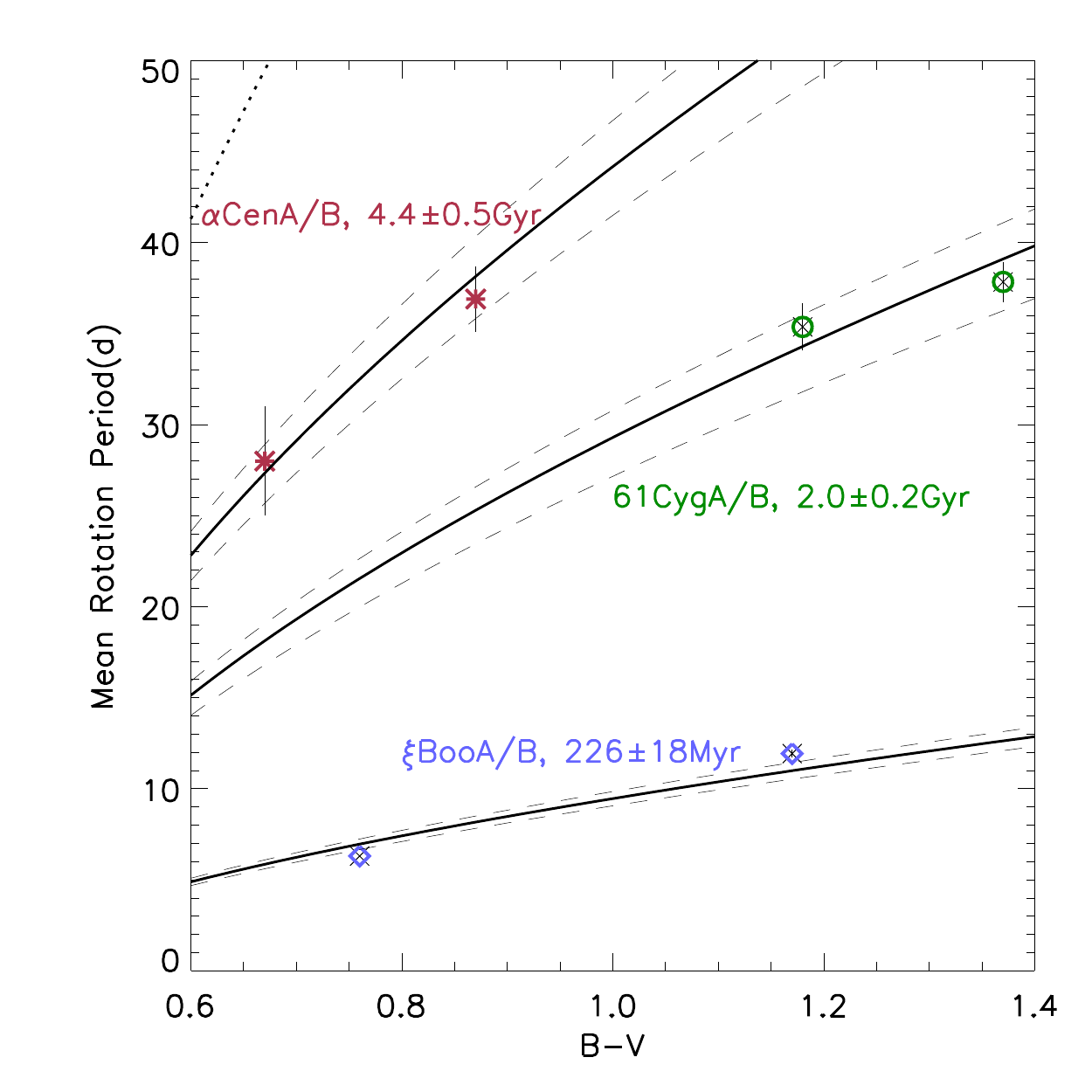} 
\caption{\label{Barnes2007_OldBinaries_Gyro} 
Rotation period versus color for three old wide binary systems. Gyrochronology provides ages for both components of both systems in agreement with each other. Figure extracted from 
\citet{Barnes2007-Gyro}. 
}
\end{figure}
%%%%%%%%%%%%%%%%%%%%%%%%%%%%%%%%%%%%%%%

%%%%%%%%%%%%%%%%%%%%%%%%%%%%%%%%%%%%%%%   FIGURE 
\begin{figure}
\center
\includegraphics[width=1.0\textwidth,scale=1.0]{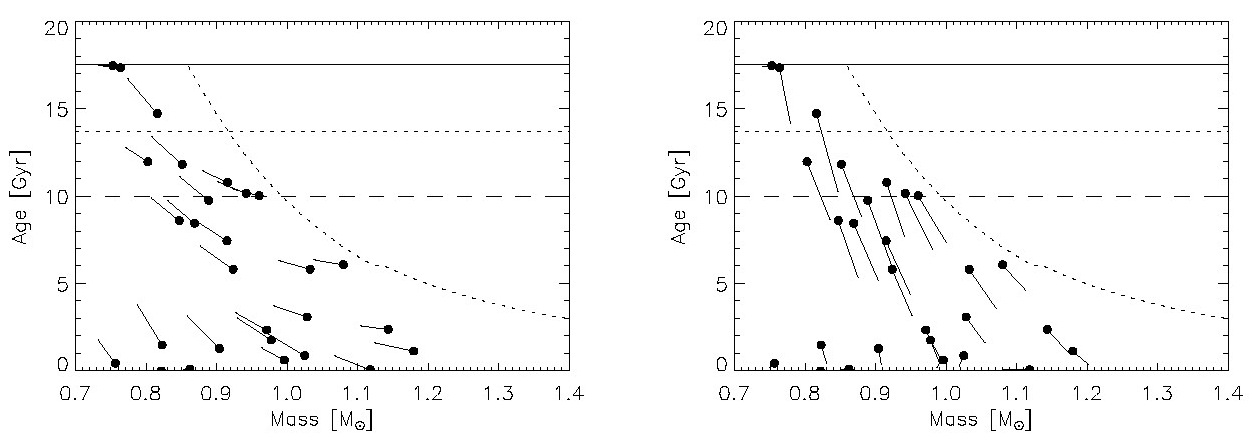} 
\caption{\label{Maxted2015_Exoplanets_Age_Gyro_Isochrones} 
%, 
Age versus mass and the effect of the helium abundance (left) and the mixing length (right). Figures from \citet{Maxted2015-ExoplanetAge-Gyro}.
}
\end{figure}
%%%%%%%%%%%%%%%%%%%%%%%%%%%%%%%%%%%%%%%

%%%%%%%%%%%%%%%%%%%%%%%%%%%%%%%%%%%%%%%   FIGURE 
\begin{figure}
\center
\includegraphics[width=1.0\textwidth,scale=1.0]{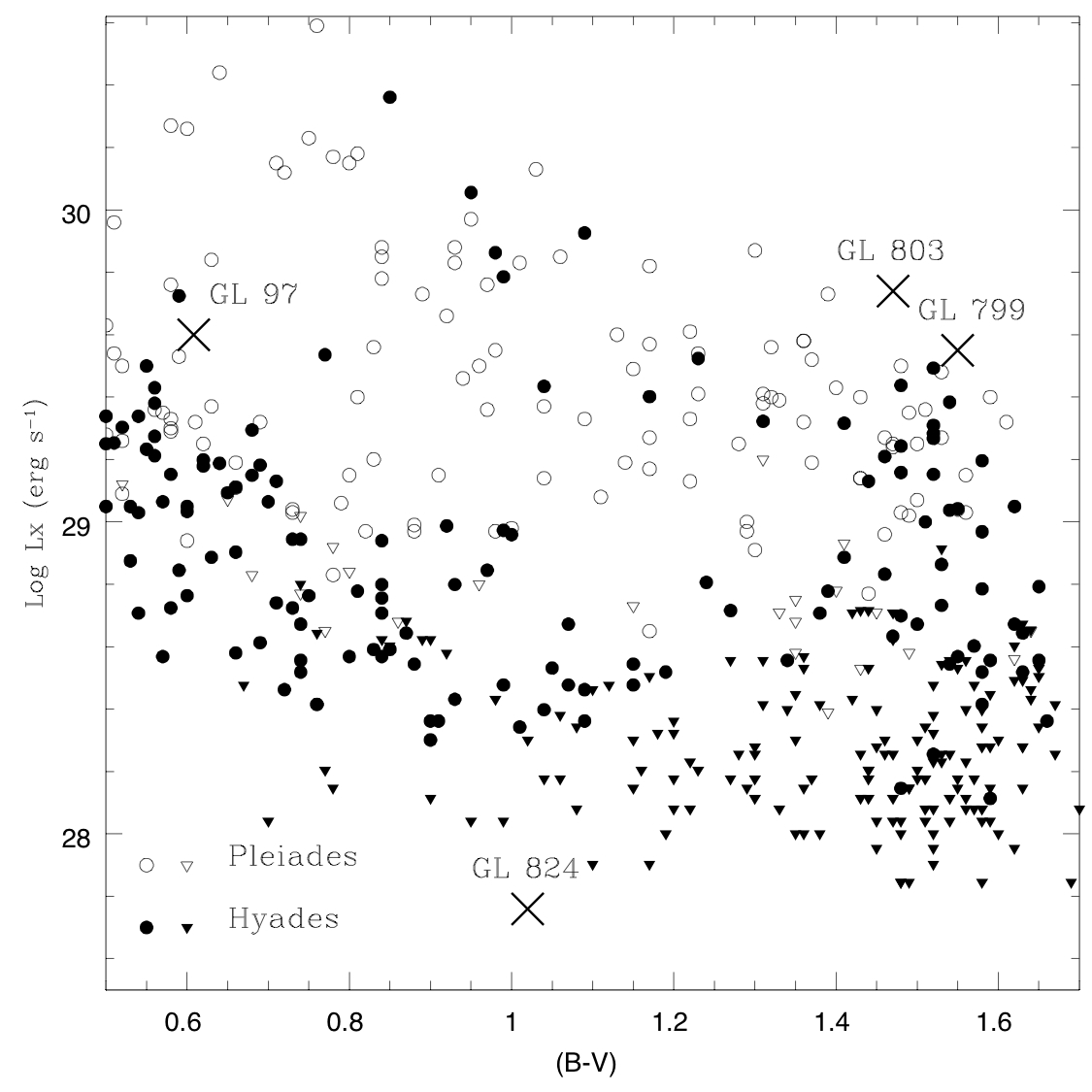} 
\caption{\label{BPMG_Lx_clusters} 
%, 
Taken from \citet{Barrado1999-BetaPic-Age}.
X-ray luminosities for Pleiades (open symbols)  and Hyades (solid symbols) members 
(125 versus $\sim$600 Myr) and a subsample of BPMG candidates.
Upper limits are represented as triangles.
}
\end{figure}
%%%%%%%%%%%%%%%%%%%%%%%%%%%%%%%%%%%%%%%

%%%%%%%%%%%%%%%%%%%%%%%%%%%%%%%%%%%%%%%   FIGURE 
\begin{figure}
\center
\includegraphics[width=1.0\textwidth,scale=1.0]{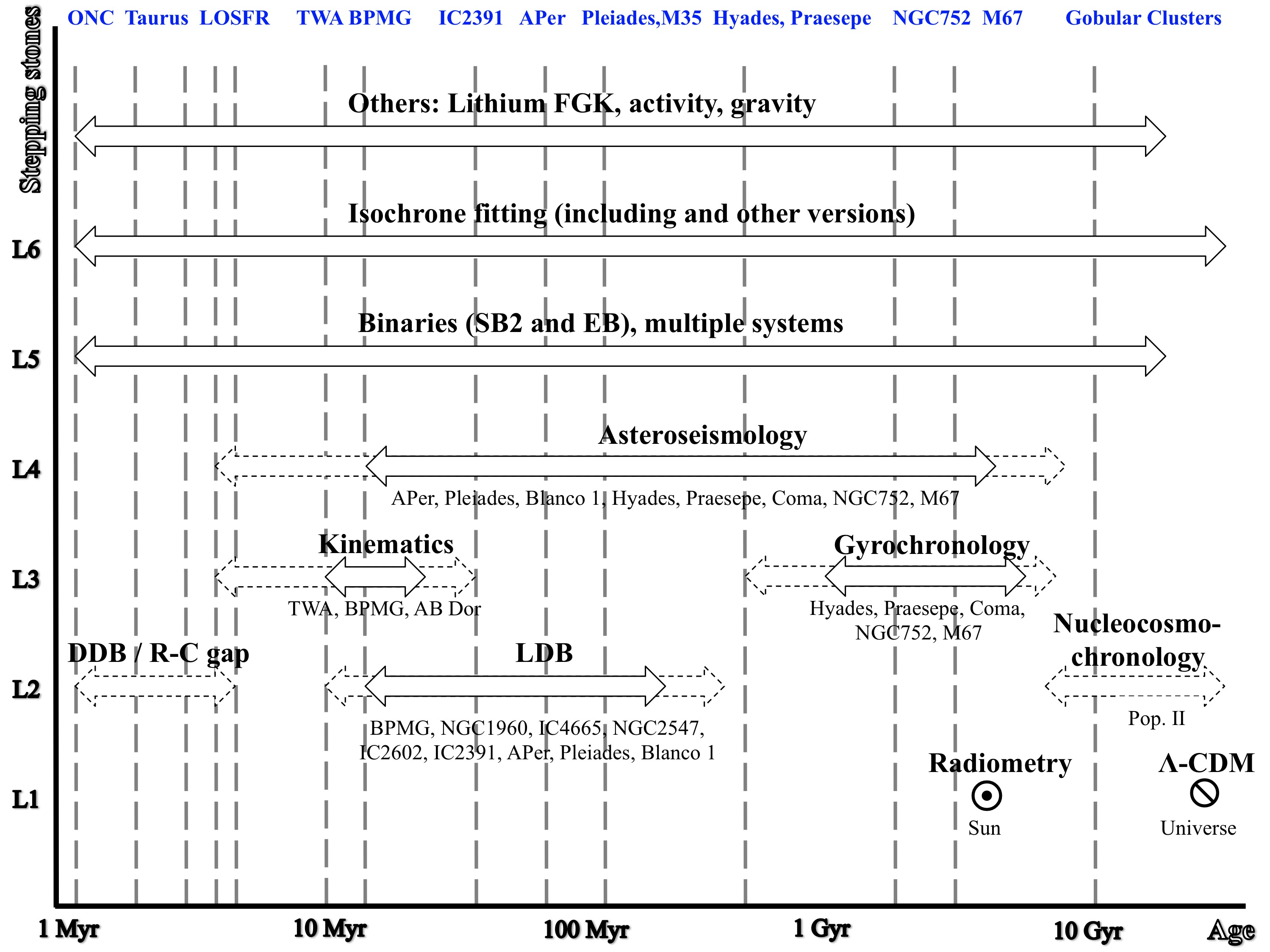} 
\caption{\label{Barrado_Age_SteppingStones_EES2015.jpg} 
%, 
A suggested age ladder or stairway. At the bottom the true age anchors: the Solar System and the universe ages. The second level in accuracy should correspond to the LDB and, once several observations issues are resolved, the nucleochronology for Population II and the deuterium depletion boundary (DDB), if this last phenomenon is, indeed, observable. Thus, different levels depend on the previous ones. We have also marked several relevant associations.
}
\end{figure}
%%%%%%%%%%%%%%%%%%%%%%%%%%%%%%%%%%%%%%%

%%%%%%%%%%%%%%%%%%%%%%%%%%%%%%%%%%%%%%%   FIGURE 
\begin{figure}
\center
\includegraphics[width=1.0\textwidth,scale=1.0]{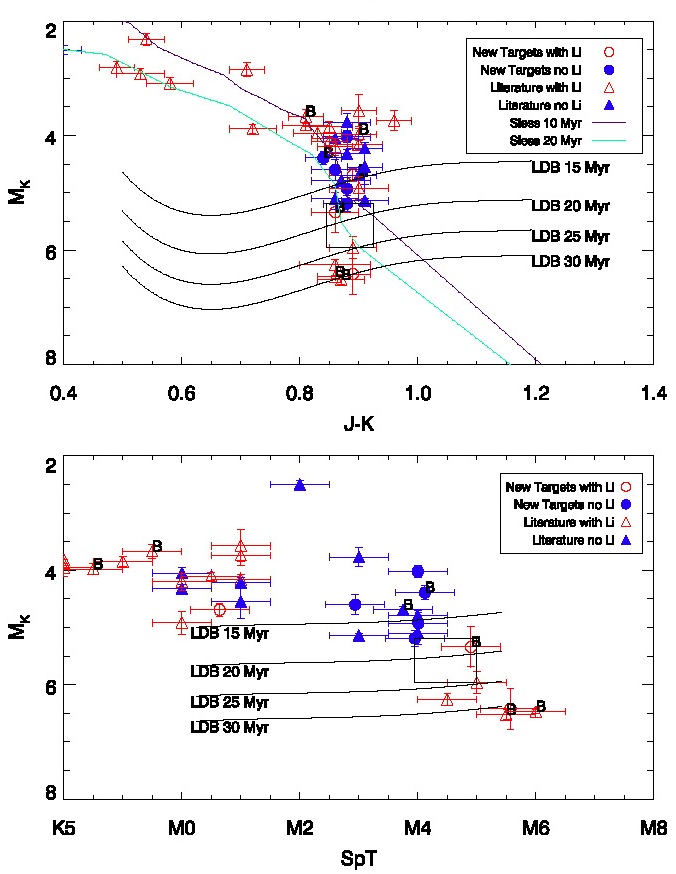} 
\caption{\label{Binks2014_LDB_BPMG_CMD} 
CMD with the age estimate for BPMG based on the Lithium Depletion Boundary. Both  panels come  from \citet{Binks2014-BPMG-LDB}.
}
\end{figure}
%%%%%%%%%%%%%%%%%%%%%%%%%%%%%%%%%%%%%%%

%%%%%%%%%%%%%%%%%%%%%%%%%%%%%%%%%%%%%%%   FIGURE 
\begin{figure}
\center
\includegraphics[width=1.0\textwidth,scale=1.0]{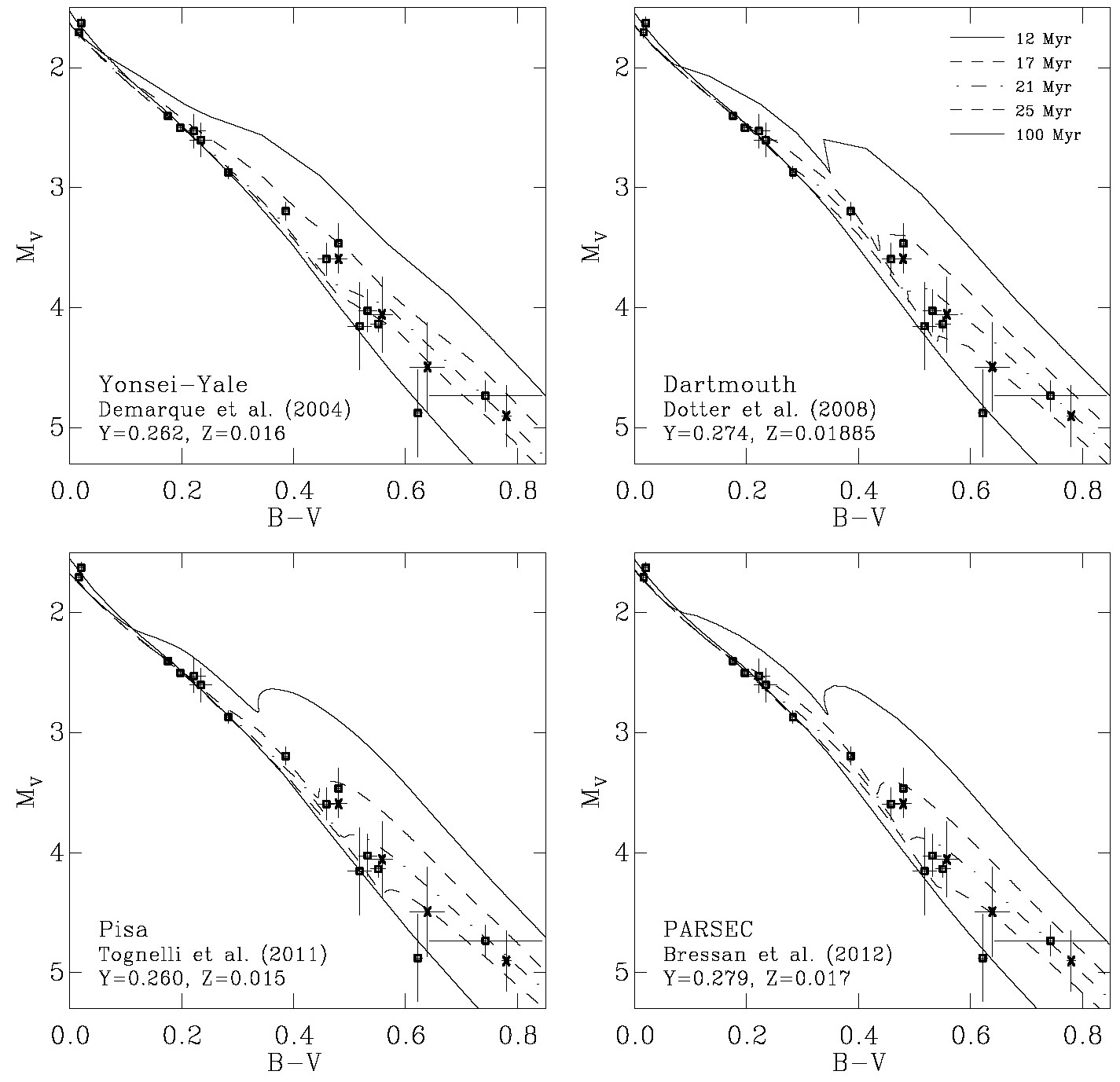} 
\caption{\label{Mamajek2014_BPMG_Age_CMD} 
%, 
Several Color-Magnitude Diagrams for the BPMG. The panels have been taken from figure from \citet{Mamajek2014-AgeBPMG}. 
}
\end{figure}
%%%%%%%%%%%%%%%%%%%%%%%%%%%%%%%%%%%%%%%

%%%%%%%%%%%%%%%%%%%%%%%%%%%%%%%%%%%%%%%   FIGURE 
\begin{figure}
\center
\includegraphics[width=1.0\textwidth,scale=1.0]{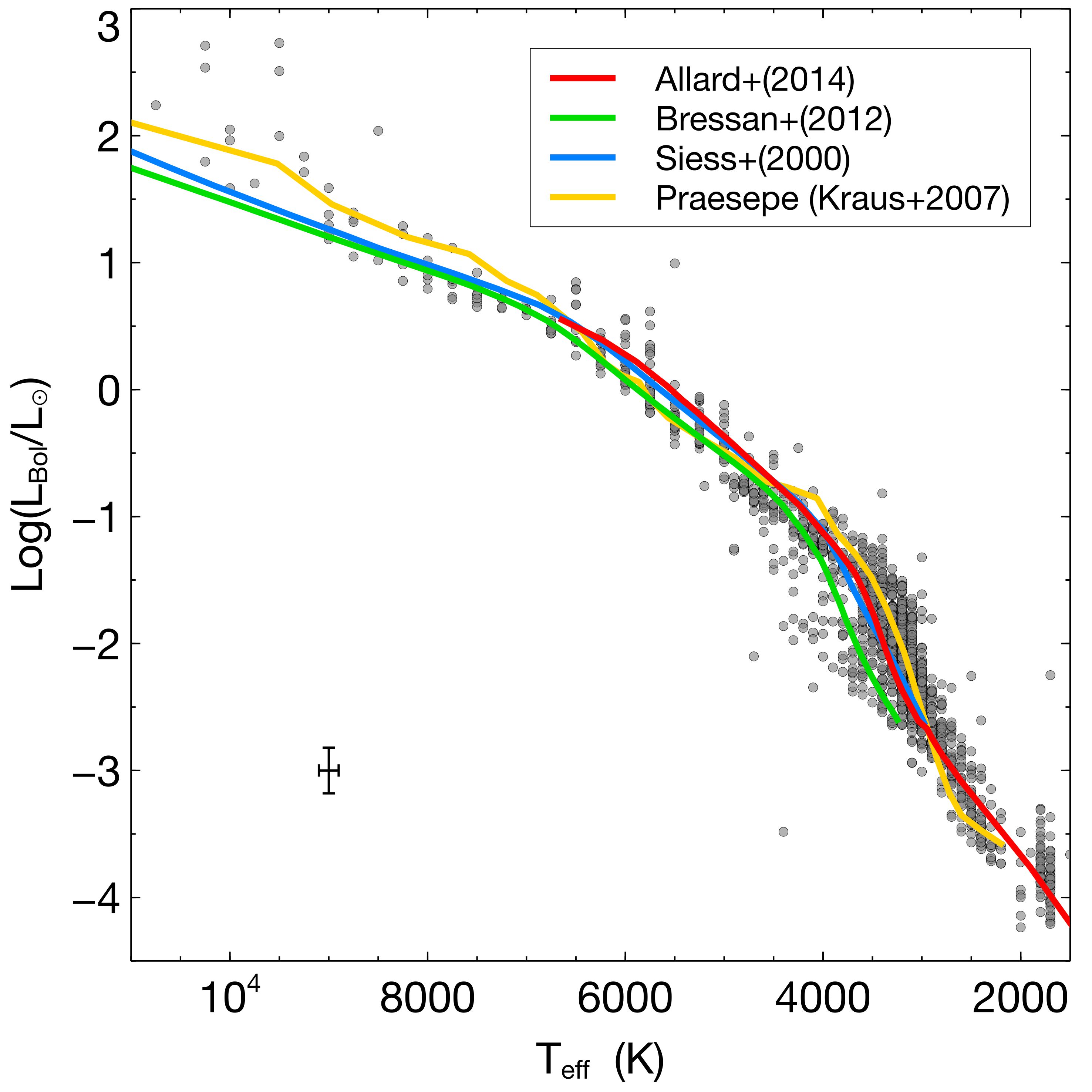} 
\caption{\label{Bouy2015_Pleiades_HR} 
%, 
HR diagram for all known members of the Pleiades, as a comparison with several models. The figure has been taken from \citet{Bouy2015.2}.
}
\end{figure}
%%%%%%%%%%%%%%%%%%%%%%%%%%%%%%%%%%%%%%%

\subsubsection{M dwarfs: the Lithium Depletion Boundary\label{LDB}}
%
%%%%%%%%%%%%%%%%%%%%%%%%%%%%%%%%%%%%%%%%%%%%%%%%%%%%%%%%%%%%%%%%%%%%%%%%%%%%%%%%%%%%%%%%%%%%%

As has been shown in Figure \ref{LDB_ALi_Teff-3clusters}, the gap for late-K and mid-M widens with age. However, there is  a limit for the cooler side, since lithium needs a temperature of about 2.5$\times$10$^6$ K to be destroyed in the stellar nucleus (\citealt{DAntona1998-Li-Burning}). In fact, for solar metallicity, 0.06 $M_\odot$ represents a hard limit and any brown dwarf less massive than this value do keep the initial lithium content. This limit is reached, depending on the model, at about 450 Myr. In addition, during the first $\sim$10 Myr a low-mass star is not hot inside and cannot destroy this element. However, for the following 10 Myr the models predict very different depletion rate, so it is safer to avoid them when deriving ages younger than 20 Myr.

Thus, there is an age range ($\sim$20-450 Myr) where a coeval population of mid- and late-M dwarfs would show a bimodal distribution of lithium: either they have or they do not, but since the destruction is very efficient the borderline is sharp (magenta, dashed line in Fig. \ref{LDB_ALi_Teff-3clusters}).

This technique was first proposed by \citet{Rebolo1992.1} and \citet{Magazzu1993.1} (see also \citealt{Pozio1991.1}), 
 and successfully applied to the Pleiades  (\citealp{Stauffer1998.2}, see also (\citealp{Basri1996.1}), 
 Alpha Persei (\citealp{Stauffer1999.1}; \citealt{Basri1999-MasAge-APer}) and IC2391 (\citealp{Barrado1999-Li-LDB-IC2391},
 see also  for an update \citet{Barrado2004-Li-Ha-IC2391}. In the case of the
 Pleiades, it yields an age of 125$\pm$8 Myr or 130$\pm$20 Myr. With preliminary results for these  three clusters
\citet{Barrado1998-BD-LDB} proposed a new age scale based on the LDB.
The LDB ages are important for several reasons: as stated before, the underlying physics is simpler than in other techniques (the dependence with the ``ingredients'' in the complex  stellar model cocktail 
--subsection \ref{subsubsec:models}-- is reduced) and can be used to validate some assumptions
(for instance, the amount of overshooting,  \citealt{Stauffer1998.2}). In fact, the LDB age scale has been proposed as a semi-fundamental scale and as a reference or calibrator for other scales (\citealt{Soderblom2014-Ages}).

Figure \ref{Stauffer1998_Pleiades_LDB_specra}, extracted from \citet{Stauffer1998.2}, clearly shows the dramatic change
in equivalent width of the LiI 6707.8 \AA{ } doublet (seen as a single line because of the spectral resolution).
 The caption explains:
{\it ``Sample spectra of Pleiades brown dwarf candidates obtained with
the Keck II LRIS. The displayed wavelength region is only a small portion
of the full spectrum, selected in order to highlight the lithium 6708 \AA{ }
region. The y-axis is correct for CFHT PL 10, while the spectra of the other two stars
are offset relative to CFHT PL 10 to avoid having the spectra overlap. The
dashed line is a spectrum of GL 65AB, a field M6--M6.5 binary, assumed to
have entirely depleted its initial lithium.''}
Thus, all that is needed is a collection of spectra for $\sim$M5-M8 members of each association.

Already in 1999, \citet{Barrado1999-Li-LDB-IC2391} concluded: {\it (1) some convective-core overshoot
is needed in evolutionary models for high-mass stars and
that (2) the amount of convective-core overshoot is not a strong
function of mass (at least in the mass range sampled at the
turnoff of these three clusters). A revised age scale for open
clusters ... would have important implications for a variety of stellar evolution
topics}.
Regarding the first point, see also \citet{Stauffer1998.2}.
Since then, a handful of open clusters and young moving groups have been targeted (in some cases reobserved and/or reanalyzed),  and their LDB age derived. Among them: 
Beta Pic MG  ($<$20 Myr, 21$\pm$4 Myr, 26$\pm$3 Myr),
NGC1960      (22$\pm$4 Myr),
IC4665       (27.7$^{+4.2}_{-3.5}$$\pm$1.1$\pm$2 Myr),
NGC2547      (35$\pm$4 Myr),   
IC2602       (46$^{+6}_{-5}$ Myr),
the Pleiades (112$\pm$5 Myr), and
Blanco 1     (132$\pm$24 Myr).
The moving groups   TWA, Octans, Eta Cha MG and AB Dor MG have also been investigated but no so far the LDB has not been reached.
Note the differences in errors and values for some cases.
Additional information can be found in the following papers:
\citet{Song2002-BPMG-LDB},  
\citet{Oliveira2003.1}, \citet{Jeffries2003-Li-NGC2547, Jeffries2005-Li-LDB-NGC2547},
\citet{Manzi2008-Li-LDB-IC4665},
\citet{Mentuch2008_TWA_BetaPic_EtaCha_MG_LDB_LithiumEvolution},
\citet{Dobbie2010.1},
\citet{Cargile2010.1},
\citet{Jeffries2013-LDB-NGC1960},
\citet{Binks2014-BPMG-LDB},
\citet{Juarez2014_Blanco1_LDB},
\citet{Malo2014_YoungAssoc_BPMG_MG_Age_LDB},
\citet{Dahm2015_Pleiades_LDB}, and 
\citet{Murphy2015_Octans_Age_LDB}.   
Recent overviews have been presented in
\citet{Barrado2011-LithiumAges},  \citet{Soderblom2014-Ages} and \citet{Jeffries2014-Ages}.
%  ``Brown Dwarfs and Very Low Mass Stars: Towards a New Age Scale for Young Open Clusters''
From the theoretical point of view, the LDB technique and its limitations are dealt in 
\citet{Burke2004_Theory_LDB_LithiumAge} and 
\citet{Tognelli2015_LDB_Age_uncertaintiesTheoretical}.

%NGC2547     \citet{Oliveira2003.1}, \citet{Jeffries2003-Li-NGC2547}, \citet{Jeffries2005-Li-LDB-NGC2547},
%IC4665      \citet{Manzi2008-Li-LDB-IC4665},
%IC2602      \citet{Dobbie2010.1},
%Blanco 1    \citet{Cargile2010.1},
%NGC1960     \citet{Jeffries2013-LDB-NGC1960},
%Beta Pic MG \citet{Binks2014-BPMG-LDB}.
%Dahm 2015 pleiades  \citet{Dahm2015_Pleiades_LDB} 
%Juarez2014 Blanco 1 \citet{Juarez2014_Blanco1_LDB}
%Malo2014 BPMG \citet{Malo2014_YoungAssoc_BPMG_MG_Age_LDB}
%Mentuch 2008 TWA BPMG, EtaCha, AB Dor  \citet{Mentuch2008_TWA_BetaPic_EtaCha_MG_LDB_LithiumEvolution}
%Murphy 2015 Octans  \citet{Murphy2015_Octans_Age_LDB}   No LDB
%\citet{Song2002_BetaPic_LDB_Lithium}  Beta Pic

The following three figures illustrate the process. Figure \ref{LDB_CMD_APer} contains a CMD for the Alpha Per cluster, modified after \citet{Stauffer1999.1}. Green solid circles are used for clusters members with lithium (for this range and age, very-low mass stars close to the substellar boundary), whereas blue empty circles represent members without lithium on their surface. The separation is very clear, specially when the color is taken into account. However, colors (therefore effective temperature) have other problems and it is better to derive the age using the location of the LDB with the magnitude or, better, bolometric luminosity (\citealt{Burke2004_Theory_LDB_LithiumAge}; \citealt{Jeffries2006-PMS-LithiumDepletion}). 
In fact, contrary to the appearance,  the location of the LDB using the magnitude/L$_\mathrm{bol}$ does not add a large error, as is shown in Figure \ref{LDB_MagAge}, where absolute magnitudes are confronted with age for the LDB. Errors are shown with grey dotted lines and they translate into small errors in age (a good precision due to small internal errors, but always referred to a specific theoretical model).
Finally, Figure \ref{Lumbol_Age_LDB_Pleiades} shows the effect of the assumed distance, always a key parameter. In the case of the Pleiades, there are significant differences between the Hipparcos distance and those derived from isochrone fitting or interferometry (subsection \ref{subsec:pleiades}).
 The $\sim$13 pc of difference could move the final age estimate by almost 20 Myr. Of course, Gaia will settle this issue for the Pleiades and for a huge amount of stellar associations, and will provide complete stellar census and very precise and {\it accurate} distances, removing this source of uncertainty. Unfortunately, Gaia will not reach deep enough to provide parallaxes for the faintest cluster members except for very nearby associations.
Note that the theoretical errors are in the range 3\%-8\% for 20-200 Myr, respectively (\citealt{Burke2004_Theory_LDB_LithiumAge}). Therefore, there is still room for improvement from the observational point of view, specially for the younger side.

From the practical point of view, the lithium feature at 6707.8 \AA{ } is very  faint (equivalent width W(Li) close to 1 \AA{ } or smaller, so good medium resolution spectra (R$\sim$2500) are needed. 
\citet{Bayo2011-IMF-C69} have shown that even with good SNR spectra at R=1250 the lithium feature can be identified.
In any case, 10m class telescopes are required, since the LDB are found at magnitudes fainter (or much fainter) than $R$=18 mag. Figure \ref{LogDist_AgeLog_LDB_Rmag} shows the interplay between age, distance and magnitude of the LDB for some clusters already investigated and a comprehensive sample of galactic clusters from \citet{Dias2002-CatalogOpenClusters} (small grey crosses).
Since measured values range from $\sim$20 Myr up to $\sim$130 Myr, there is still a significant age range  where  this method can be applied, specially for the older end. Eventually, 30m class telescopes will be needed to enlarge the sample where this technique can be realistically applied, but spectrographs with a multiobject capability have enough room to enlarge our current database.

%%%%%%%%%%%%%%%%%%%%%%%%%%%%%%%%%%%%%%%%%%%%%%%%%%%%%%%%%%%%%%%%%%%%%%%%%%%%%%%%%%%%%%%%%%%%%
%
\subsection{Asteroseismology\label{subsec:asteroseismology}}
%
%%%%%%%%%%%%%%%%%%%%%%%%%%%%%%%%%%%%%%%%%%%%%%%%%%%%%%%%%%%%%%%%%%%%%%%%%%%%%%%%%%%%%%%%%%%%%

Stars are essentially unstable during their life-time, they are in a perpetual albeit normally slow evolution. However, they undergo phases with rapid change and/or experience pulsating episodes. Figure 1 in \citet{Lebreton2014-Asteroseismology}, in these series, illustrates a Hertzprung-Russell diagram with different types of pulsation (a review on asteroseismology can be found in \citealt{Cunha2007-AsteroseismologyInterferometry-Review}). As a matter of fact, helioseismology has been a terrific tool to understand the internal structure of the Sun. The technique has been extended to other stars thanks to the arrival of very accurate photometry produced by spaceborne instruments: CoRoT, MOST and Kepler. Eventually CHEOPS, TESS and PLATO will add new information.

\citet{Lebreton2014-Asteroseismology} discussed in depth asteroseismology and its application to the age-dating. We would like to emphasize few issues here, related to exoplanetary studies
(see also subsection \ref{gyrochronology}). The properties of the planetary systems uncovered so far (an ever increasing amount thanks to the Kepler satellite and the ground-based surveys based on photometric and spectroscopic searches) depend strongly on the properties of the host, the star. And, of course, age is paramount together with the stellar mass. A typical example is the multiple system of HR8799, first discovered by \citet{Marois2008-HR8799-Exoplanets}.
The ages used in \citet{Marois2008-HR8799-Four-Exoplanets} are either 30$^{+20}_{-10}$ Myr or 60$^{+100}_{-30}$ Myr, which in both cases produce masses of the companions well inside the planetary domain. On the other hand, \citet{Moya2010-HR8799-Age} (see their Table 1 with estimates in the literature and also \citealt{Moya2010-HR8799-Pulsating}) conclude that the age is closer to 1 Gyr, which would mean that the companions are much more massive and in fact they would be brown dwarfs. However, some ambiguity remains since some models are still compatible with young ages. The bottom line, in any case, if the lack of a one-to-one relation, of a unique solution even with large errors.

Another example is provided by the giant star KIC8219268 (the Kepler Object of Interest KOI2133, aka Kepler 91).
\citet{LilloBox2014.Kepler91_REB} confirmed the planetary nature of the eclipsing companion by REB modulation of the Light Curve (Reflection, Ellipsoidal and Beaming). They also used the solar-like oscillations to determine the stellar properties via asteroseismology and derive accurate planetary parameters. Accurate radial velocity data were used later on to re-confirm the planetary nature in an independent manner (\citealt{LilloBox2014.Kepler91_RV}). Note, however, the very rich wealth of data available in this case (exquisite photometry from Kepler which includes transits, REBs and asteroseismology; high-spectral resolution spectroscopy, high spatial imaging), which is not the normal case.

%%%%%%%%%%%%%%%%%%%%%%%%%%%%%%%%%%%%%%%%%%%%%%%%%%%%%%%%%%%%%%%%%%%%%%%%%%%%%%%%%%%%%%%%%%%%%
%
\subsection{Kinematics: the role of the Moving Groups\label{subsec:kinematics}}
%
%%%%%%%%%%%%%%%%%%%%%%%%%%%%%%%%%%%%%%%%%%%%%%%%%%%%%%%%%%%%%%%%%%%%%%%%%%%%%%%%%%%%%%%%%%%%%

It has been already shown how isochrones can be fitted to the data of stellar associations, including those  with a low number of {\it bona fide} members, the so called moving groups (see, for instance, the list by \citealt{Shkolnik2012-MG-25pc}). In principle, to be truly coeval, members of a moving group should be chemically homogeneous and should have been born nearby to each other. The case of UMaG has been already discussed (in subsection \ref{interferometry}). Other very well known MG is the TW Hya Association (TWA), first defined by \citet{Kastner1997-TWA} ($10^{+10}_{-7}$ Myr, \citealt{Barrado2006-Age-2M1207}). Another very interesting MG is associated to  $\beta$ Pic, a bright star which contains not only a very complex and beautiful circumstellar disk discovered with the IRAS satellite (\citealt{Aumann1984-IRASexcess-MS}, \citealt{Backman1993-BetaPic-VegaPhenomenon}), but also a planetary companion (\citealt{Lagrange2009-BetaPic-Exoplanet}).

The Beta Pic moving group (BPMG) is about a 20$\pm$10 Myr (\citealt{Barrado1999-BetaPic-Age, Barrado2001-BetaPic-Age}). It contains several tens of members (\citealt{Song2003-TWA-BPMG-Tucana}). Other age estimates are 12$^{+8}_{4}$ Myr, based on isochrone fitting (\citealt{Zuckerman2001-BPMG-Age}, see their figure 1),   21$\pm$4 Myr by using the Lithium Depletion Technique (\citealt{Binks2014-BPMG-LDB}), as we have already seen, or 23$\pm$3 Myr (\citealt{Mamajek2014-AgeBPMG}, where their table 1 summarizes other age estimates since 1999). 

As a matter of fact, \citet{Mamajek2014-AgeBPMG} also computed the trajectories of members and computed backwards their position into the past.
This age estimate is called the ``traceback age''. A similar method is the ``expansion age'', first computed by \citet{Blaauw1952-Age-Expanding-ScoCen,Blaauw1952-Age-ZetaPer-OB}.
A detailed description of kinematics ages in moving groups can be found in section 3.2 of  \citet{Soderblom2014-Ages}.
A example is presented in  Figure \ref{Mamajek2014_BPMG_Age_Traceback}, where both panels have been taken from  \citet{Mamajek2014-AgeBPMG}. For panel at the left the original quotation says:
{\it ``1$\sigma$ dispersions in X; Y; Z coordinates ($\sigma$X, $\sigma$Y , $\sigma$Z) as
a function of time in the past, assuming linear trajectories. The
quadrature sums of the X- and Y - dispersions ($\sigma$$_{XY}$) and X-, Y -
and Z- dispersions ($\sigma$$_{total}$) are also plotted. Linear trajectories in
Z are obviously the poorest approximation (contrast with dispersion
measured for epicyclic orbit in Fig. 4). The $\sigma$$_{XY}$ dispersion
may be the most useful overall metric of the group's size using
the linear trajectory technique."} 
On the other hand, the one corresponding to the panel on the right is:
{\it "Distribution of BPMG members in the XY plane now
(filled triangles) and 12 Myr ago (open circles) using epicycle
orbit approximation. The dispersion in the X and Y directions
are plotted now and 12 Myr ago. The trajectory for the star $\beta$
Pic itself is plotted as a solid arc, and labelled with a '$\beta$'. The
reference frame has its origin at the Sun's current position, but
is co-moving with the LSR of \citet{Schorich2010-LSR}."} They conclude that none of the adopted 
kinematic assumption can produce an age in agreement with small errors, value which also should be compatible with other constraints. However, this is not the first time a  kinematic age has been determined for BPMG (see \citealt{Ortega2002-BPMG-Origin, Ortega2004-BPMG-Origin}; \citealt{Song2003-TWA-BPMG-Tucana}; \citealt{Makarov2007-Origin-MG-BPMG}).  
Despite this computation the general conclusion of  \citet{Soderblom2014-Ages}, specifically for BPMG and TWA, is that not reliable kinematic age estimate has been derived.

Another interesting case is the Castor moving group.
It was presented in \citet{Barrado1998-CastorMG} with an age estimate of 200$\pm$100 Myr, 
and since then it has been listed in several works,
 specially some connected to kinematics and age in the solar neighborhood
(additional analyses and possible members in \citealt{Montes2001-SingleMembers-MG}, \citealt{Caballero2010-Castor-WideBinary}, and \citealt{Shkolnik2012-MG-25pc}). However,
 \citet{Mamajek2013FomalhautC}, by using new data, have questioned the reality of this MG
 (i.e. rejecting the possibility of having a collection of coeval stars born at the same place).
 They conclude: {\it "Despite these stars (the Fomalhaut system, Vega, LP 944-20,
and the Castor system) being young and having somewhat
similar velocities, their velocities are well-constrained enough
and different enough that it is clear that they were not in the
vicinity of one another even in the recent past, let alone a couple
of Galactic orbits ago. We conclude that the CMG is comprised
of stars from different birth sites rather than a coeval system,
and hence “membership” to the CMG does not provide useful
age constraints for the Fomalhaut system (or Vega, LP 944-20,
Castor, or other CMG members)."}
Certainly, science is about a healthy skepticism and reanalysis, getting better and better data in order to improve our interpretation of reality. The Castor MG could be, indeed, several distinct groups or lack any real connection among the proposed members
(but a counterexample seems to the UMaG).
 In any case, the problem remains: how a handful of young stars have been born and where? How many at the same time? A similar case is presented by "isolated" Classical TTauri stars (\citealt{delaReza1989-IsolatedTTauri}) and young M dwarfs in the solar neighborhood or as interlopers in clusters (\citealt{Oppenheimer1997.1}; \citealt{Barrado2004-Li-Ha-IC2391}; \citealt{Shkolnik2011-YoungM-GALEX};   \citealt{Rodriguez2013-YoungM-GALEX}). As in the cases of other proposed MGs, 
 the data, specially the values provided by Gaia, will judge soon enough.

Indeed, the Gaia potential here is immense. The analysis of the Gaia data products will allow the discovery of a significant number of moving groups and the analysis in depth of those already known. The very precise and accurate positions, distances and proper motions will produce exquisite motions across the Galaxy and the possibility of tracking back their trajectories to the formation location and, thus, the moment when it happened.  Note, however, that radial velocity matching the proper motions will still be needed.
Surveys like Gaia-ESO (\citealt{Gilmore2012.GaiaESO}; \citealt{Randich2013.GaiaESO}) represent valuable steps, although survey instruments will be, eventually, mandatory (such as LAMOST, \citealt{Cui2012-LAMOST}, or  WHT/WEAVE, \citealt{Dalton2012-WEAVE}).

%%%%%%%%%%%%%%%%%%%%%%%%%%%%%%%%%%%%%%%%%%%%%%%%%%%%%%%%%%%%%%%%%%%%%%%%%%%%%%%%%%%%%%%%%%%%%
%
\subsection{Gravity indicators\label{subsec:gravity}}
%
%%%%%%%%%%%%%%%%%%%%%%%%%%%%%%%%%%%%%%%%%%%%%%%%%%%%%%%%%%%%%%%%%%%%%%%%%%%%%%%%%%%%%%%%%%%%%

There are several spectral features which are very sensitive to the surface effective gravity.
Figure 7 of \citet{Huelamo2009-WD-EB} shows a fit to the WD component of the triple system discussed in
subsection \ref{whitedwarfs} and \ref{eclipsingbinaries}. The helium lines have been used to derive a temperature and a gravity and from these values an age. On the other hand, cooler spectral types display other gravity-sensitive features. \citet{Prisinzano2012-SpectraClassification-PMS-NGC5630} have applied the Ca I triplet or 6102, 6122 and
6162 \AA{ }, very strong in giant stars, to PMS stars.
In the  case of M  and ultracool dwarfs (L, T and Y spectral types, \citealt{Kirkpatrick1999-Ldwarf},  \citealt{Burgasser1999-Tdwarf} and \citealt{Cushing2011-Ydwarfs}), they exhibit alkaline lines (the doublets at NaI 5889.95 and 5895.92 \AA, KI 7664.91 and 7698.98 \AA, NaI 8183.26 and 8194.8 \AA, for instance) which can dominate the optical spectrum.

Str\"omgren photometry, a  medium-width narrow-band photometric system, can also be used to estimate gravities,
specially for stars hotter than the Sun (\citealt{Alexander1986-Stromgren-Interpretation}; \citealt{Napiwotzki1993-Stromgren-TempGravity}). A practical application, for the case of Vega-like stars with A spectral type, can be found in \citet{Song2001-VegaType-Ages-Stromgren}.

%%%%%%%%%%%%%%%%%%%%%%%%%%%%%%%%%%%%%%%%%%%%%%%%%%%%%%%%%%%%%%%%%%%%%%%%%%%%%%%%%%%%%%%%%%%%%
%
\subsection{Secondary age scales\label{subsec:secondaryagescales}}
%
%%%%%%%%%%%%%%%%%%%%%%%%%%%%%%%%%%%%%%%%%%%%%%%%%%%%%%%%%%%%%%%%%%%%%%%%%%%%%%%%%%%%%%%%%%%%%

When discussing the different classification of the age-dating techniques, we defined the secondary indicators in subsection \ref{subsubsec:practical}. They depend on previous age estimates and therefore are subject to larger absolute errors. Note, however, that in principle relative errors could be smaller and the relative sorting in age  (in some cases free of the ``curse of the models'') could be as good as in the case of the primary indicators.

\subsubsection{Gyrochronology\label{gyrochronology}}
%
%%%%%%%%%%%%%%%%%%%%%%%%%%%%%%%%%%%%%%%%%%%%%%%%%%%%%%%%%%%%%%%%%%%%%%%%%%%%%%%%%%%%%%%%%%%%%

Arguably, the gyrochronology or evolution of rotation is the best age indicator for a single, solar-type and cooler star, if the mass of a star is known (mostly from the spectral type). Main sequence stars lose angular momentum due to stellar winds. In fact, \citet{Kraft1967-Rotation-Age} showed that F and G stars have a correlation between activity and rotation and that this correlation depends on the age. Few years later,  \citet{Skumanich1972-ActivityRotationLithium-Decay} pointed out that activity, rotation and lithium abundance decay with age for solar-like stars older than the Pleiades, following a linear trend in logarithmic scale for rotation and activity. These behaviours have since then being seen in less massive stars.

Indeed, the distribution of rotational velocities (the projected $vsini$ or the rotational period when available) in clusters of different ages do display a clear trend. 
Figure \ref{Kovacs2015_GyroAges} belongs to \citet{Kovacs2015-Gyro}. It can be appreciated that the age tendency is present. However. quasi-coeval clusters, such as Blanco 1, M35 and M45 (The Pleiades) on one hand, or M44 (Praesepe) and the Hyades, on the other, differ in their distributions. In any case, even for clusters as old as the Hyades (about 600 Myr), there is a significant scatter for the same color. It is true, however, that components of wide, physically associated binaries give the same age using gyrochronology (Figure \ref{Kovacs2015_GyroAges}, extracted from \citealt{Barnes2007-Gyro}). 
The original captions reads:
{\it ``Color-period diagram for three wide binary systems, $\xi$ Boo A/B,
61 Cyg A/B, and $\alpha$ Cen A/B. Rotational isochrones are drawn for ages of 226 Myr,
2.0 Gyr, and 4.4 Gyr, respectively, and the errors are indicated with dashed lines.
Note that for all three wide binary systems, both components give substantially the
same age. The dotted line corresponds to the age of the universe.''}
As remarked by \citet{Soderblom2010_AgeReview}, 500 Myr seems to be dividing line for this method and perhaps it is safer to apply it to older (or even significantly older) stars.

The K2 phase of the Kepler satellite is already observing several cornerstone clusters of very different ages, from very young to similar to the Sun  (Upper Sco, Pleiades, M35, Hyades, Praesepe, M67). This amazing database will be crucial to test the validity of the gyrochronology and to define the limits in mass and age where it can be applied.

%\subsubsection{The evolutionary status of  exoplanets\label{subsec:Exoplanets}}
%
%%%%%%%%%%%%%%%%%%%%%%%%%%%%%%%%%%%%%%%%%%%%%%%%%%%%%%%%%%%%%%%%%%%%%%%%%%%%%%%%%%%%%%%%%%%%%

We have already mentioned how to derive ages for the stellar host of a planetary system and, hence, help to estimate the properties of the planets within it. \citet{Maxted2015-ExoplanetAge-Bayesian} have used a Bayesian approach in order to derive ages based on fits with theoretical models (\ref{subsec:isochrone}). They have also derived ages based on gyrochronology (subsection \ref{gyrochronology} and \citealt{Maxted2015-ExoplanetAge-Gyro}).

Figure \ref{Maxted2015_Exoplanets_Age_Gyro_Isochrones} comes from \citet{Maxted2015-ExoplanetAge-Gyro}, with a caption stating:
{\it ``Change in the best-fitting masses and ages of transiting exoplanet host stars due to a change in the 
assumed helium abundance or mixing length parameter. Dots show the best-fitting mass and age for
 the default values of Y and $\alpha$$_{MLT}$ and lines show the change in mass and age due to
an increase in helium abundance $\Delta$Y = +0.02 (left panel) or a change in mixing length parameter
$\Delta$$\alpha$$_{MLT}$= +0.2 (right panel). Horizontal lines indicate the age of the Galactic disc (dashed), 
the age of the Universe (dotted) and the largest age in our grid of stellar models (solid). The curved
dotted line shows the terminal age main sequence (TAMS) for stars with solar composition."}.
From these panels it is clear the role that both the chemical composition and the physics inside the models are playing. \citet{Maxted2015-ExoplanetAge-Gyro} concluded that the gyro-ages are significantly younger than isochrone ages for half of their sample. Second order effect might be at play (magnetism, tidal interactions or other), but until we completely understand all sides of stellar evolution, ages would, somehow, remain elusive.

\subsubsection{Stellar activity: from the corona to the photosphere\label{stellaractivity}}
%%%%%%%%%%%%%%%%%%%%%%%%%%%%%%%%%%%%%%%%%%%%%%%%%%%%%%%%%%%%%%%%%%%%%%%%%%%%%%%%%%%%%%%%%%%%%

As explained in the previous subsection (\S\ref{gyrochronology}), stellar activity is linked to rotation and decays with age. However, if the situation with rotation is not as clear as one might expect, activity is even more complex. To begin with, the activity level of the Sun is not constant during a solar cycle, about 22 years if we take into account the magnetic polarity  (either in X-rays from the corona, H$\alpha$ emission coming from the chromosphere or spottiness and plages in the photosphere). Moreover, since the discovery of the sunspot at the beginning of the XVII century (by Thomas Harriot, Johannes and David Fabricius,  Galileo Galilei and   Christoph Scheiner, although there are historical previous sightings), its has been shown that there are long term variations with a minimum activity between  1645 and 1715 (Maunder minimum, see the long term Mt. Wilson monitoring, with an update in \citealt{Schroder2013-MtWilson-ActivityMonitoring}). This phenomena seem to be present in other solar-like stars. In any case, even for coeval stars the situation might be more complex than for the Sun.

Figure \ref{BPMG_Lx_clusters} shows the coronal activity, using X-rays as a proxy, for two open clusters (the Pleiades and the Hyades) and few members of the BPMG (the diagram comes from \citealt{Barrado1998-Activity-Hyades-Coma} and the BPMG where analyzed in \citealt{Barrado1999-BetaPic-Age}). The first feature that strikes the eye is certain dependence with age (Pleiades members, younger, are more active {\it on average} than their Hyades counterparts). However, the spread for a given color (a stellar mass) is so large that the loci of both clusters are intermingled.

Down in the stellar photosphere the situation is analogous. \citet{BarradoMartin03.1} compared the chromospheric emission (using H$\alpha$) for late-type stars  belonging to young associations (due to accretion) and open clusters (activity). The presence of flares complicates even more the situation. Clearly, we are dealing with a qualitative indicator (an accretor, a possible young or old star), not a quantitative one. The same can be said for spottiness or other activity indicators  (Mg II h and k at 2797 and 2803 \AA{ } in the UV; CaII H and K at 3968.49 and 3933.82; H$\beta$ at 4861.3 \AA; Mg I  triplet at 5167, 5172, and 5183 \AA; He I D1 at 5895.92, D2 at 5889.95, and D3 at 5876  \AA;  CaII IRT at 8498, 8542, and 8662 \AA; HeI 10830 \AA. See
\citealt{Montes1997-LibrarySpectra-StelalrActivity}; \citealt{Montes1998-SpectraFGKM};
\citealt {LopezSantiago2010-ActivityRotationKinematicsAge}).
The most widely used and better calibrated is the $R'_{HK}$ index, after the removal of the photospheric contribution (see \citealt{Noyes1984_Rotation_ActivityPeriod} for details). In any case, even with the careful calibration and acquisitions of data along several decades, its quality as an age indicator is, at best, only good from a qualitatively perspective, as the analysis by \citet{Mamajek2008-Age-Activity-SolarType} indicates.

Thus, the situation is the same for these additional activity indicators. Therefore, as in many other age-dependent features, activity is a useful technique from the statistical point of view, valid in conjunction with other methods and/or when analysing a cohort of members belonging to an association. When dealing with individual objects, it should be handled with extreme care.

%%%%%%%%%%%%%%%%%%%%%%%%%%%%%%%%%%%%%%%%%%%%%%%%%%%%%%%%%%%%%%%%%%%%%%%%%%%%%%%%%%%%%%%%%%%%%  SECTION
%%%%%%%%%%%%%%%%%%%%%%%%%%%%%%%%%%%%%%%%%%%%%%%%%%%%%%%%%%%%%%%%%%%%%%%%%%%%%%%%%%%%%%%%%%%%%
%%%%%%%%%%%%%%%%%%%%%%%%%%%%%%%%%%%%%%%%%%%%%%%%%%%%%%%%%%%%%%%%%%%%%%%%%%%%%%%%%%%%%%%%%%%%%
%
\section{Stepping stones and the age stairway\label{sec:examples}}
%
%%%%%%%%%%%%%%%%%%%%%%%%%%%%%%%%%%%%%%%%%%%%%%%%%%%%%%%%%%%%%%%%%%%%%%%%%%%%%%%%%%%%%%%%%%%%%
%%%%%%%%%%%%%%%%%%%%%%%%%%%%%%%%%%%%%%%%%%%%%%%%%%%%%%%%%%%%%%%%%%%%%%%%%%%%%%%%%%%%%%%%%%%%%
%%%%%%%%%%%%%%%%%%%%%%%%%%%%%%%%%%%%%%%%%%%%%%%%%%%%%%%%%%%%%%%%%%%%%%%%%%%%%%%%%%%%%%%%%%%%%

As we have seen, some age scales are more accurate than other (or more precise) and few have a simple underlying physics. None can be used for all stellar masses and all age values. Some, in fact, are very restrictive in their applicability. What it seems to be clear is that some methods are more reliable than others. Thus, starting with our stellar anchors (the age for the Sun and the limit imposed by the Big-Bang), it would be possible to built a step-by-step age stairway. Coming back to the ``jenga'' analogy, to remove some pieces from the bottom to locate them higher up in the pile. Or, perhaps, to discover we need to start over again. As already stated in \citet{Mamajek2008-Ages}: ``{\it we should aim for consistency. A given property (or properties) allow us to sort a set of stellar associations from the youngest
to the oldest, even if we can not derive absolute ages... Eventually, we should be able to construct different ages scales which should be consistent with each other, and should produce absolute as well as relative ages}''.

Figure \ref{Barrado_Age_SteppingStones_EES2015.jpg}, based on the results discussed here and on previous reviews (\citealt{Mermilliod2000.4}; \citealt{Mamajek2008-Ages}; \citealt{Soderblom2010_AgeReview}; \citealt{Soderblom2014-Ages}; \citealt{Jeffries2014-Ages}),  displays one possible 
age stairway, based on different techniques and the degree of confidence. The first level, the true anchors we have at our disposal, does not provide a reference for ages younger than the Sun, so we have to rely on the most trustable. During the last 20 year a consensus is being built around the LDB ages. However, it is only valid  at best for ages in the range 10-450 Myr. The deuterium depletion boundary can play a similar role for younger associations (for a search in young associations, see \citealt{Cody2011-DeuteriumPulsation-Spitzer}), as well as the radiative-convective gap. Population II stars can be dated using several radiative heavy isotopes ($^{232}$Th, $^{238}$U), but the method is far from useful as yet. On the third level expanding ages (kinematics) in young moving groups are essentially independent of models, and Gaia data will be crucial to verify whether they can be used and trusted. On the other hand, gyrochronology can fill the age gap between $\sim$500 Myr and the oldest stars, specially for $\tau$$>$1 Gyr. Several very well observed clusters (three with similar ages such as the Hyades, Praesepe and Coma, and NGC752 and M67) are being used to define the rotation decay with age. Kepler K2 has, in fact, taken data for some of them. Asteroseismology provides a detailed knowledge of the stars, but it is strongly dependent on models and this is the reason why it has been located in the next level. Again, Kepler data (and eventually TESS, CHEOPS and PLATO) will have a lot to say, specially in clusters. Complementary radial velocity data will be very handy. Moreover, eclipsing binaries, with accurate radial velocities will be very helpful (specially multiple systems and/or in associations of different ages), since the masses and evolutionary status of the components could cover any point in the parameter space (high-, solar and low-mass stars or even brown dwarfs; any luminosity class and even white dwarfs). After all these processes have been understood, the evolutionary models can be recalibrated and the isochrone fitting performed. Finally, other methods  (lithium depletion in FGK stars, stellar activity, gravity indicators and so on) could safely be applied. Note that the diagram hides a dependency on mass, since each method is valid for a specific mass range and this fact has to be taken into account in this quest for a definitive global age scale.

In order to calibrate properly these techniques, using the ones at lower levels, and to provide a consistent picture, the overlapping age ranges are very important. 
Figure  \ref{Barrado_Age_SteppingStones_EES2015.jpg} also includes a collection of young star forming regions, moving groups and clusters. These associations have several excellent qualities 
(proximity, number of members, composition, age, proper motion, reddening, etc) and have been used recurrently in the literature. 
On top of it, some fall on these overlapping age ranges and are pivotal to transfer age estimates from one level to the next. As an example, the old open cluster M67 and the radiometrical age from the Sun and the gyrochronology, or the BPMG for the lithium depletion, the kinematics and, perhaps, the asteroseismology. Thus, some associations  are truly cornerstones or we believe they will become so.
New data and analyses would make them even more useful. Some examples follow.

%%%%%%%%%%%%%%%%%%%%%%%%%%%%%%%%%%%%%%%%%%%%%%%%%%%%%%%%%%%%%%%%%%%%%%%%%%%%%%%%%%%%%%%%%%%%%
%
\subsection{A rising ``star'': Lambda Orionis star forming region\label{subsec:lambdaorionis}}
%
%%%%%%%%%%%%%%%%%%%%%%%%%%%%%%%%%%%%%%%%%%%%%%%%%%%%%%%%%%%%%%%%%%%%%%%%%%%%%%%%%%%%%%%%%%%%%

As we have seen, very young stars pose specific problems because not only the complexity of the phenomenology they can display (intense activity, fast evolution, circumstellar disks, inhomogeneous intracloud extinction, and so on), but also for the lack of an age anchor (see figure \ref{Barrado2015-AgeAnchors} and section \ref{sec:anchors}). Gaia will not be very helpful here, specially for the low-mass end, since it will not go deep enough and the youngest regions are better observed in the near-infrared, which is attenuated by the absorption to a lesser extent than the optical bands. Projects like DANCE (\citealt{Bouy2013.1}), with a multi-wavelength approach, will be a significant step forward.

One very interesting star forming region is associated to the massive star $\lambda$ Orionis.
It is located at about 400 pc (\citealt{Murdin77.1}). \citet{Duerr82} identified three associations within a very large bubble seen in H$\alpha$, later on photometric and spectroscopically characterized by 
\citet{Dolan1999-Li-C69, Dolan2001-Spatial-C69, Dolan02-Photometry-C69}. Deeper photometry and spectroscopy well inside  the substellar domain have been published by 
\citet{Barrado2004-BD-C69, Barrado2007-LOri167, Barrado2007-Accretion-C69} and \citet{Bayo2011-IMF-C69, Bayo2012-Li-Rotation-Activity-C69}.

The Lambda Orionis star forming region (LOSFR) includes associations in very distinct evolutionary status: Collinder 69 at the center, including $\lambda$ Ori; Barnard 30 on the H$\alpha$ rim, and Barnard 35 mid way between both, together with other younger population recently identified in the area (\citealt{Konig2015-SOri-LOSFR-C69-YSO-WISE}). The ages are in the range 1 to 5 Myr approximately, or even younger for some areas in the rim.
The diversity of data already collected in the the different parts would be very helpful to provide additional nails to the age stairway for the younger steps.

%%%%%%%%%%%%%%%%%%%%%%%%%%%%%%%%%%%%%%%%%%%%%%%%%%%%%%%%%%%%%%%%%%%%%%%%%%%%%%%%%%%%%%%%%%%%%
%
\subsection{Stellar associations: Beta Pic Moving Group\label{subsec:BPMG}}
%
%%%%%%%%%%%%%%%%%%%%%%%%%%%%%%%%%%%%%%%%%%%%%%%%%%%%%%%%%%%%%%%%%%%%%%%%%%%%%%%%%%%%%%%%%%%%%

So far we have discussed very young star forming regions, open clusters with a large diversity of ages (from 30/50 --depending on the age scale-- to 4000 Myr) and old globular clusters. Unfortunately, there is no nearby stellar association with an age between 10 and 50 Myr. This very important gap is filled with moving groups. We have already discussed some age-related properties for few of them. O.J. Eggen carried out an intense search for kinematic groups 50 years ago, some if not most are in fact not coeval, but the blooming came with the initial discovery of apparently isolated TTauri stars (\citealt{delaReza1989-IsolatedTTauri}) and the identification of the moving group associated to the Classical TTauri TW Hya, known as TWA (\citealt{Kastner1997-TWA}). TWA has been one of the {\it prima donna} in this show. Another has been the BPMG, identified  a little bit later on (\citealt{Barrado1999-BetaPic-Age}). Since then, a cascade of identifications  of new MGs and additional candidate members have been produced
(see, for instance, \citealt{Zuckerman2001-BPMG-Age, Zuckerman2004-MG-YoungStarsNearSun}; \citealt{Torres2001-GAYA-MG, Torres2008-SACY}). A complete description can be found in \citet{Kastner2016-Nearby-MG-Ages}. Certainly, Gaia will identify new members of these  groups, as well as discover new associations (as well as discarding candidate members or even some proposed groups, as we have already mentioned).

Together with TWA, the BPMG plays an important role because of its membership list, proximity and youth. Since 
\citet{Jura1998-TWHya-BetaPic-DebrisDisks} suggested that the star $\beta$ Pic was young, a number of age estimates have been published, as listed in table 1 of \citet{Mamajek2014-AgeBPMG}. 

One of the most recent values have been derived by \citet{Binks2014-BPMG-LDB}. The original caption of Figure \ref{Binks2014_LDB_BPMG_CMD} reads:
{\it ``Locating the LDB in 3 separate colour (or spectral-type) vs. magnitude diagrams. New members from Table 1 and objects from the literature are indicated. Absolute magnitudes are calculated from 2MASS K and a trigonometric parallax where available or a kinematic distance otherwise. Known, unresolved binaries are marked with 'B'. Black lines represent constant luminosity loci from Chabrier \& Baraffe (1997) where Li is predicted to be 99\% depleted at the ages indicated. The green and maroon lines are 10 and 20 Myr isochrones from \citet{Siess2000.1}. The rectangle in each diagram represents the estimated LDB location and its uncertainty, based on the faintest Li-depleted member and the brightest Li-rich member (but excluding the unresolved binary at M$_K$$\sim$5.3).}"

Another recent example comes  from \citet{Mamajek2014-AgeBPMG}. Likewise, the caption of their figure 6, include here as
Figure \ref{Mamajek2014_BPMG_Age_CMD}, states: 
{\it ``$M_V$, $(B-V)$ CMDs of the A-, F- and G-type 
  BPMG members compared against the Yonsei-Yale (Y2; \citealt{Demarque2004-YaleTracks}, top left),
  Dartmouth   (\citealt{Dotter2008-DarthmouthTracks}, top right),
  Pisa (\citealt{Tognelli2011-PisaTracks}, bottom left) and
   PARSEC (\citealt{Bressan2012-PARSECtracks}, bottom right) model isochrones.
In all panels the upper continuous line represents 
the position of the single-star  sequence for the often quoted age of 12 Myr.
Below this, the dot-dash and 
bounding dashed isochrones represent the position based on the  LDB age of $21\pm4$ Myr
according to \citet{Binks2014-BPMG-LDB}. Finally, the lower continuous line denotes the position for an age
of 100 Myr. The squares represent the 'classic' sample of members as defined by
\citet{Zuckerman2004-MG-YoungStarsNearSun} 
whereas the crosses denote  additional members from \citet{Malo2013-NeabyMG}."}

Both techniques are conceptually very different but the final results agree quite well with each other. However, if we survey different results in the last 20 years, we see significant differences, specially before it was recognised that the most massive star, $\beta$ Pic, was young. For completeness, we have added several verbatim conclusions from a collection of works:

\begin{itemize}
\item \citet{Lanz1995-BetaPicAge}:
 {\it `` ... the star is either a pre-main-sequence (PMS) star
nearing the zero-age main sequence (ZAMS), or it is a main-sequence star older than 0.3 Gyr.''}
\item \citet{Brunini1996-BetaPicAge}: {\it `` ... argues in favour of a large age for $\beta$ Pic. However, the estimation of stellar ages employing cometary fluxes should be treated with caution, on account of the diversity of possible planetary systems.''}
\item \citet{Barrado1999-BetaPic-Age, Barrado2001-BetaPic-Age}:
 {\it ``The estimated age for b Pic is then $20\pm10$ Myr, where the uncertainty in the age arises primarily from possible errors in the pre-main-sequence isochrones and in the conversion from color to effective temperature.''}
\item \citet{Malo2014_YoungAssoc_BPMG_MG_Age_LDB}
{\it ``We find that the inclusion of the magnetic field in evolutionary models increase the isochronal age estimates for the K5V-M5V stars. Using these models and field strengths, we derive an average isochronal age between 15 and 28 Myr and we confirm a clear Lithium Depletion Boundary from which an age of $26\pm3$ Myr is derived, consistent with previous age estimates based on this method.''}
\item \citet{Binks2014-BPMG-LDB}:
 {\it ``The LDB age of the BPMG is $21\pm4$ Myr and insensitive to the choice of low-mass evolutionary models. This age is more precise, likely to be more accurate, and much older than that commonly assumed for the BPMG.}''
\item \citet{Mamajek2014-AgeBPMG}:
 {\it ``The results from recent LDB and isochronal age analyses are now in agreement with a
median BPMG age of 23$\pm$3 Myr (overall 1$\sigma$ uncertainty, including $\pm$2 Myr statistical
and $\pm$2 Myr systematic uncertainties).}''
\end{itemize}

Going back to the recollection of  \citet{Kastner2016-Nearby-MG-Ages}:
{\it `` ... They conclude that the age of the $\beta$PMG
should be revised upwards, from the widely quoted $\sim$12 Myr (\citealt{Zuckerman2001-BPMG-Age};
\citealt{Torres2006-SACY}) to $\sim$23 Myr (which is, ironically, closer to the original estimate by
\citealt{Barrado1999-BetaPic-Age}). However, this refinement in the age of the $\beta$PMG is
perhaps less interesting than the conclusion by \citet{Mamajek2014-AgeBPMG} that, at least in
application to the $\beta$PMG stars, the Li depletion boundary and isochronal age estimation
techniques are superior to kinematic methods of age determination.''}

All in all, after every source of uncertainties has been taken into account, it seems that the most advisable behaviour is to round up the age estimate and to keep the 20 Myr, and to include a generous error-bar. How to estimate the uncertainty from a realistic point of view is, indeed, a complicated matter. Hopefully, Gaia will contribute to this discussion and help us to reduce the errors and to get a better, accurate age.

%%%%%%%%%%%%%%%%%%%%%%%%%%%%%%%%%%%%%%%%%%%%%%%%%%%%%%%%%%%%%%%%%%%%%%%%%%%%%%%%%%%%%%%%%%%%%
%
\subsection{The cornerstone: the Pleiades\label{subsec:pleiades}}
%
%%%%%%%%%%%%%%%%%%%%%%%%%%%%%%%%%%%%%%%%%%%%%%%%%%%%%%%%%%%%%%%%%%%%%%%%%%%%%%%%%%%%%%%%%%%%%

One very relevant cluster, if not  the most important, is the Pleiades, located at about 130 pc and with an age of 125 Myr.
The Pleiades is known since Antiquity (it is mentioned several times by Homer and the Babylonian named it MUL.MUL) and since then it has been a astronomical milestone.
 We have already discussed the age of this association as derived with different methods, and described in detail the LDB age (subsection \ref{LDB}). The Pleiades distance is another contentious issue and two very divergent values based on different methods.
The  parallax from Hipparcos (\citealp{Perryman97}) gives and updated distance 
of   $120.2\pm1.9$ pc (\citealt{vanLeeuwen2009.1}), whereas
isochrone fitting furnished a value of  $133.5\pm1.2$  pc (\citealt{Pinsonneault1998.1}).
 More recently, 
\citet{Melis2014.1} have derived $136.2\pm1.2$ pc based on an accurate  parallax for four {\it bona fide}  members obtained with the VLBI.
Note that these values should, in principle, correspond to the distance to the cluster center,
 whose  core radius should be  around 3 degrees, which corresponds to 5-6 pc.

 There is an extraordinary amount of works devoted to the Pleiades in the literature but, despite this fact, new cluster members have been
uncovered (\citealt{Sarro2014.1};  \citealt{Bouy2015.2}). 

 Figure \ref{Bouy2015_Pleiades_HR}, taken from  \citet{Bouy2015.2}, displays the HRD for the Pleiades.
The original caption says: {\it ``Hertzsprung-Russell diagram of the Pleiades (black dots) with the
\citet{Allard2014.1} (red), \citet{Bressan2012-PARSECtracks} (light green) and \citet{Siess2000.1} (light blue) models, as well as Praesepe's sequence as reported in
\citet{Kraus2007-PraesepeComa} (green).''}
It contains 2109 members and 812 are new, as derived from  proper motions. As a matter of fact, even a 
reanalysis of the Tycho catalogue (\citealt{Hog2000.1}), which covers the Pleiades bright end of the cluster, has revealed 83 stars with high membership probability.

In any case, the confrontation between the photometric and astrometric database created by the DANCE project 
(\citealt{Bouy2013.1}) and theoretical models shows that there are significant differences (see, for instance, their figure 7). This fact should affect, certainly, to age determinations. Since age scales depend, one way or another, to the Pleiades chronology, we should be aware of this essential problem. Empirical cluster sequences, such as those displayed in their figure 6, should help in order to alleviate this situation. Comparison with ``twin'' clusters with similar ages, such as M35 (also observed with Kepler K2) and Blanco 1, would also be very productive and interesting.

%%%%%%%%%%%%%%%%%%%%%%%%%%%%%%%%%%%%%%%%%%%%%%%%%%%%%%%%%%%%%%%%%%%%%%%%%%%%%%%%%%%%%%%%%%%%%
%
\subsection{Solar-age cohort: M67 \label{subsec:M67}}
%
%%%%%%%%%%%%%%%%%%%%%%%%%%%%%%%%%%%%%%%%%%%%%%%%%%%%%%%%%%%%%%%%%%%%%%%%%%%%%%%%%%%%%%%%%%%%%

We have already met the M67 open cluster  (NGC~2682) in the context of the WD dating.
 It is relatively close (m-M=9.60,
\citealt{Nissen1987-M67}, and well populated. Its members are
distributed in relatively small area on the sky ($\sim1$ square deg),
making photometric studies
easier than in the case of very compacted clusters or dispersed
ones. But the most interesting characteristics are, perhaps, its
age, close to that of the Sun, and its metallicity, also solar.
Using different methods, the age of the cluster has been
estimated in the range  3-5 Gyr  (\citealt{Yadav2008-M67-Catalog}).
Therefore, this cluster can be used to compare how different
properties, such as rotation, stellar activity, lithium
abundances, behave in stars  of different mass with the solar
age, helping to understand the evolutionary status and structure
of the Sun.  

In particular, lithium abundances of solar-type stars
belonging to this cluster
have been studied in detail (\citealt{Hobbs1986-Lithium-M67} ; \citealt{GarciaLopez1988-Li-M67} ; 
\citealt{Balachandran1995-Li-F-M67}; \citealt{Pasquini2008-Lithium-M67})
 presented a very detailed study of the lithium in
this cluster, including the effect of rotation in tidally locked binary systems evolving off the main-sequence (\citealt{Barrado1997-Lithium-Binarity-M67}). As we have seen, M67 is within one of the the Kepler K2 field, so eventually accurate rotational periods will be available. Thus, it will add key information for the gyrochronology.
It is also very well suited for asteroseismological studies (although its members are quite faint). 
Thus, its role as a fundamental calibrator is  assured.

This is just a sort list, but stellar associations such as ONC ($\sim$1 Myr), IC2602/IC2391 ($\sim$50 Myr), the Hyades ($\sim$625 Myr), NGC752 ($\sim$1.6 Gyr) or Ruprecht 147 (aka NGC 6774, $\sim$4 Gyr), among others (see Figure \ref{Barrado_Age_SteppingStones_EES2015.jpg}) would serve, as we have pointed out, as stepping stones towards a age scale system. As a summary, we need all of them to cross this ``dangerous'' waters and reach the other side, the ``promised land'' of accurate ages.

%%%%%%%%%%%%%%%%%%%%%%%%%%%%%%%%%%%%%%%%%%%%%%%%%%%%%%%%%%%%%%%%%%%%%%%%%%%%%%%%%%%%%%%%%%%%%  SECTION
%%%%%%%%%%%%%%%%%%%%%%%%%%%%%%%%%%%%%%%%%%%%%%%%%%%%%%%%%%%%%%%%%%%%%%%%%%%%%%%%%%%%%%%%%%%%%
%%%%%%%%%%%%%%%%%%%%%%%%%%%%%%%%%%%%%%%%%%%%%%%%%%%%%%%%%%%%%%%%%%%%%%%%%%%%%%%%%%%%%%%%%%%%%
%
\section{Conclusions\label{sec:conclusions}}
%
%%%%%%%%%%%%%%%%%%%%%%%%%%%%%%%%%%%%%%%%%%%%%%%%%%%%%%%%%%%%%%%%%%%%%%%%%%%%%%%%%%%%%%%%%%%%%
%%%%%%%%%%%%%%%%%%%%%%%%%%%%%%%%%%%%%%%%%%%%%%%%%%%%%%%%%%%%%%%%%%%%%%%%%%%%%%%%%%%%%%%%%%%%%
%%%%%%%%%%%%%%%%%%%%%%%%%%%%%%%%%%%%%%%%%%%%%%%%%%%%%%%%%%%%%%%%%%%%%%%%%%%%%%%%%%%%%%%%%%%%%

%%%%%%%%%%%%%%%%%%%%%%%%%%%%

We have revisited methods most frequently used  to derive stellar ages and discuss some of the pros and many drawbacks. After this overview, there are different factors which should be taken into account:

  \begin{enumerate}
     \item  Anchors: We have very few absolute values for the age scale. Depending on how strict we are, perhaps only one.     
     \item Most phenomena we are dealing with are outside the age range defined by the anchors. We, thus, are to certain extent extrapolating. Beware!                                          
     \item  Primary indicators: Different evolutionary codes which produce different sets of models. Even for the same set, chemical composition and the detailed physics can modify the results.                                                    
     \item  Primary indicators: Conversions observation-theory, a thorny problem.                                     
     \item  Primary indicators: Can we assume coevality? Stellar associations show an apparent age spread but this might be related with second-order parameters (rotation, activity, magnetic fields, accretion, binarity, etc), which can modify the stellar properties.                                            
     \item  Secondary indicators: Do we really know the ages of the “well-known” SFR and clusters? They are used to calibrate secondary (or empirical) indicators and  if they are biased we can miss the target completely.
     \item  Secondary indicators: Do we really understand the properties we use, such as stellar activity?  
     \item  Secondary indicators: Again, coevality. Can we assume it on members of cluster, star forming regions and, specially, moving groups?                                                  
     \item  Consistency: Different masses and ages are derived by different methods.                               
     \item  Consistency: Each age value is linked to models and to specific scales. This is something which cannot be forgotten when comparing results.                 
  \end{enumerate}

Almost finishing, we would like to list few additional suggestions and caveats:

  \begin{enumerate}
     \item  Models are complex beasts with a lot of physics inside, never forget.  
     \item  Be realistic with error-bars. It is always safer.                             
     \item  Keep in mind the difference between precision and accuracy.                              
     \item  Search ``below the carpet''. Any estimate has interesting details and subtleties which are not always obvious.                               
     \item  From a more general perspective, it is advisable to read papers, specially old ones.                        
     \item   Give credit to previous results, even if they look ``old''.                       
     \item  Be skeptical. This is one of the most important tools scientists have.                                           
  \end{enumerate}

All these caveats and limitations should emphasize that still there is a lot of work to be done, many problems await resolutions. But this situation should not hide the fact  the extraordinaries advances we, as a community, have achieved. We know many things about stellar evolution and its scales, now we have to improve the details. We have already have sketched, in Figure \ref{Barrado_Age_SteppingStones_EES2015.jpg}, how to carry out a complete program to create a sound age stairway with reliable age scales for different masses and evolutionary stages.

Just to finish, a quote from
% Archimedes of Syracuse, written in the III century BCE: {\it ``Those who claim to discover everything but produce no proofs of the same may be confuted as having actually pretended to discover the impossible''.}\footnote{``On spirals'', translated by T. L. Heath, ``The Works of Archimedes'', University (1897).}
 Jean-Baptiste Delambre, from what is said to be the first modern history of astronomy: {\it ``The history owns nothing to the dead except the truth''}\footnote{``L'historien ne doit aux morts que la v\'erite'', from ``Histoire de l'astronomie moderne'',  1821.} (\citealt{Delambre1821-HistoryModernAstronomy}).
After all, science is itself a stairway were all steps, any contribution, even if they are very short, are very important.

\begin{acknowledgements}
Many thanks to both the SOC and the LOC: E. Moraux, C. Charbonnel, Y. Lebreton, F. Martins, A. Robin, M.-H. Sztefek, F. Maia, I. Joncour, and  B. Dintrans. V. Trimble and A. Bayo have made very useful suggestions. This work has been supported by  Spanish grant  AYA2012-38897-C02-01.
\end{acknowledgements}

% \newpage
$\,$
% \newpage

%%%%%%%%%%%%%%%%%%%%%%%%%%%%%%%%%%%%%%%%%%%%%%%%%%%%%%%%%%%%%%
%%%%%%%%%%%%%%%%%%%%%%%%%%%%%%%%%%%%%%%%%%%%%%%%%%%%%%%%%%%%%%
%%%%%%%%%%%%%%%%%%%%%%%%%%%%%%%%%%%%%%%%%%%%%%%%%%%%%%%%%%%%%%
%%%%%%%%%%%%%%%%%%%%%%%%%%%%%%%%%%%%%%%%%%%%%%%%%%%%%%%%%%%%%%
%%%%%%%%%%%%%%%%%%%%%%%%%%%%%%%%%%%%%%%%%%%%%%%%%%%%%%%%%%%%%%
\bibliographystyle{aa} %aa.bst
%\bibliography{0_LDB}
\bibliography{00_bibliography}

%%%%%%%%%%%%%%%%%%%%%%%%%%%%%%%%%%%%%%%%%%%%%
%%%%%%%%%%%%%%%%%%%%%%%%%%%%%%%%%%%%%%%%%%%%%
%
%  Figures
%
%%%%%%%%%%%%%%%%%%%%%%%%%%%%%%%%%%%%%%%%%%%%%
%%%%%%%%%%%%%%%%%%%%%%%%%%%%%%%%%%%%%%%%%%%%%
%%%%%%%%%%%%%%%%%%%%%%%%%%%%%%%%%%%%%%%%%%%%%
%%%%%%%%%%%%%%%%%%%%%%%%%%%%%%%%%%%%%%%%%%%%%
%%%%%%%%%%%%%%%%%%%%%%%%%%%%%%%%%%%%%%%%%%%%%
\clearpage

%%%%%%%%%%%%%%%%%%%%%%%%%%%%%%%%%%%%%%%
%%%%%%%%%%%%%%%%%%%%%%%%%%%%%%%%%%%%%%%
%%%%%%%%%%%%%%%%%%%%%%%%%%%%%%%%%%%%%%%
%%%%%%%%%%%%%%%%%%%%%%%%%%%%%%%%%%%%%%%
%%%%%%%%%%%%%%%%%%%%%%%%%%%%%%%%%%%%%%%
%%%%%%%%%%%%%%%%%%%%%%%%%%%%%%%%%%%%%%%
%%%%%%%%%%%%%%%%%%%%%%%%%%%%%%%%%%%%%%%
%%%%%%%%%%%%%%%%%%%%%%%%%%%%%%%%%%%%%%%
%%%%%%%%%%%%%%%%%%%%%%%%%%%%%%%%%%%%%%%
%%%%%%%%%%%%%%%%%%%%%%%%%%%%%%%%%%%%%%%

\end{document}